\documentclass[3p, review]{elsarticle}
\pdfoutput=1
\newcommand{\abbreviations}[1]{%
\begingroup\def\thefootnote{}\footnotetext{\textit{Abbreviations:}#1}\endgroup}

\usepackage{epsfig,epsf}
\usepackage{graphicx,tabularx}
\usepackage{epstopdf}
\usepackage{amsmath, amssymb}
\usepackage{bm}
\usepackage{bbm}
\usepackage{subcaption}
\usepackage{enumitem}
\usepackage[nottoc,numbib]{tocbibind}
\usepackage{color}

\newcommand{\resp}[1]{\textcolor{black}{#1}}

\allowdisplaybreaks[1]

\title{Nanocrystal Programmable Assembly Beyond Hard Spheres (or Shapes) and Other (Simple) Potentials}
\author[1]{Alex Travesset\corref{cor1}}%
\affiliation[1]{organization={Iowa State University and Ames Lab},
addressline={Department of Physics and Astronomy},
postcode={50010},
city={Ames, IA},
country={USA}}

\begin{document}
\begin{frontmatter}

\begin{abstract}

Ligands are the key to almost any strategy in the assembly of programmable nanocrystals (or nanoparticles) and must be accurately considered in any predictive model. Hard Spheres (or Shapes) provide the simplest and yet quite successful approach to assembly, with remarkable sophisticated predictions verified in experiments. There are, however, many situations where hard spheres/shapes predictions fail. This prompts three important questions: {\em In what situations should hard spheres/shapes models be expected to work?} and when they do not work, {\em Is there a general model that successfully corrects hard sphere/shape predictions?} and given other successful models where ligands are included explicitly, and of course, numerical simulations, {\em can we unify hard sphere/shape models, explicit ligand models and all atom simulations?}. The Orbifold Topological Model (OTM) provides a positive answer to these three questions. In this paper, I give a detailed review of OTM, describing the concept of ligand vortices and how it leads to spontaneous valence and nanoparticle ``eigenshapes'' while providing a prediction of the lattice structure, without fitting parameters, which accounts for many body effects not captured \resp{by} (two-body) potentials. I present a thorough survey of experiments and simulations and show that, to this date, they are in full agreement with the OTM predictions. I conclude with a discussion on whether NC superlattices are equilibrium structures and some significant challenges in structure prediction.

\end{abstract}
\end{frontmatter}
\abbreviations{NC, Nanocrystal; HS, Hard Sphere; BNSL, Binary Nanocrystal Super Lattice; PF, Packing Fraction; OPM, Optimal Packing Model; OTM, Orbital Topological Model; OCM, Optimal Cone Model; MOLT, Molecular Theory; WHAM, Weighted Histogram Analysis Method; PMF, Potential of Mean Force; fcc, face centered cubic; bcc, body centered cubic; bct, body centered tetragonal.}
\tableofcontents
\section{Introduction}

Nanocrystals (NCs), also referred to as nanoparticles, are conceptualized as "Programmable Atom Equivalents"~\cite{Macfarlane2011}, i.e. ``large" atoms with many degrees of freedom: size, shape, chemical composition\cite{Solomon2007}, and type and structure of capping ligands\cite{Bolesa2016}. Motivation to understand and predict NC assembly is driven by both fundamental science and new applications as described in many previous references, for example \cite{Kovalenko2015,Begley2019,Kagan2016,Boles2016,LiSun2022,Jansen2023}. 

\begin{figure}
    \centering
    \includegraphics[width=1\textwidth]{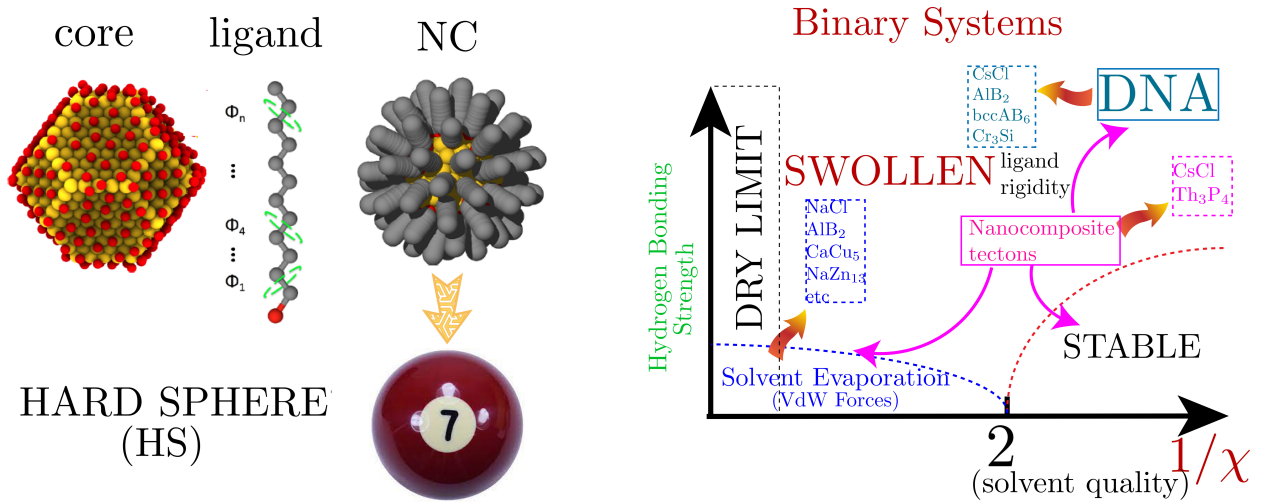}
    \caption{Example of a NC: core consists of Au(yellow) and S(red) \resp{with the ligand a simple hydrocarbon. schematic depiction of the NC modeled} as hard spheres (billiard balls). On the right is the relation between different assembly strategies in binary systems as discussed in the text.  The bccAB$_6$ phase is also labeled Cs$_6$C$_{60}$ in many references. }\label{fig:intro}
\end{figure}

Many strategies have been developed to assemble NCs into materials. Most of them, but not all, as with liquid crystals\cite{Ghosh2022} for example, consist in functionalizing NC cores with appropriate ligands, as depicted in Fig.~\ref{fig:intro}. Ligands are necessary because the bare interactions are not specific enough for a controlled assembly, so their role is to make NCs stable in a solution and enable assembly into a material in a controlled and reproducible manner. There are many strategies in which the assembled NC material is swollen; that is, the solvent is part of the final structure. This is the case for DNA assembly\cite{Mirkin1996, Alivisatos1996, Park2008, Nykypanchuk2008}, Nanocomposite Tectons\cite{Santos2019}, Interpolymer complexation\cite{Nayak2019} and many others. In other cases, the end process in assembly is a solvent free structure, that is, a dry system. This is the case in solvent evaporation\cite{Whetten1996,Korgel1998,
Shevchenko2002} or tuning ionic strength\cite{Zhang2017a}. 

Binary systems, i.e consisting of two NC species, have very sophisticated phase diagrams, so they provide an excellent paradigm in which to put model predictions to a test. As an example, Fig.~\ref{fig:intro} shows a schematic representation of the different phases parameterized by two coordinates: the strength of hydrogen bonds and the quality of the solvent. In a good solvent ($\chi > \frac{1}{2}$) and weak hydrogen bonds, NCs are stable: NCs repel each other and remain in a liquid-like state. As the strength of the hydrogen bonds increases, the ensuing attraction drives the assembly into several phases for DNA assembly\cite{Macfarlane2011} and different ones for nanocomposite tectons\cite{Santos2019}, with the differentiating factor being the flexibility of the ligand\cite{Xia2022a}. Yet, another set of phases are obtained in poor solvent and no hydrogen bonds\cite{Shevchenko2006, Shevchenko2006a,Ye2015}. The example in Fig.~\ref{fig:intro} illustrates that different strategies are related and may give rise to new opportunities, such as, for example, dry systems with strong hydrogen bonds.

This paper focuses on the assembly of solvent-free structures. Herein, unless otherwise stated, swollen systems will not be specifically discussed. \resp{Hard Sphere (HS) or hard shape} models have played a central role in describing assembly with significant predictive power, which I review in Sect.~\ref{Sect:HS}. There are other models that explicitly take into account ligands. Particularly relevant for this study are the Optimal Packing Model (OPM)\cite{Landman2004} and the Optimal Cone Model (OCM)\cite{Schapotschnikow2009}. Additionally, numerical simulations have been extensively applied to understand NC assembly and are discussed in great detail in this paper. Yet, simulations by themselves are not enough: At the end of the day, they must be integrated and provide validation for an overarching model that identifies the few relevant parameters that control the dynamics and equilibrium of the system.

\resp{HS or hard shapes do} provide an excellent model for the NC core; see Fig.~\ref{fig:intro}, but it is far from intuitively obvious why they would provide a fair representation of the entire NC, as the ligands appear to be soft and very compressible. There are many other reasons to be skeptical about \resp{hard shape} models. The internal energy of $N$ \resp{hard shape} particles is given by the expression of the ideal gas and the free energy is of the order of $k_B T$ 
\begin{equation}\label{Eq:hs_thermo}
    U=\frac{3}{2} N k_B T \quad , \quad F \approx N k_B T \quad \mbox{NC systems as HS systems}
\end{equation}
with a possible kinetic energy contribution in the internal energy for rotation,
which implies that the internal energy is always constant and phase transitions are entropically driven, with  cohesive energies of the order of $k_B T$ per NC. Numerical simulations, see the discussion in Sect.~\ref{Sect:Simuls}, show that Eqs.~\ref{Eq:hs_thermo} are in complete disagreement for any real NC system. In fact, typical internal energies and free energies for a solid phase are 
\begin{equation}\label{Eq:real_thermo}
    U \approx  -1200 N k_B T \quad , \quad F \approx -500 N k_B T \quad \mbox{ real NC systems},
\end{equation}
that is, almost three orders of magnitude larger than Eq.~\ref{Eq:hs_thermo} (and with opposite signs). In fact, the main contribution to the entropy comes from ligand conformations, while translational (and rotational) contributions, which are the only ones that the \resp{hard shape} models account for, are almost negligible. Given these obvious
discrepancies, one might be tempted to abandon \resp{hard shape models} altogether. Yet, these models are extremely valuable, with a number of sophisticated predictions verified in actual experiments. As happens in many areas of physics, there are aspects that are independent of the details of the microscopic (more properly, nanoscopic) interactions. Basically, \resp{hard shape} models are relevant because they capture the underlying universality of packing with elegant and minimal simplicity\cite{Torquato2018}; that is, model-independent results that optimize the available space.

Therefore, HS \resp{or hard shape} provide a correct first-order approximation to NC assembly structure prediction. It would seem obvious that the next order would simply consist of adding reasonable enthalpic terms. For example, Ref.~\cite{ZhouArya2022} has successfully characterized all possible binary NC structures at an interface (2D) with the addition of relative interaction energy between NCs and the solvent interface. DNA with double-stranded (rigid) ligands is another example in which this approach is very successful\cite{Lin2017}. However, as shown in more detail below, see Fig.~\ref{fig:exp_2d_3d}, when ligands are flexible (and their size is not negligible compared to the core), experiments reveal a dependence of nearest-neighbor lattice constants with the NC coordination, i.e. the number of nearest-neighbors, which in turn points to the presence of many-body effects, which may suggest the need to go beyond pair potentials, requiring a model where ligands are described in a more explicit form.

Ligands are explicitly modeled in OXM models (X = P, C) and therefore provide more detail than HS models. The OPM is very successful in describing the lattice constants in single-component fcc (or bcc) lattices\cite{Landman2004}, while the OCM accurately describes the equilibrium nearest-neighbor separation of two or three isolated NCs\cite{Schapotschnikow2009}. However, the closest-neighbor NC separations fall in between OPM and OCM in most other experimental situations\cite{Boles2016,Coropceanu2019,Moretti2023}. OXM models account for the ligand conformations, so at this point in the discussion they may appear to be conceptually different and unrelated to the HS models.

The previous discussion provides the motivation for this paper: {\em How to bring together HS models, the} OXM (X=P,C) {\em models and all atom simulations} with the goal of developing a general framework for NC prediction. In this paper, I make the case that the Orbifold Topological Model (OTM)\cite{Travesset2017a, Travesset2017b}, provides such a unified model. I will show that at this point in time there is not a single experiment in disagreement. Yet, there are still challenges and aspects that need to be developed towards a completely predictive framework, which are discussed throughout the paper.

The end result of a NC assembly experiment is usually a lattice of NCs, i.e. a superlattice, see simulation snapshot in Fig.~\ref{fig:simul_sc} or electron microscopy images in Fig.~\ref{fig:alb2_experiment}. Ref.~\cite{Dshemuchadse2022} provides a detailed account of the structures, relates them to other soft- and hard-matter systems, and describes strategies to expand their range. Superlattices are basically lattices of crystals (or crystals of crystals), as the NC cores themselves are already crystals. Predicting the superlattices that define equilibrium for given NCs is a fundamental goal. A more modest problem, yet still very important, is given a superlattice consisting of several NCs species, predict its structure: lattice constants, packing fraction, ligand arrangements, etc. {\em the OTM provides a general solution to this problem.}, independent of the microscopic interactions details. To address the more difficult problem of predicting what superlattices define equilibrium, that is, what superlattices should be observed in an actual experiment, a free energy is needed, which OTM does not, in principle, provide. I will discuss exciting recent developments that bring us closer to achieving this goal.

\resp{The main goal of this paper is to present the OTM in the broadest possible context. For this purpose, I provide a number of examples that are worked in full detail, as a way to illustrate how the formulas are applied in actual predictions, which hopefully will facilitate applications of the OTM. I also devote some significant effort in outlining relevant open questions.} The flow \resp{and logic} of the paper starts by discussing \resp{hard shape} models and their triumphs, providing, of course, a detailed account of the failures that motivate the OTM. Then, I introduce the OTM and provide its validation through experiment and simulations while pointing out its most outstanding challenges. Over the years, different papers have reviewed the role of ligands in NC assembly and this is a very large field. Unavoidably, there are many papers that I will for sure unjustly not cite for which I apologize, and different subjects that are not covered here. I refer to Ref.~\cite{Cademartiri2015, Angioletti-Uberti2012,Boles2016,LeeYoon2019,Grzelczak2019,Lee2022,Bhattacharjee2023} for reviews that cover overlapping and complementary topics.

\begin{figure}[htb]
    \centering
    \includegraphics[width=1\textwidth]{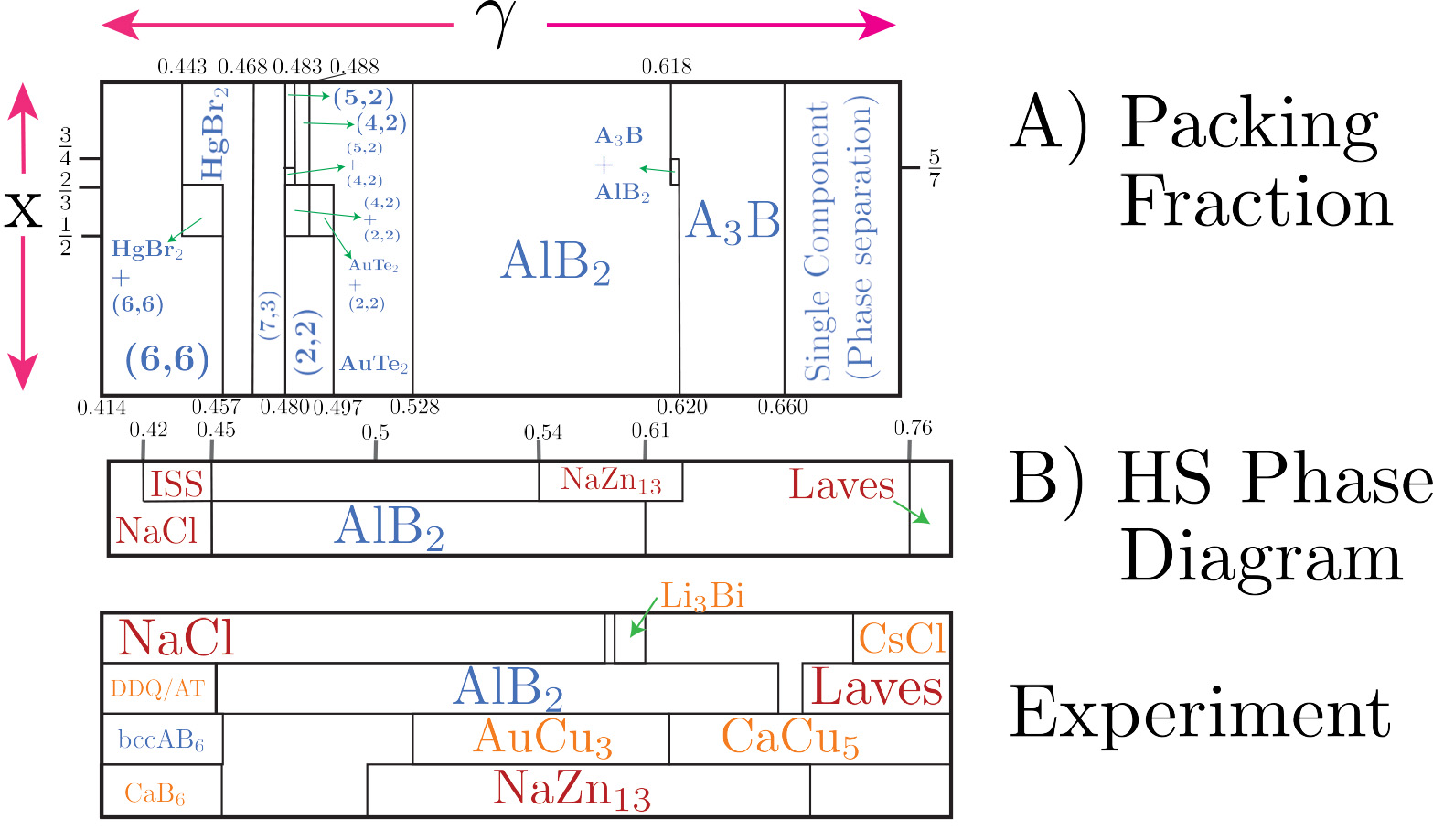}
    \caption{Optimal PF phases are adapted from Ref.~\cite{Hopkins2012}(in blue). The HS Phase diagram is estimated from Ref.~\cite{Dijkstra2015}(in red, phases not present in PF). The experimental data is estimated from the data gathered in Ref.~\cite{Travesset2017a} (Orange for phases not present in neither HS Phase diagram or PF). variables $\gamma,x$ defined in Eq.~\ref{Eq:hs:def_gamma}. Naming of the phases is the same as in the orginal references.  }\label{fig:HS_phase_diagrams}
\end{figure}

\section{Packing and Hard Sphere/Shape Models}\label{Sect:HS}

\subsection{Summary of HS models}\label{SubSect:HS:Models}

I describe a few aspects of HS that are of particular relevance for this paper. I refer to Ref.~\cite{Royall2023}  for a recent authoritative review of the state of the art and many other aspects. 

The Packing Fraction (PF) ($\eta$) is defined by
\begin{equation}\label{Eq:hs:def_eta}
    \eta_{HS} = \frac{\mbox{volume occupied by particles}}{\mbox{total volume}} \stackrel{s.c. \ spheres}{=} \frac{\pi}{6} \frac{N}{V} d_{HS}^3 \ ,
\end{equation}
where the sub-index HS is to emphasize that PF is computed within a HS model. The rhs of the equation illustrates the case of a single component system consisting of $N$ HS of diameter $d_{HS}$ within a volume $V$. In other cases, for example, binary systems of spherical particles, there are two other thermodynamic coordinates: $\gamma$(relative particle size) and $x$(stoichiometry), defined as
\begin{equation}\label{Eq:hs:def_gamma}
    \gamma = \frac{d_{HS}^B}{d_{HS}^A} < 1 \quad , \quad  x = \frac{N^B}{N^A+N^B}
\end{equation}
where $d_{HS}^i, N^i$ $i=A,B$ are the diameters and numbers of the large(A) and small(B) particles. 

There are two types of problems that are particularly relevant for NC assembly:
\begin{enumerate}[label=\Alph*]
\item{) Maximum PF ($\eta$) for given parameters, e.g ($\gamma, x$)}
\item{) Phase diagram as a function of $\eta$(or pressure, which is the conjugate thermodynamic variable) and other parameters, e.g ($\gamma, x$).}
\end{enumerate}

For single component spheres the solution to problem A is degenerate, as both fcc and hcp have the highest PF $\eta_{HS}=\frac{\pi\sqrt{2}}{6}\approx 0.74048$. Solution for problem B) contains liquid and crystalline phases at a transition $\eta_{HS} \approx 0.5$, glass phases for increasing PF. Near the highest PF, fcc is favored over hcp\cite{Mau1999, Elser2014}. The case of binary systems of spherical particles is illustrated in Fig.~\ref{fig:HS_phase_diagrams} for $\gamma > 0.41$. The phases with the highest PF~\cite{Filion2009,Hopkins2012} are many and diverse. The phase diagram, obtained by minimizing the free energy (or maximizing the entropy, as the internal energy is given by the monovalent ideal gas formula, see Eq.~\ref{Eq:hs_thermo},) is (very) schematically constructed from the data provided by Marjolein Dijkstra in Ref.~\cite{Dijkstra2015}. 

It would seem reasonable to expect that close to the maximum PF, the solutions of problems A and B would be equivalent. In other words, the maximum PF phases would have the lowest free energy. However, this expectation is not fulfilled. This is clear for binary systems in Fig.~\ref{fig:HS_phase_diagrams}, as only the AlB$_2$ phase appears as a solution of both problems A and B. Further evidence is provided from other rigorous results with more complex shapes\cite{Haji-Akbari2009, Agarwal2011, VanAnders2015, Cersonsky2018, Geng2019, Geng2021}. 

The dichotomy between the minimum of free energy and the highest PF has motivated to consider the assembly of HS in terms of ``eigenshapes''. That is, the system is described by an effective shape that completely packs space. This shape is obviously different from the actual HS, and there are ways to actually calculate it\cite{Vo2022}. In the following, I show that it is possible to generalize this idea when ligands are included, although in that case, the eigenshape is not a property of the NC only and depends on the environment.

\begin{figure}
    \centering
    \includegraphics[width=1\textwidth]{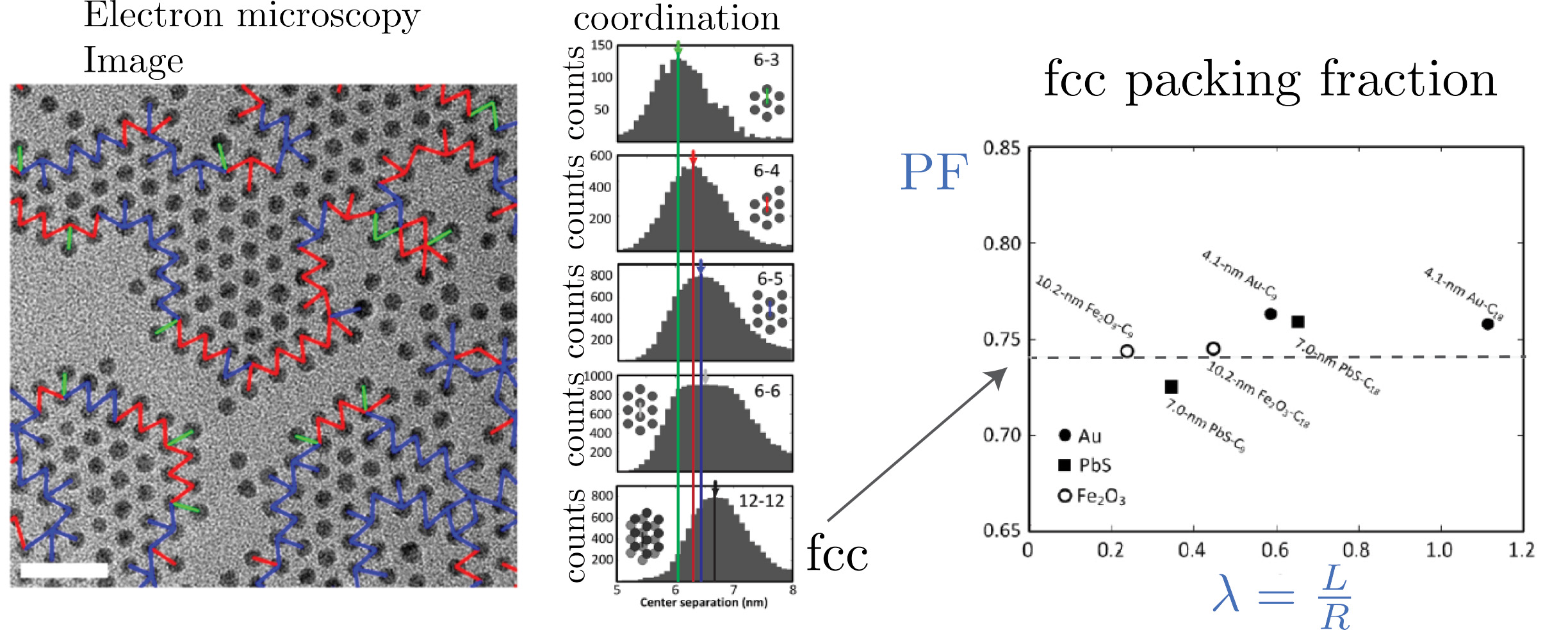}
    \caption{Electron microscopy image of a single component system (4.1 nm Au-C18 NC) deposited on a substrate. From there, NCs with coordination 3,4,5,6 are identified and histogrammed from where the nearest neighbor separation is obtained. It is apparent that the nearest neighbor distance increases with coordination and that for coordination 6 the results are almost equivalent with fcc.
    Also shown are the results of a similar experiment with fcc (coordination 12). The PF of many different single component NCs is completely consistent with the fcc HS PF. Results are from Ref.~\cite{Boles2015} (permission from D. Talapin).}\label{fig:exp_2d_3d}
\end{figure}

\subsection{Mapping NC to HS}\label{SubSect:HS:Mapping}

\begin{figure}[htb]
    \centering
    \includegraphics[width=1\textwidth]{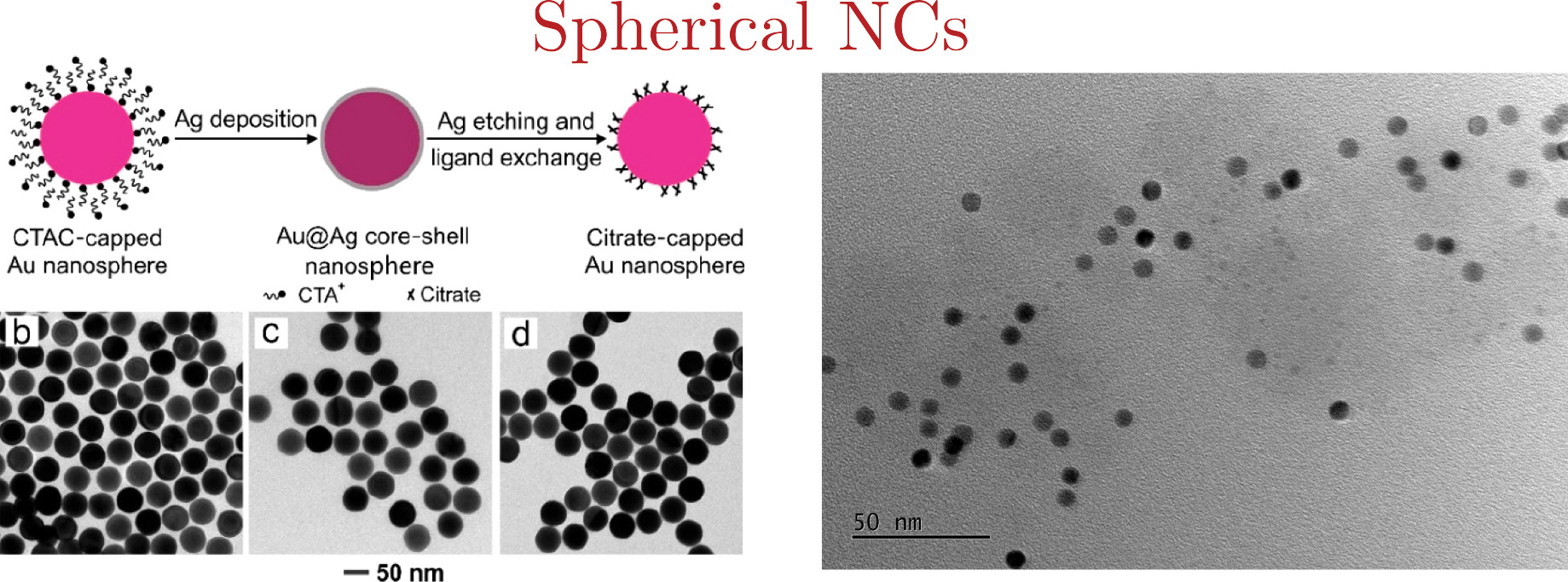}
    \caption{Spherical NCs are possible to synthesize. These images are from Ref.~\cite{ZhouHuo2018}. On the right 6 nm spherical NCs from the Zhou lab. Provided by Shan Zhou and printed with permission. }\label{fig:spherical}
\end{figure}

NC core shapes are usually polyhedra, although spherical cores do exist; see Fig.~\ref{fig:spherical}. In some cases \resp{NCs with polyhedral cores may be approximated} as spherical because the flexibility of the ligand helps round off edges and corners. For a spherical NC, the first issue is {\em what is the correct effective HS diameter, which must include a contribution from the ligands}.

Early studies \cite{Shevchenko2006, Shevchenko2006a, Boles2016} defined the NC HS diameter as the lattice constant of the resulting hexagonal single-component lattice, deposited on a solid substrate. Due to the interactions of the bottom NC ligands with the solid substrate, and the unconstrained ligands at the top, this definition is not optimal; this is apparent from Fig.~\ref{fig:exp_2d_3d}, where the maximum of the histograms for coordination 6 is broad and slightly below the one for fcc. In Ref.\cite{Zha2020} we showed that the \textit{Optimal Packing Model} (OPM) \cite{Landman2004} gives the best possible definition. Therefore, the HS diameter $d_\text{HS}$ of a spherical NC with a core radius $R$ and grafting density $\sigma$ is defined as
\begin{equation}\label{Eq:hs:OPM}
\tau(\lambda,\xi) = (1+3\xi\lambda)^{1/3} 
\end{equation}
with
\begin{equation}
\tau \equiv \frac{d_\text{HS}}{2R} \quad , \quad \xi \equiv \frac{\sigma}{\sigma_\text{max}} \quad \mbox{and} \quad \lambda \equiv \frac{L}{R}
\end{equation}
where $\sigma_\text{max}=\frac{1}{A_0}$ is the maximum grafting density of the core (or $A_0$ is the ligand footprint, which is $d^2$ in the language of polymer theory\cite{RubinsteinBook2003}) and $L$ is the maximum (extended) ligand length ($R_{max}$ in the language of polymer theory) shown schematically in Fig.~\ref{fig:OTM_example}. The parameter $\lambda$ is sometimes referred as the ``softness''. In the language of polymer physics\cite{RubinsteinBook2003}, the OPM Eq.~\ref{Eq:hs:OPM} defines the radius of a spherical brush where the ligands are in a non-solvent condition. If the spherical NC is divided as non-overlapping cones, see a cone shown in Fig.~\ref{fig:OTM_example}, then the OPM results follows by assuming that the ligand entirely fills out its own available cone. 

A very important property that cements Eq.~\ref{Eq:hs:OPM} as the correct definition of the HS diameter is that the resulting PF is exactly the same as the corresponding HS PF. This is proved as follows: The amount of space occupied by matter for a single NC is (note that $A_0 L$ is the volume occupied by one ligand)
\begin{equation}\label{Eq:hs:vol_matter}
V_{matter} = V_{core}+V_{ligands}=\frac{\pi}{6} (2R)^3+ 4\pi R^2 \sigma A_o L = \frac{\pi}{6} (2R)^3\left(1+3\xi\lambda\right) = \frac{\pi}{6} (2R)^3 \tau^3 = \frac{\pi}{6} d_{HS}^3, 
\end{equation}
that is, the matter, i.e. the NC, occupies the same volume as a HS of diameter $d_{HS}$\cite{Boles2015,Travesset2017b}. Therefore for a BNSL with $n_A,n_B$ NCs in the Wigner cell, the PF is
\begin{eqnarray}\label{Eq:hs:OPM:pf}
    \eta = \frac{V_{matter}}{V}=\frac{V_A + V_B}{V_{WS}} \stackrel{superlattice}{=} \frac{\pi}{6}\frac{ n_A (d^A_{HS})^3 + n_B(d^A_{HS})^3}{V_{WS}} \stackrel{HS \ PF \ definition}{=} \eta_{HS}(\gamma) \ ,
\end{eqnarray}
where $V_{WS}$ is the volume of the Wigner-Seitz cell.
This can easily be generalized to any other configuration. 

If one accepts the OPM as the correct definition of the HS, the structure of the phase: lattice constant, nearest-neighbor separations, PF or ligand conformations are quantitatively predicted, see Sect.~\ref{SubSect:HS:describe:exp} for an example worked out in full detail. The other conclusion that follows is that if the NCs are packed as HS, where the nearest-neighbor distances are determined by $d^A_{HS},d^B_{HS}$, there will be a fraction $1-\eta_{HS}$ of empty space within the superlattice. \resp{In SubSect.~\ref{Subsect:incompressibility} I discuss where this free space is located. Regardless, this free space may be used for NCs to be closer} than the distances that follow from HS. The closest approach that could be theoretically achieved is if ligands completely fill any empty space, which would require a NC diameter given by
\begin{equation}\label{Eq:hs:OCM}
d_{OCM} = \eta^{1/3}_{HS} d_{HS} < d_{HS} \ \rightarrow \eta_{OCM}=1 > \eta_{HS} \quad ,
\end{equation}
which defines the Optimal Cone Model (OCM) diameter\cite{Schapotschnikow2009}. As I show below, the actual effective NC diameter lies in between the HS (or OPM) and OCM values, see also the simulations in Fig.~\ref{fig:bnsl_summary}. \resp{Rigorously, $\eta=1$ is not attainable. For example, with a simple model where the core is a sphere of radius $R$ and the monomers forming the ligands are small spheres of radius $R_m << R$, see Fig.~\ref{fig:incompres}, the maximum achievable PF is 
$\eta = \eta_{fcc}+(1-\eta_{fcc})\eta_{fcc} \approx 0.94$. The issue of the maximum PF will be re-analyzed in the discussion of the incompressibility condition SubSect.~\ref{Subsect:incompressibility}
mean field models, see also SubSect.~\ref{Subsect:MOLT}.}

The generalization of Eq.~\ref{Eq:hs:OPM} to NC cubes has been developed\cite{HallstromMe2023}. If the core has edge length $l_E$, the \resp{equivalent hard cube} edge $l_{HS}$ that includes the contribution of the ligands is given as
\begin{equation}\label{Eq:hs:cube:edge_hs}
    l_{HS}=l_E\left(1+6\xi\lambda\right)^{1/3} \ ,
\end{equation}
note the appearance of a factor $6$, where the parameters $\xi$ and $\lambda$ are defined in the same way as Eq.~\ref{Eq:hs:def_gamma}:
\begin{equation}\label{Eq:hs:cube:xi_lambda}
\xi = \frac{\sigma}{\sigma_{Max}} \quad , \quad \lambda = \frac{L}{l_E} .
\end{equation}

The validity of OPM and OCM as the length of the ligand increases, reaching the polymer limit, is discussed in SubSect.~\ref{SubSect:Simuls:polymer}. Related models have been developed in this limit: Ref.~\cite{Midya2021} is, in fact, a generalization of OCM for spherical NCs functionalized with long polymers. The $n$ monomers of each ligand are divided as $n=n_{dry}+n_{inter}$, so that $n_{dry}$ of them are within a size $h_{dry}$ and are described by the OPM formula Eq.~\ref{Eq:hs:OPM}, and $n_{inter}$ are shared among all nearest-neighbor NCs by an ``interpenetrating'' layer $h_{inter}$, where each ligand has the ideal chain size $\propto n^{1/2}_{inter}$ within. The model assumes an interpenetration layer of dimension $h_{inter}$. With respect to the structure of the ligands, the model is equivalent to OCM if $h_{inter}/h_{dry}<<1$ or, what is the same if the vortices, see Fig.~\ref{fig:simul_clusters} are effectively two-dimensional. However, the lattice constant (or nearest-neighbor distance) is entirely determined by space filling constraints, \resp{i.e, the incompressibility condition Eq.~\ref{Eq:incompress:sc}, which is discussed this further in SubSect.~\ref{Subsect:incompressibility}}. NC lattice constants for short and moderately long ligands are consistent with OCM and space-filling constraints for low coordination only. Simulations in Ref.~\cite{HallstromMe2023} have analyzed the overlap region for cubic NCs, showing a finite overlap region so that $h_{inter}$ is finite, although the ligands are too short to have a measurable effect.

\subsection{AlB$_2$ is a paradigmatic HS phase}\label{SubSect:HS:describe:exp}

\begin{figure}[htb]
    \centering
    \includegraphics[width=1\textwidth]{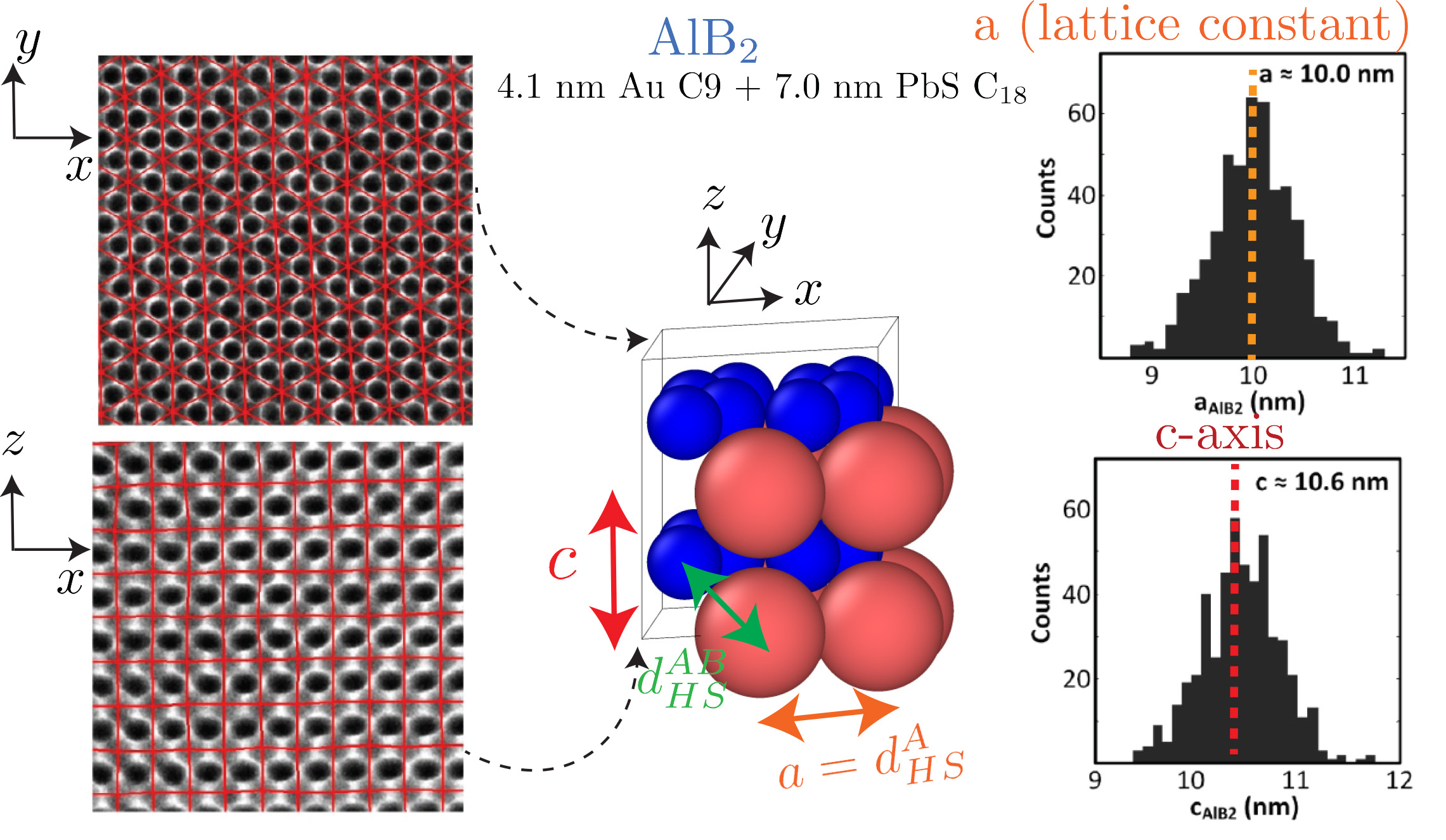}
    \caption{Electron Microscopy images of the AlB$_2$ BNSL as reported in Ref.~\cite{Boles2015} for the $xy$ plane(top) and the $zx$ plane bottom. The AlB$_2$ lattice (for the corresponding $\gamma=0.57$) is shown with spheres not at scale for proper visualization. The measurement of NC distances is performed and histogrammed (on the right) and in this way, the lattice constant $a$ and c-axis $c$ are determined. (permission from D. Talapin)}\label{fig:alb2_experiment}
\end{figure}

It is illustrative to show that experimental measurements allow a very detailed comparison with theoretical predictions.
Fig.~\ref{fig:alb2_experiment} shows electron microscopy images for the AlB$_2$ BNSL that were reported in Ref.~\cite{Boles2015}. As discussed, the HS diameters given by Eq.~\ref{Eq:hs:OPM} are
\begin{eqnarray}\label{Eq:hs:alb2:opm}
    d_{HS}^A = 10.04 \mbox{ nm} & 2R =7.0 \mbox{ nm }, n=18 \nonumber\\
    d_{HS}^B = 5.75 \mbox{ nm} & 2R = 4.1 \mbox{ nm }, n=9 \ ,
\end{eqnarray}
here $n$ is the number of carbons in the ligand chain. The maximum extension $L$ of the ligands are given (in nm) either 
\begin{equation}\label{Eq:hs:L:estimate}
    L = 
    \left\{
    \begin{array}{c c}
    0.12 (n+1) & \text{from Ref.~\cite{Boles2015}}  \\
    0.128 n + 0.2 & \text{from Ref.~\cite{Travesset2017a}}
    \end{array}
    \right. \ .
\end{equation}

The second formula has a better theoretical justification and for ligands like polystyrene it gives more accurate results, but for simple hydrocarbon ligands both give completely consistent results. In this review I will follow the first formula for hydrocarbon ligands as it seems to be more widely used in experiments. The HS diameters have used $\xi=1$, consistently with the grafting density reported in the experiments\cite{Boles2015}. Therefore, this BNSL occurs at a value of $\gamma$, see Eq.~\ref{Eq:hs:def_gamma}
\begin{equation}
\gamma = \frac{d_{HS}^B}{d_{HS}^A}=0.57
\end{equation}
\resp{The AlB2 BNSL has space group symmetry is P6/mmm(191), with Wyckoff positions 1a and 2d. In cartesian coordinates this amounts to the the large A NC(red) and a small B NC(blue) NC in Fig.~\ref{fig:alb2_experiment} located at}
\begin{eqnarray}\label{Eq:hs:alb2:wyckoff}
\mbox{Position A}&:& (0,0,0), \quad \resp{\mbox{Wyckoff position 1a} \left(0, 0, 0\right)}\\\nonumber
\mbox{Position B}&:& (0, \frac{a}{\sqrt{3}},\frac{c}{2}), \quad \resp{\mbox{Wyckoff position 2d} \left(\frac{1}{3},\frac{2}{3},\frac{1}{2}\right)}
\end{eqnarray}
and also, as the figure illustrates, for this value of $\gamma$, the A and B NCs are in contact (or ``kissing'') so the distance between A and B is given by $d^{AB}_{HS}=\frac{1}{2}\left(d_{HS}^A+ d_{HS}^B \right)$. Then from Fig.~\ref{fig:alb2_experiment} and Eq.~\ref{Eq:hs:alb2:opm} and \ref{Eq:hs:alb2:wyckoff}, we have a pure theoretical prediction, because I have assumed that the NCs are HS (through the OPM formula) and the geometrical relations of the BNSL without fitting or measured experimental parameters. It is
\begin{eqnarray}\label{Eq:hs:alb2:c_a}
   && \mbox{(Theory), HS through OPM}  \nonumber \\*
  c = 2\sqrt{(d^{AB}_{HS})^2-\frac{(d_{HS}^A)^2}{3}} &=&  10.7 \mbox{ nm}  \quad \\* \nonumber
  a = d_{HS}^A &=& 10.0 \mbox{ nm} \\* \nonumber
  & & \mbox{(Experiment), Electron Microscopy, Fig.~\ref{fig:alb2_experiment}}\\* \nonumber
   c&=&10.6 \mbox{ nm}  \quad  \\* \nonumber
    a&=&10.0 \mbox{ nm} \ ,
\end{eqnarray}
and I compare these results with the ones obtained from the experiment, the orange and red dashed drawn lines from the histogram obtained from electron microscopy in Fig.~\ref{fig:alb2_experiment}.

The agreement between theory and experiment is within an accuracy of less than 1$\%$. The experimental PF follows from the measured lattice constants and the amount of matter, i.e. core and ligands within the BNSL unit cell. It gives
\begin{equation}
    \eta_{exp} = \frac{\mbox{Volume Matter}}{\sqrt{3} a^2 c/2}=\eta_{HS}(\gamma=0.57)=0.78 \ ,
\end{equation}
where the last identity follows because the general result Eq.~\ref{Eq:hs:OPM:pf} applies in this case. The important point here is that the available experimental data allows precise and quantitative comparisons with theoretical models. One may conclude from this example that HS provides a successful description to the point that HS is all we need. Certainly, if HS models were always to work this well, there is no need to work much further, perhaps adding an enthalpic term as outlined in the introduction. Yet, the same methods applied to other BNSLs, as shown, for example, in Fig.~\ref{fig:exp_theory_comp} exhibit clear discrepancies between HS and experiments. What the results in Fig.~\ref{fig:alb2_experiment} tell us is that these discrepancies reflect real physical effects and motivate us to develop a theory that goes beyond HS.

\subsection{How well HS predictions compare to experiments}\label{SubSect:HS:Experiments}

The definition of the HS diameter in SubSect.~\ref{SubSect:HS:Mapping} combined with the detailed phase diagrams (or packing fractions) from SubSect.~\ref{SubSect:HS:Models} allow rigorously establishing the range of validity of the HS to predict the assembly of NC by solvent evaporation. Although perfectly spherical NCs are available, see Fig.~\ref{fig:spherical}, most NCs are polyhedral in nature, yet ligands help round off edges and are well described as spherical, see SubSect.~\ref{SubSect:otm:beyond_spheres} and SubSect.~\ref{Subsect:MOLT} for a more detailed discussion. Therefore, failures of HS models will highlight missing physics and help develop accurate models that minimally extend HS.

{\em 2D single-component spherical NCs}, see Fig.~\ref{fig:exp_2d_3d}, show that the nearest-neighbor separation strongly depends on coordination: The maximum of histograms, which are identified as the nearest-neighbor distance, increases with coordination, demonstrating a clear breakdown of the HS description. The simulations to be presented in Sect.~\ref{Sect:Simuls} confirm these results; see Fig.~\ref{fig:substrate} and the subsequent discussion. Additional confirmation results are provided in Ref.~\cite{Moretti2023} and include polystyrene ligands, which shows that the breakdown of the HS model is not specific to a particular ligand or NC.

In {\em 3D single component spherical NCs} an fcc phase should be expected. For short ligands $\lambda < \lambda_c\approx 0.7$, experimental results are in agreement\cite{Whetten1996, Whetten1999}. However, for longer ligands $\lambda > \lambda_c$, bcc is found to be more stable. This is explained by the argument put forward in Ref.~\cite{Landman2004}, where next to nearest distances are reached for sufficiently long ligands, making the actual number of nearest neighbors from 12(fcc) to 14(bcc) (8 nearest neighbors plus 6 next to nearest neighbors), the condition that ligands are long enough to reach next to nearest neighbors leads to the equation
\begin{equation}\label{Eq:exp_comp:bcc_fcc}
    \frac{2}{\sqrt{3}} d_{HS} = 2R+2L \rightarrow \frac{2}{\sqrt{3}}\tau(\lambda_c,\xi) = 1+\lambda_c
\end{equation}
with solution $\lambda_c\approx 0.7$ if $\xi=1$, see Eq.~\ref{Eq:hs:OPM} for definitions of the different parameters. At a quantitative level, the fcc lattice constants are accurately predicted by the HS formula Eq.~\ref{Eq:hs:OPM}. However, bcc requires a small correction, departing from HS to slightly compressible potentials, so that its PF becomes the same as fcc, that is,
\begin{equation}\label{Eq:exp_comp:bcc}
    d_{bcc} = \left(\frac{\eta_{HS}(bcc)}{\eta_{HS}(fcc)}\right)^{1/3} d_{HS} = \left(\frac{3\sqrt{3}}{4\sqrt{2}}\right)^{1/3} d_{HS} = 0.972081 
    \cdot d_{HS}
\end{equation}
as confirmed from both experiment\cite{Bian2011,Goodfellow2015, Weidman2016} and simulations\cite{Zha2018}(see Fig.~\ref{fig:simul_sc} and Fig.~\ref{fig:bnsl_polymer} and discussion).  \resp{Eq.~\ref{Eq:exp_comp:bcc} has a nice interpretation that is provided further below, as it reflects an underlying incompressibility condition, see Eq.~\ref{Eq:incompress:bcc_to_fcc}.} In summary, whenever single components assemble into NCs with short ligands $\lambda < \lambda_c$ HS predictions match both qualitatively (the fcc phase) and quantitatively (lattice constants) HS predictions. For longer ligands $\lambda > \lambda_c$ HS fail: the stable phase is bcc and with slightly smaller nearest-neighbor distance given by Eq.~\ref{Eq:exp_comp:bcc}. Still, regardless of what is the equilibrium state, experiments, see Fig.~\ref{fig:exp_2d_3d}, show that when NC coordination is 12, the PF takes on the HS fcc value.

I should mention that body-centered tetragonal (bct)\cite{Whetten1999, Landman2004, Lokteva2019}, hcp\cite{Stoeva2003, Lokteva2019} and \resp{C14}\cite{Hajiw2015} lattices have been reported with single component NCs. It is possible to continuously transform a bcc to fcc through an intermediate bct, so at this point the evidence\cite{Weidman2016} is that such bct structures are not equilibrium phases but metastable intermediates that remain stable only for certain solvent contents, a point elaborated further when discussing MOLT in SubSect.~\ref{Subsect:MOLT}. The appearance of the \resp{C14} phase is attributed to polydispersity: NCs with different diameters spontaneously phase separate into a binary system, that is, crystal fractionalization\cite{Cabane2016, Lindquist2018}, since the \resp{C14} lattice is basically a single component $\gamma=1$ MgZn$_2$\cite{Lai2019a}. Here I mention another possibility where the 4 coordinated NC within the \resp{C14} may develop the four-valence eigenshape in Fig.~\ref{fig:eigen}, a point that can experimentally verified by analysis of lattice constants. Whether such an MgZn$_2$ structure with single-component NCs would be stable remains to be proved with detailed calculations.

\begin{figure}[htb]
    \centering
    \includegraphics[width=1\textwidth]{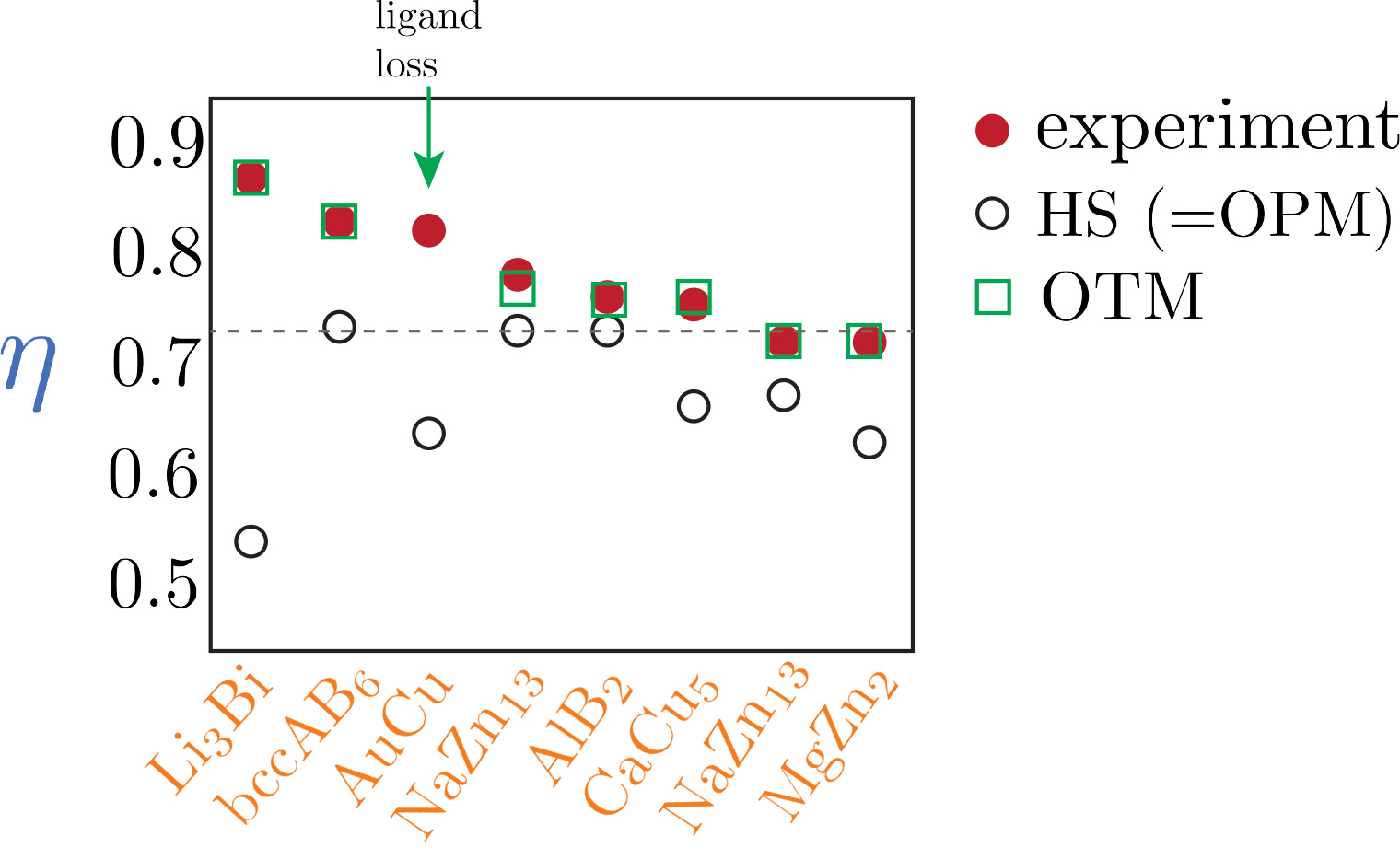}
    \caption{PF computed from experiment\cite{Boles2015} compared with HS and with  OTM\cite{Travesset2017b,Travesset2017a}. The case of AuCu is discussed in the text.}\label{fig:exp_theory_comp}
\end{figure}

For {\em NC mixtures} consisting of two species, that is, \textit{ binary NC systems}, the predictions depend on both $\gamma$ and $x$, see the definitions in Eq.~\ref{Eq:hs:def_gamma}. I have already discussed a BNSL, see Fig.~\ref{fig:alb2_experiment}, where AlB$_2$ is successfully described by HS. Yet, this is the exception rather than the rule. At the qualitative level, Fig.~\ref{fig:HS_phase_diagrams} shows that some of the phases present in the HS phase diagram: NaCl, AlB$_2$, NaZn$_{13}$, MgZn${_2}$(Laves) and bccAB$_6$(for HS exists at a lower value $\gamma$ and is not shown in the figure) are also found in experiments. There are, however, several missing phases: CaCu$_5$, CaB$_6$, CsCl and Li$_3$Bi. 

\begin{figure*}[htb]
\centering
\begin{subfigure}{.45\linewidth}
\centering
\includegraphics[width=\textwidth]{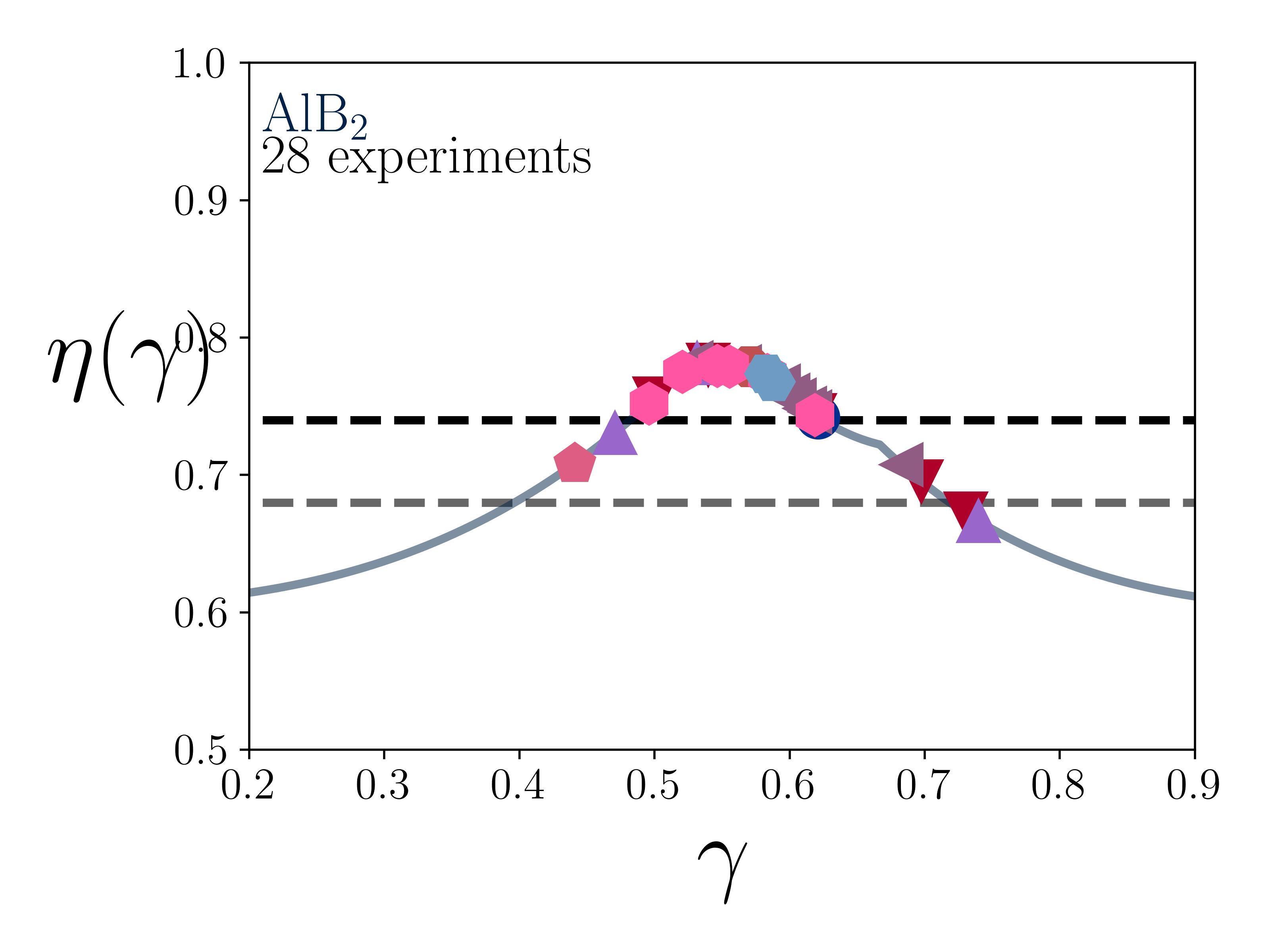}
\caption{AlB$_2$ BNSL}
\end{subfigure}
\centering
\begin{subfigure}{.45\linewidth}
\centering
\includegraphics[width=\textwidth]{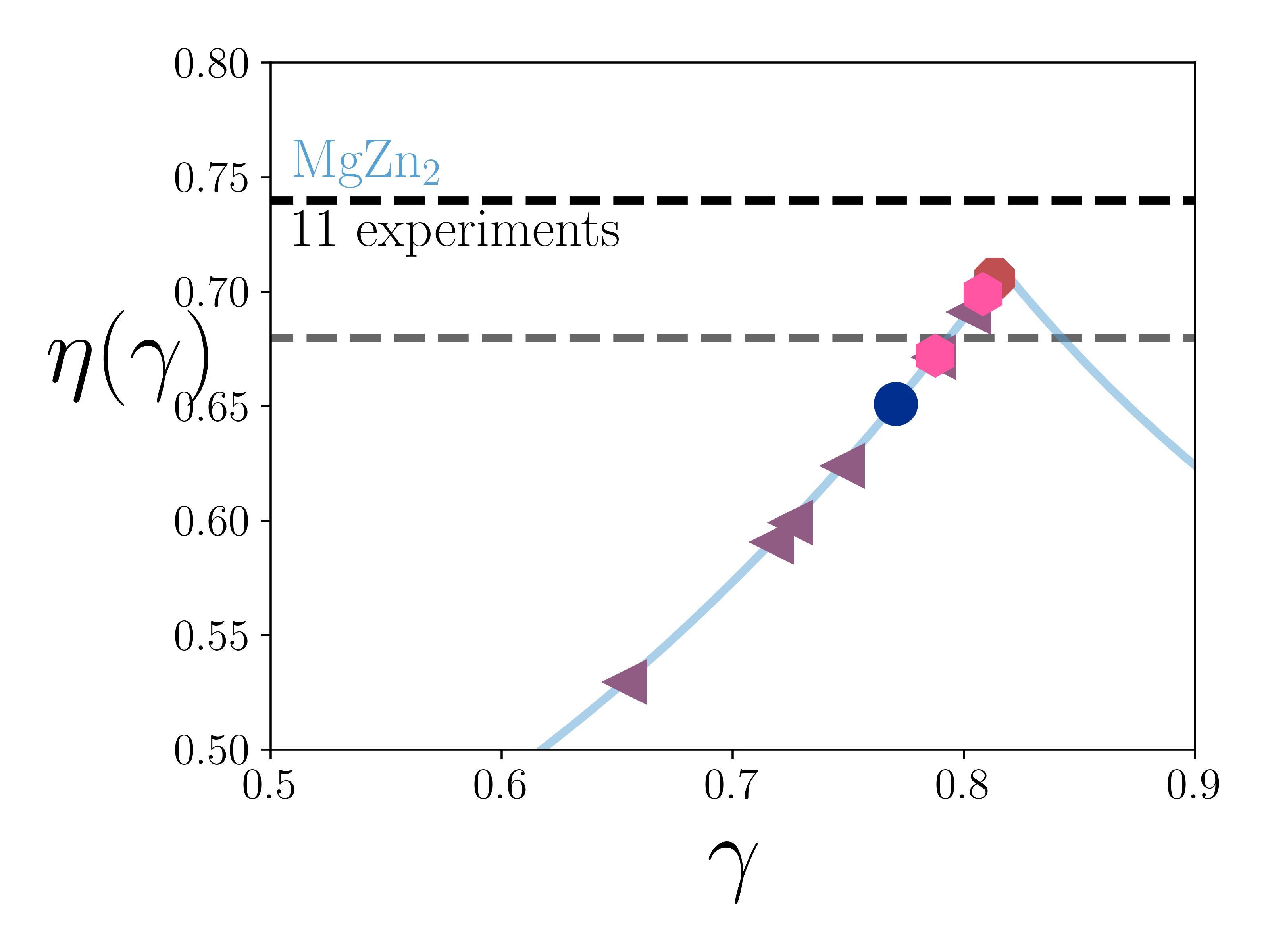}
\caption{MgZn$_2$ BNSL}
\end{subfigure}
\caption{Calculation of maximum HS PF as a function of $\gamma$ and experiments where the BNSL is found. The dashed lines are the fcc and bcc PF, respectively. The experiments are from Ref.~\cite{Shevchenko2006a, BodnarchukKovalenkoHeissTalapin2010, EversNijsetal2010, Wei2015, Boles2015, ChenOBrien2008, YangWeiBonvillePileni2015}(AlB$_2$) and
Ref.~\cite{Boles2015, Shevchenko2006a, EversNijsetal2010, ChenOBrien2008, Wei2015}(MgZn$_2$). There is some uncertainty in the determination of $\gamma$ from the experiments, as the grafting density (needed in Eq.~\ref{Eq:hs:OPM}) was not available, so $\xi=1$ was assumed in those cases.}
\label{fig:pf_bnsl_examples}
\end{figure*}

At a quantitative level, Fig.~\ref{fig:exp_theory_comp} compares the measured PF in experiments\cite{Boles2015} with the predictions of HS. Agreement is obtained for AlB$_2$ only and one of the NaZn$_{13}$ BNSLs. For the other cases, actual PF fractions predicted from HS are underestimated and in some cases, like MgZn$_2$ or Li$_3$Bi, the disagreement is quite large. Furthermore, the actual critical values of $\gamma$ where the phases are stable are also in disagreement, although experimentally there is significant uncertainty.

Regardless of actual phase diagrams or optimal PF, what has made HS models so appealing is the correlation between the maximum PF and the values of $\gamma$ (see Eq.~\ref{Eq:hs:def_gamma}) where the BNSL is experimentally realized. The calculation of the MgZn$_2$ BNSL HS PF is as follows. When $\gamma<<1$, the most favorable packing consists of the A particles in contact, each with coordination (or kissing number 4) and the B-particles filling the interstitials. As the diameter of B particles increases ($\gamma$ becomes larger), there is a critical value $\gamma_c=\sqrt{\frac{2}{3}}$, where the B particles in the pyrochloric structure, see Fig.~\ref{fig:mgzn2}, kiss each other, and then, A-A contacts cannot hold any longer and are replaced by the B-B contacts. The resulting PF is given by
\begin{equation}\label{Eq:hs:mgzn2:pf_all}
    \eta_{HS}(\gamma) = \left\{
    \begin{array}{cc}
    \frac{\sqrt{3} \pi}{16}(1+2\gamma^3) & \gamma \le \gamma_c \\
    \frac{\sqrt{2}\pi}{24}(\frac{1}{\gamma^3}+2) & \gamma \ge \gamma_c
    \end{array}
    \right.
\end{equation}
Note that for $\gamma=0$, the above expression reduces to $\frac{\sqrt{3}}{16}\pi\approx 0.34087$. which is the PF of the diamond lattice. Eq.~\ref{Eq:hs:mgzn2:pf_all} is plotted in Fig.~\ref{fig:pf_bnsl_examples}, where the maximum PF corresponds to the cusp at $\gamma=\gamma_c$.

Similarly to Eq.~\ref{Eq:hs:mgzn2:pf_all}, the PF for all other BNSLs are calculated. Fig.~\ref{fig:pf_bnsl_examples} illustrates the two representative examples that I mostly use in this paper: AlB$_2$ and MgZn$_2$. For AlB$_2$ all experiments
occur around the PF maximum, and the values of $\eta$ are high. Fig.~\ref{fig:exp_theory_comp} shows that the PF (and lattice constants) for AlB$_2$ are consistent with the HS predictions, making stronger the case already made from Fig.~\ref{fig:alb2_experiment} that NC pack as HS in this BNSL.  MgZn$_2$ still shows a correlation between the maximum PF and $\gamma$ where the experiments are found. However, there are quite a few cases in which the resulting PF is too low $\eta < 0.65$, so these BNSL cannot be stable because there is too much free space available. Another conspicuous aspect is that experiments are only found on the left of the PF maximum, a point that is quite important and will be elaborated further in the following. Therefore, we must conclude that HS models describe AlB$_2$ well but break down when applied to MgZn$_2$. Similar considerations follow for all other reported BNSLs\cite{Travesset2017a}.

\begin{table}[htb]
\centering
\begin{tabular}{c c c}
\hline
Edge length (nm) & \multicolumn{2}{c}{Lattice constant (nm)} \\
& Experiment & Theory (Eq.~\ref{Eq:hs:cube:edge_hs})  \\ \hline
9.8(9) & 11.1 & 11.2 \\
9.3(8) & 10.8 & 10.7  \\
8.8(7)  & 10.3 & 10.2 \\
5.6(4) & 7.2 & 7.0  \\ \hline
\end{tabular}
\caption{Comparison between experimental results and theoretical prediction Eq.~\ref{Eq:hs:cube:edge_hs} (with $\sigma_{Max}=2.7$ chains/nm$^2$), see discussion in Ref.~\cite{HallstromMe2023} for complete details. The experimental grafting density is $\sigma=0.81$ chains/nm$^2$.}\label{tab:lat_constant}
\end{table}

In addition to spherical-like NCs, there are other systems in which experiments can be compared against HS predictions. In the context of Perovskite NCs, BNSL with cubic NCs have been thoroughly investigated\cite{Cherniukh2021, Cherniukh2021a, CherniukhMe2022, HallstromMe2023}. Single-component NC cubes are packed as simple cubic superlattices (except for small NCs whose nuances are discussed in Ref.~\cite{Boehme2023}), in agreement with \resp{hard shape} models\cite{John2004, Damasceno2012}  with lattice constants predicted from Eq.~\ref{Eq:hs:cube:edge_hs}\cite{HallstromMe2023}, see Table~\ref{tab:lat_constant}. However, analysis of BNSLs that include both cubes and spheres show clear breakdowns from \resp{hard shapes}\cite{Cherniukh2021, HallstromMe2023}. A similar situation takes place for truncated tetrahedral NCs\cite{Boles2014, WangChen2022}, where experiments report a monoclinic and two diamond phases. The \resp{hard shape} phase diagram of truncated tetrahedra\cite{Damasceno2012b} contains many phases including diamond, but does not include the monoclinic phase.

The conclusions of the extensive comparison between \resp{hard shapes} and experiment are that there are examples where NCs behave as \resp{hard objects}, like fcc or the AlB$_2$ BNSL, but often the \resp{hard shape} description breaks down. It should therefore become apparent that ligands often cannot be coarse-grained into a hard shape because conformations play a fundamental role. Yet, despite its obvious shortcomings, \resp{hard shapes} do provide a simple and very valuable starting point that will be generalized into a complete framework for structure prediction.

\begin{figure}[htb]
    \centering
    \includegraphics[width=1\textwidth]{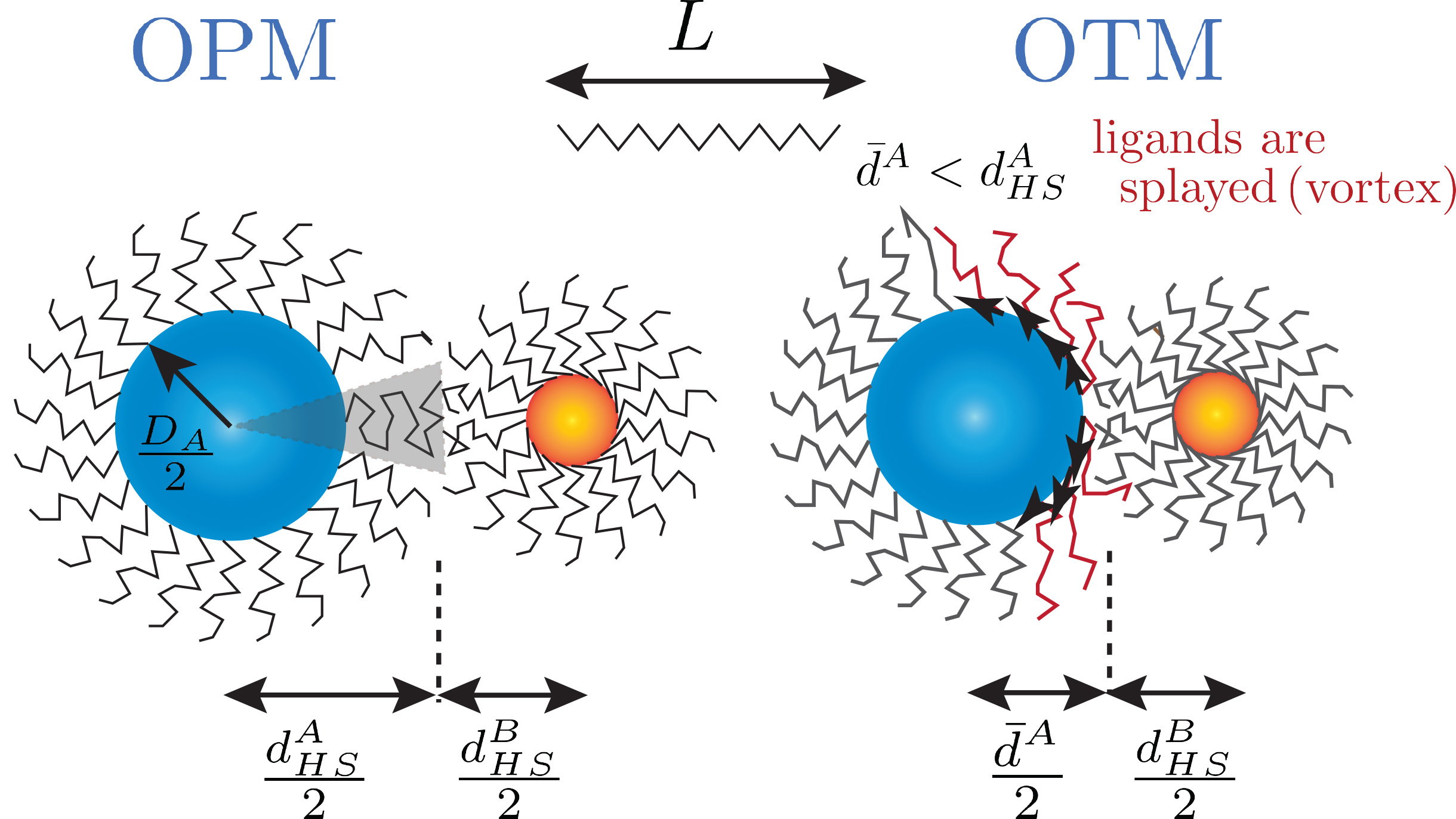}
    \caption{Example of OPM vs OTM: Within OPM each ligands fills out its own cone, as shown with the shaded volume. The only way to get NCs closer is for ligands to splay and creating a vortex. Note that once ligands splay, they occupy space belonging to other cone ligands.}\label{fig:OTM_example}
\end{figure}

\section{The Orbifold Topological Model (OTM)}\label{Sect:OTM}

\subsection{A case for OTM}

I have introduced two models, the OPM see Eq.~\ref{Eq:hs:OPM} and the OCM see Eq.~\ref{Eq:hs:OCM}. Both define an effective NC diameter. As discussed in SubSect.~\ref{SubSect:HS:Experiments}, the OPM defines the HS diameter of a NC and leave a significant amount of ``empty'' space; see also Eq.~\ref{Eq:hs:OPM:pf}, and the OCM defines a situation where the NCs are packed in such a way that the ligands fill out the entire available space. When analyzing experimental results, there are situations where OPM or OCM work, but neither of them provides a complete solution. Basically, what is needed is a new model that agrees with the OPM and OCM in the limits when they are correct.

Before dwelling into how OTM provides a general solution that reduces to OPM and OCM in the appropriate limits, it is worth describing the effect of potentials that go beyond the \resp{hard shape} discussed in SubSect.~\ref{SubSect:HS:Experiments}. 
For example, Refs~\cite{Travessetpnas2015, HorstTravesset2016} (also \cite{Travesset2014,Calero2016,Tkachenko2016}) and~\cite{Lacour2019} have discussed softening the potential from \resp{hard shape} to an inverse repulsive power law. We recently provided a detailed determination of the phase diagram of binary particles that interact with Lennard-Jones potentials\cite{Ren2020}. These systems have rich phase diagrams, but when applied to NC prediction do not provide any real improvement over the simpler \resp{hard shape} models and, furthermore, require defining additional parameters that are not obvious how to relate to the actual physical properties of the ligands. In summary, going beyond \resp{hard shape} by simply modifying a two-body potential may provide an ad-hoc solution for a particular case but leads to models that have less predictability. As I elaborate further, this is hardly surprising, as ligand deformations preclude a description of NC interactions in terms of two-body potentials.

\begin{figure}
    \centering
    \includegraphics[width=1\textwidth]{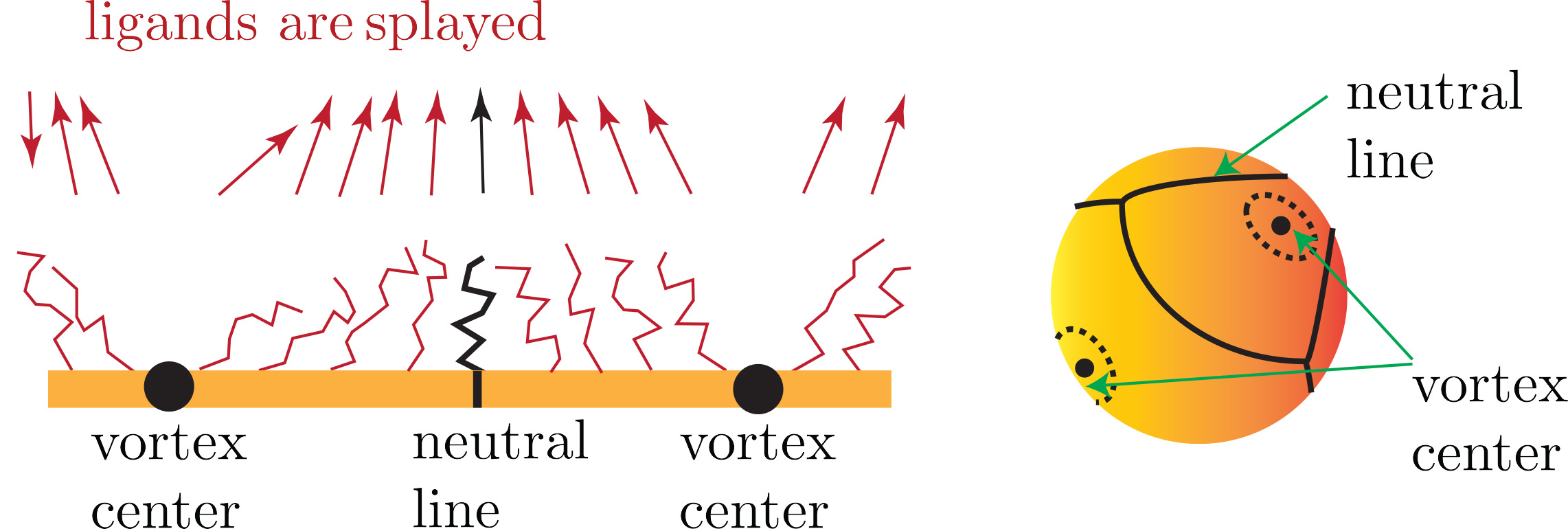}
    \caption{Two neighboring vortices must contain a neutral line where ligands are perpendicular and escape to the third dimension. On the surface of a sphere, the neutral lines define regions surrounding each vortex.}\label{fig:OTM_neutral}
\end{figure}

\subsection{The formalism}\label{Subsect:otm:formalism}

I deal with spherical NCs first. For a given arrangement, I first assume that the entire system consists of HS, with diameter given by OPM Eq.~\ref{Eq:hs:OPM}, which pack optimally. That means that a given BNSL will have a PF exactly equal to $\eta_{HS}(\gamma)$, see Fig.~\ref{fig:pf_bnsl_examples} and Eq.~\ref{Eq:hs:def_gamma}, Eq.~\ref{Eq:hs:OPM:pf}. Then, the coordination of a given NC within the nanostructure is defined as all of the NCs that are in actual contact. Mathematicians define the coordination of a particle as the ``kissing number''. It is important to emphasize that the coordination is a rigorous and unambiguous quantity for any arrangement because NCs are assumed to be HS. Thus, the coordination is the number of nearest neighbors for a HS arrangement where the particles touch each other, and in the same way the number of next-to-nearest and higher-order neighbors are rigorously defined.

The main idea in OTM is that at low coordination, large ligand deformations, consisting of splayed ligands as shown in Fig.~\ref{fig:OTM_example} and Fig.~\ref{fig:OTM_neutral}, the vortices, are possible. 
The Orbifold Topological Model is then derived from two principles\cite{Travesset2017a, Travesset2017b}:
\begin{enumerate}
\item{For a given lattice, the minimum free energy occurs at the lattice constant with maximum HS PF.
}
\item{If the coordination is six or more, then NCs can only pack as HS, no large ligand deformations are allowed.}
\end{enumerate}

The first assumption is justified by the notion that assembly into a dry state is driven by attractive interactions, and therefore the lowest free energy will optimize the number of ligand monomers at a given unit volume, consistent with existing constraints.

The second assumption reflects that in order to break the HS description and get NCs closer than their respective HS separation, ligands must splay as shown in Fig.~\ref{fig:OTM_example}. These splayed configurations were dubbed vortices in the original papers. Vortices do interact repulsively within a given NC; This is illustrated in Fig.~\ref{fig:OTM_neutral} where, mapping the ligands into spins, they must twist for an energy cost, similarly as domain walls in ferromagnets.  Therefore, it becomes energetically costly for a given NC to have a large number of vortices. Also, because ligands within vortices need to occupy neighboring cones, see Fig.~\ref{fig:OTM_example}, volume constraints make a large number of vortices very challenging to realize. The conclusion is that if $N_v$ defines the number of vortices for a given NC, configurations with $N_v \ge 6$ are forbidden. There is nothing ``magical'' about the impossibility of $N_v \ge 6$ vortices: As we have seen already from experiments in Fig.~\ref{fig:exp_2d_3d} and will see with numerical simulations, the nearest neighbor distance increases with coordination and a putative NC configuration with $N_v \ge 6$ would have a nearest neighbor distance that is very close to $N_v=0$, as it is demonstrated experimentally in Fig.~\ref{fig:exp_2d_3d}. Also, we have already seen a similar situation where the bcc structure develops a small correction, see Eq.~\ref{Eq:exp_comp:bcc} for 8 nearest neighbors. I will use a notation as in Fig.~\ref{fig:OTM_example}:
\begin{eqnarray}\label{Eq:OTM:def_bar}
    d^i_{HS} &=& \mbox{ HS diameter of NC i} \nonumber\\
    \bar{d}^i &=& \mbox{ OTM diameter of NC i} \ .
\end{eqnarray}
$d^i_{HS}$ is defined from Eq.~\ref{Eq:hs:OPM}. The formula for $\bar{d}^i$ is more complex and examples will be derived in the following. 

For those familiar with the theory of topological defects\cite{ChaikinBook2003}, it is useful to think of a given NC as a skyrmion: Each ligand chain grafted to the NC deﬁnes a vector whose origin is at the NC and whose end is at the center of the terminal ligand group.  Thus, the ligands define a mapping $S^2 \rightarrow S^2$ which enables to define a Skyrmion. Because of the way the ligands are attached, this Skyrmion has always a topological charge $q_S=-1$. Still, by focusing on the projection of the ligand vector to the sphere surface, see Fig.~\ref{fig:OTM_example}, the order parameter defines a vortex. It cannot be stressed enough that the vortices are not topological, as they may contain singular points where the order parameter escapes to the third dimension, the neutral lines see Fig.~\ref{fig:OTM_neutral}, and as a result, there are no topological constraints on the total number of vortices $N_v$. Ligands are ﬂexible, so the concept of Skyrmion topological charge $q_S$ is only approximate, so it seems more appropriate to denote it as a soft Skyrmion.

\begin{figure}
    \centering
    \includegraphics[width=1\textwidth]{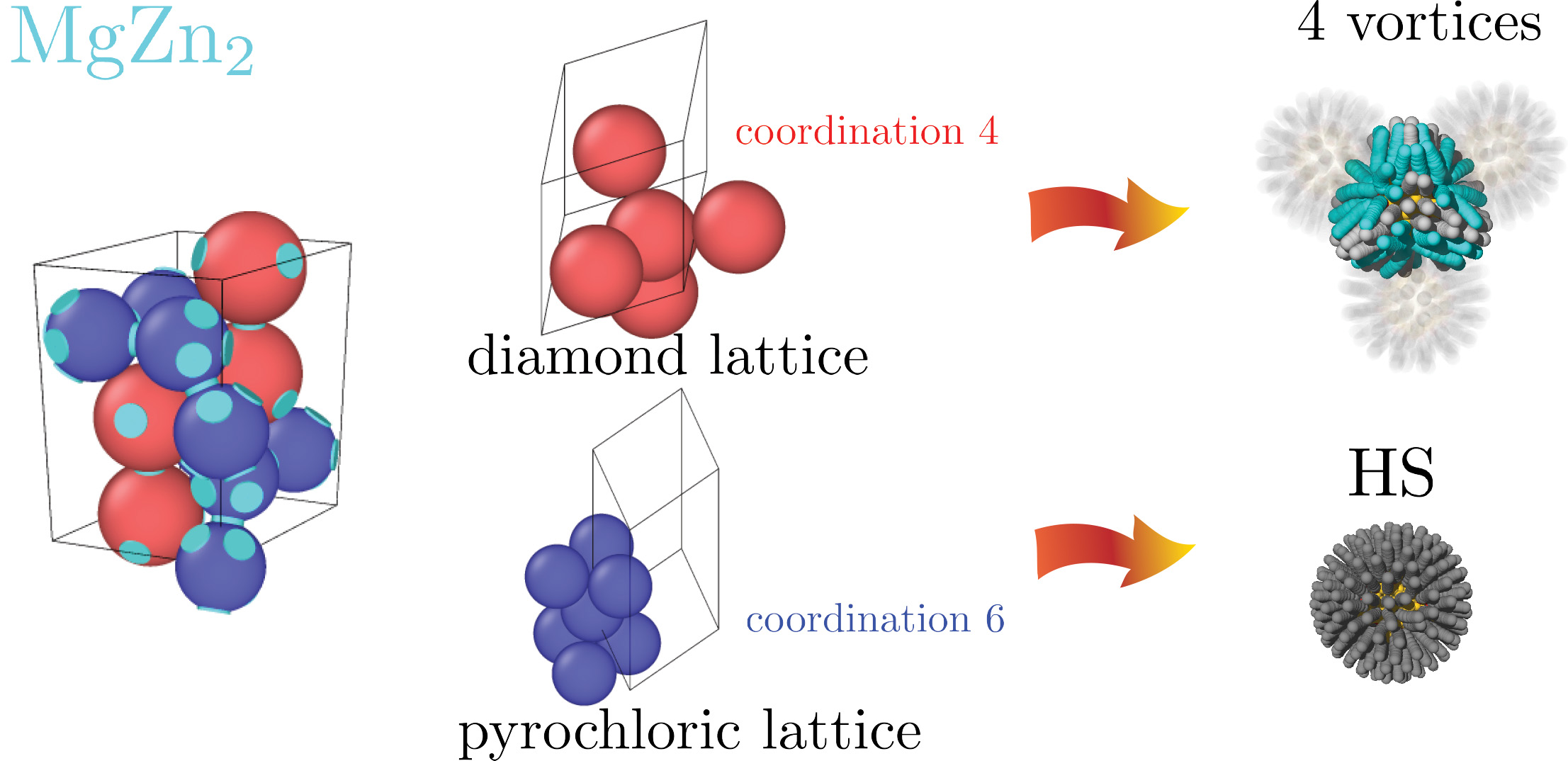}
    \caption{OTM applied to the MgZn$_2$ BNSL, which consists of two interpenetrating lattices, here shown at $\gamma=\gamma_c$. The large A NCs(red) for a diamond lattice, while the smaller B ones(blue) a pyrochloric lattice. For $\gamma<\gamma_c$ the large NCs are in contact and have coordination 4, while for $\gamma > \gamma_c$ the smaller are in contact and have coordination 6. According to OTM, only for $\gamma<\gamma_c$ vortices, see Fig.~\ref{fig:OTM_example}, are allowed.}\label{fig:mgzn2}
\end{figure}

As a concrete example, I discuss the application of OTM to MgZn$_2$. This BNSL is described by space group P63/mmc (194) with a unit cell with $n_A = 4$ in Wyckoﬀ positions 4f and $n_B = 8$ in positions 2a and 6h; see Fig.~\ref{fig:mgzn2}. The maximum of $\eta_{HS}(\gamma)$, see Fig.~\ref{fig:pf_bnsl_examples}b), is reached for $\gamma=\gamma_c=\left(\frac{2}{3}\right)^{1/2}\approx 0.8165$. 
At $\gamma < \gamma_c$ each A-NC is in contact with four other A-NCs while for $\gamma > \gamma_c$ each B-NC with 6-B-NCs. 

For $\gamma<\gamma_c$ each NC has coordination 4 so $N_v=4$ is possible. Let us obtain the formula for $\bar{d}^A$ (see Eq.~\ref{Eq:OTM:def_bar} for the notation definition). We can define the following quantities
\begin{eqnarray}\label{Eq:otm:defs}
    \gamma &=& \frac{d^B_{HS}}{d^A_{HS}} \quad \mbox{ Same as Eq.~\ref{Eq:hs:def_gamma}} \nonumber \\
    \bar{\gamma} &=& \frac{d^B_{HS}}{\bar{d}^A}
\end{eqnarray}
Following assumption I, it should be that
\begin{eqnarray}\label{Eq:otm:mgzn:Eq1}
    \bar{\gamma} = \gamma_c = \frac{d^B_{HS}}{\bar{d}^A} \rightarrow \bar{d}^A = \frac{d^B_{HS}}{\gamma_c} = \frac{\gamma}{\gamma_c} d^A_{HS} < d^A_{HS} \ .
\end{eqnarray}
The condition $\bar{\gamma} = \gamma_c$ cannot be hold indefinitely. At $\gamma=\gamma^{c}$ defined as
\begin{equation}\label{Eq:otm:mgzn:Eq2}
    \gamma^c = \frac{\sqrt{12}\gamma_c}{2\sqrt{11}-\sqrt{12}\gamma_c} = \frac{\sqrt{2}}{\sqrt{11}-\sqrt{2}}\approx 0.7434
\end{equation}
B-NCs, which are HS, come into contact with A-NCs and the condition $\bar{\gamma}=\gamma_c$ cannot be held anymore, so B-particles need to be displaced. Therefore, it is 
\begin{equation}\label{Eq:otm:mgzn:Eq3}
    \bar{d}^A=\left(\frac{3}{11}\right)^{1/2}(1+\gamma) d^A_{HS} \quad \mbox{ for } \gamma < \gamma^c \ .
\end{equation}
The PF becomes
\begin{eqnarray}\label{Eq:otm:mgzn:Eq4}
    \eta(\gamma) \equiv \eta_{OTM}(\gamma)=\left\{
    \begin{array}{cc}
    \eta_{HS}(\gamma) & \gamma \ge \gamma_c \\
    \left(\frac{1/\gamma^3+2}{1/\gamma_c^3+2}\right)\eta_{HS}(\gamma_c) & \gamma^c \ge \gamma < \gamma_c \\
    \pi \frac{(\sqrt{11})^3}{48} \frac{1+2\gamma^3}{(1+\gamma)^3} & \gamma <  \gamma^c
    \end{array}
    \right.
\end{eqnarray}
where the explicit expression for $\eta_{HS}(\gamma)$ is given in Eq.~\ref{Eq:hs:mgzn2:pf_all}. Therefore, the MgZn$_2$ BNSL has an OTM branch only for $\gamma<\gamma_c$. For $\gamma > \gamma_c$ NCs pack as HS.

It is really important to stress that OTM predictions like Eq.~\ref{Eq:otm:mgzn:Eq1},\ref{Eq:otm:mgzn:Eq2},\ref{Eq:otm:mgzn:Eq3} and Eq.~\ref{Eq:otm:mgzn:Eq4} do not require additional or fitting parameters and are independent of the microscopic potentials defining the interaction details. Derivations for all other BNSLs follow the same steps and are provided in Ref.~\cite{Travesset2017a, Travesset2017b}. AlB$_2$, see Fig.~\ref{fig:pf_bnsl_examples}, for example, does not have any OTM branch, which explains why NCs pack as HS, as it has already been extensively shown. 

\begin{figure}[htb]
    \centering
    \includegraphics[width=1\textwidth]{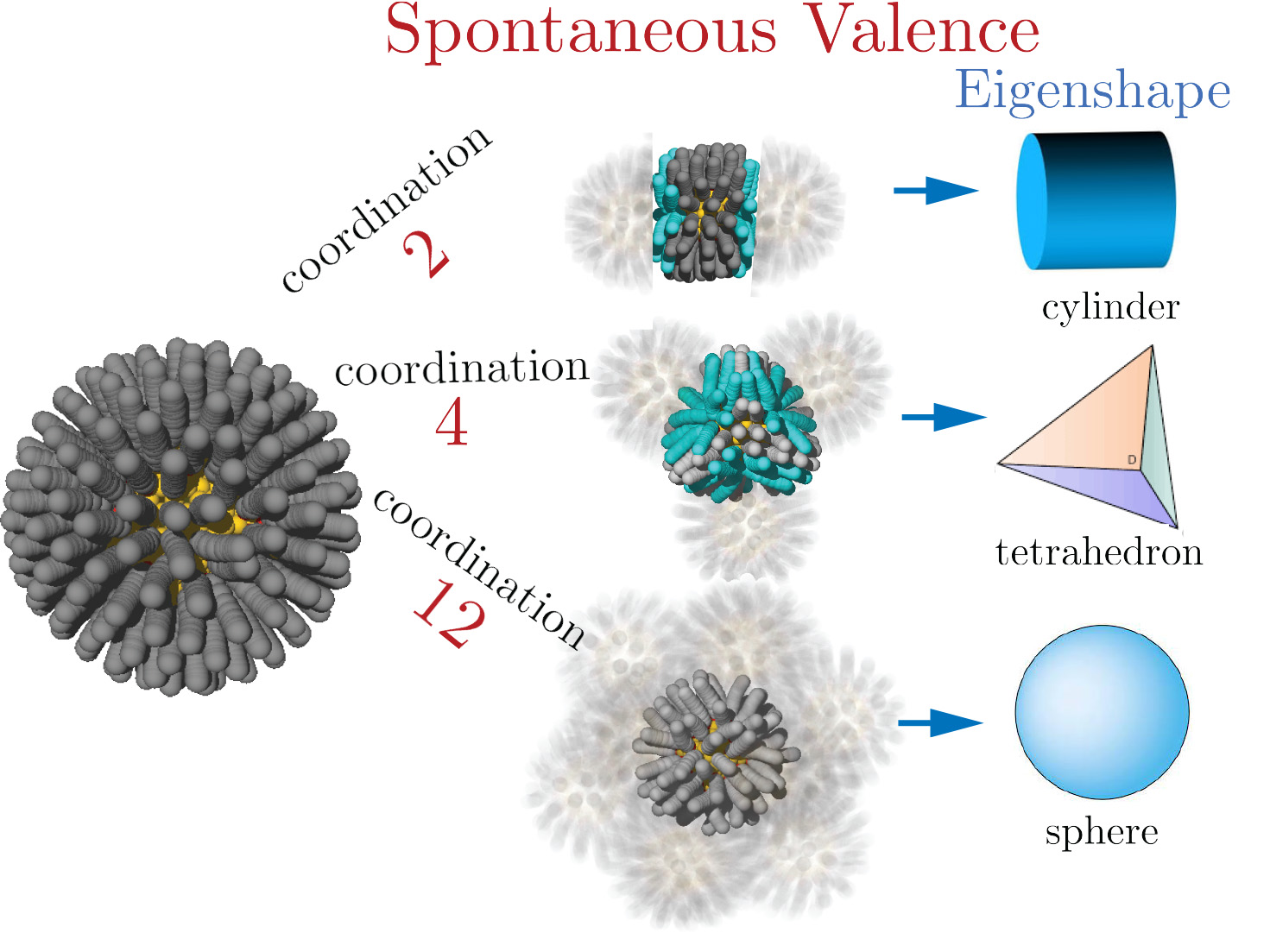}
    \caption{Simulations from Ref.~\cite{Waltmannn2019},  discussed in Sect.~\ref{Sect:Simuls}, demonstrating the NC spontaneous valence depending on the environment, as predicted from OTM. A NC with coordination 2 is described with a cylindrical ``eigenshape'', a NC with coordination 4 with a tetrahedron and coordination 12 is a sphere. Nearest neighbor NCs are shown with transparencies for proper visualization. Other eigenshapes are described in Ref.~\cite{Travesset2017a}. }\label{fig:eigen}
\end{figure}

Fig.~\ref{fig:mgzn2} shows that for $\gamma < \gamma_c$ the NC within a MgZn$_2$ BNSL effectively becomes a tetrahedron. That is, the NC shape after assembly, the ``eigenshape'', is different than the ``bare'' spherical shape. This ability to spontaneously develop a valence is determined by the environment, i.e. the NC coordination, is a general result illustrated in Fig.~\ref{fig:eigen}, where different valences are shown with the help of simulation results that will be presented later on. It should be emphasized that even for a given coordination valences are not unique: there are other valences for coordination 4, for example, see Ref.~\cite{Travesset2017a}.

\begin{figure}
    \centering
    \includegraphics[width=1\textwidth]{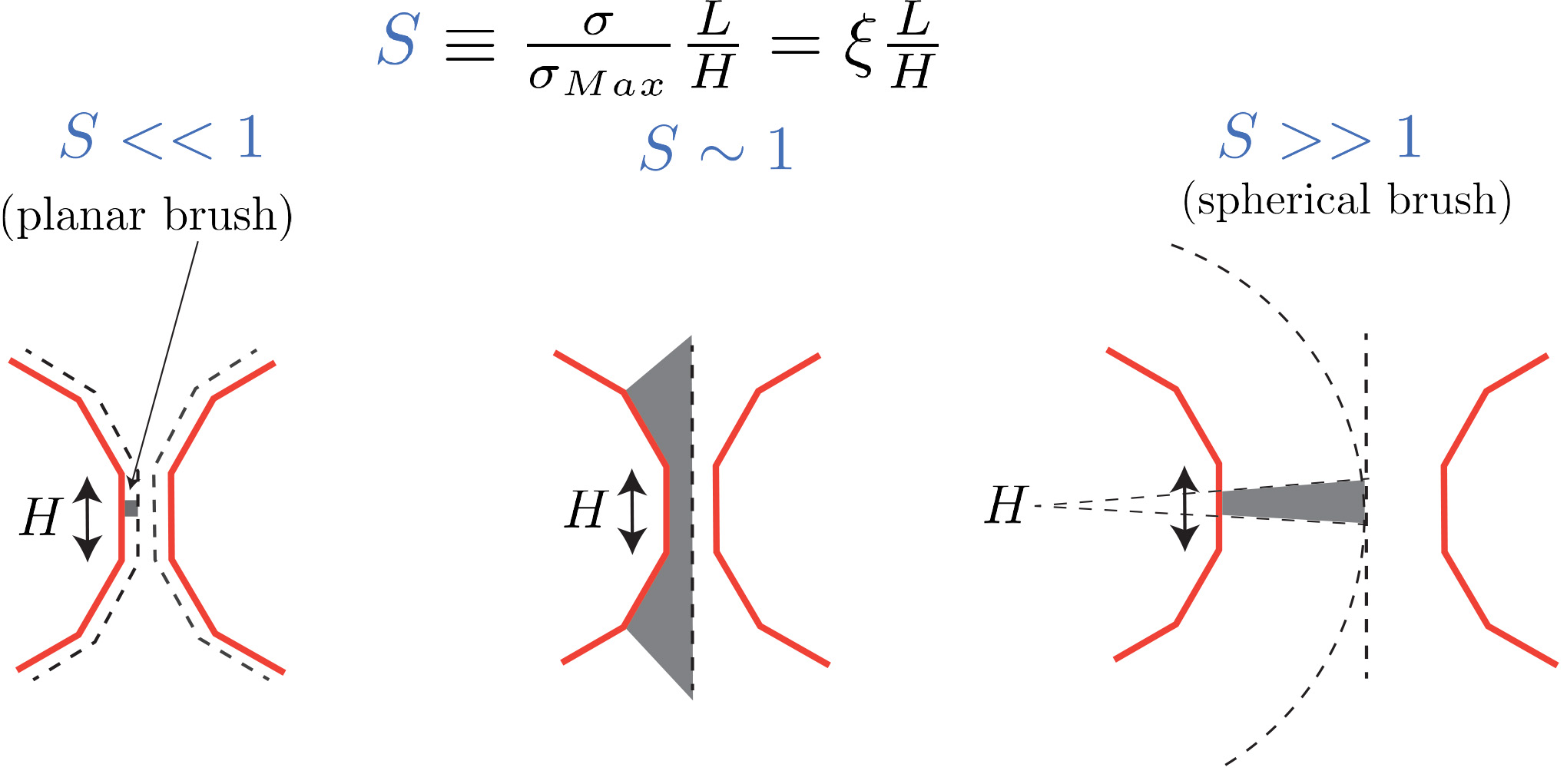}
    \caption{The parameter $S$ defined in Eq.~\ref{Eq:otm:s_param} and illustration of the different limits. In the planar brush limit, each face defines a planar brush. In the spherical limit, the polyhedral nature of the NC core becomes irrelevant. The limit $S\sim 1$ defines a nontrivial situation where ligands lead to complex eigenshapes.}\label{fig:brush}
\end{figure}

\subsection{OTM beyond quasi-spherical NCs}\label{SubSect:otm:beyond_spheres}

Although spherical NCs have been made\cite{ZhouHuo2018}, see Fig.~\ref{fig:spherical}, the NC cores used in assembly experiments are not spherical. \resp{A rigorous theory describing shape within OTM remains to be developed, but there are a number of general statements that can be made, which are quantitatively tested in cubic NCs. A general NC} has the shape of a polyhedra consisting of vertices, edges, and faces. The faces have the shape of a polygon. Generally, not all polygons that define the faces have the same size or shape, but I will consider $H$ the characteristic size of the largest polygon. The relevant parameter is
\begin{equation}\label{Eq:otm:s_param}
    S \equiv \frac{\sigma}{\sigma_{Max}}\frac{L}{H} =  \xi \frac{L}{H}  \ .
\end{equation}
Then there are three situations, depicted in Fig.~\ref{fig:brush}
\begin{equation}
    S \quad \left\{
    \begin{array}{cc}
    >> 1 & \mbox{ Spherical limit} \\
    \sim 1 & \mbox{ OTM} \\
    << 1 & \mbox{ Planar limit}
    \end{array}\right.
\end{equation}
In the spherical limit \resp{$S>>1$}, shape becomes irrelevant and the core is effectively a sphere, see Fig.~\ref{fig:brush}, so the OTM for spherical NCs as described previously apply, \resp{see also the discussion of the polymer limit}. In the planar limit \resp{$S<<1$}, each of the NC faces consist of planar brushes\cite{RubinsteinBook2003}, therefore the NC may be coarse grained as a HS with the same shape as the core, with an additional enthalpic interaction. If the ligand is long enough to be considered a polymer, a rigorous method to quantitatively describe the shape is provided from the methods in Ref.~\cite{Dimitriyev2021}. Therefore, $S \sim 1$ represents the case in which quantitative solutions are more difficult to obtain. It also happens to be the most common situation in NC assembly where $H$ is in the nanometer scale and the ligand $L$ size is typically of the same order.

\begin{figure}
    \centering
    \includegraphics[width=1\textwidth]{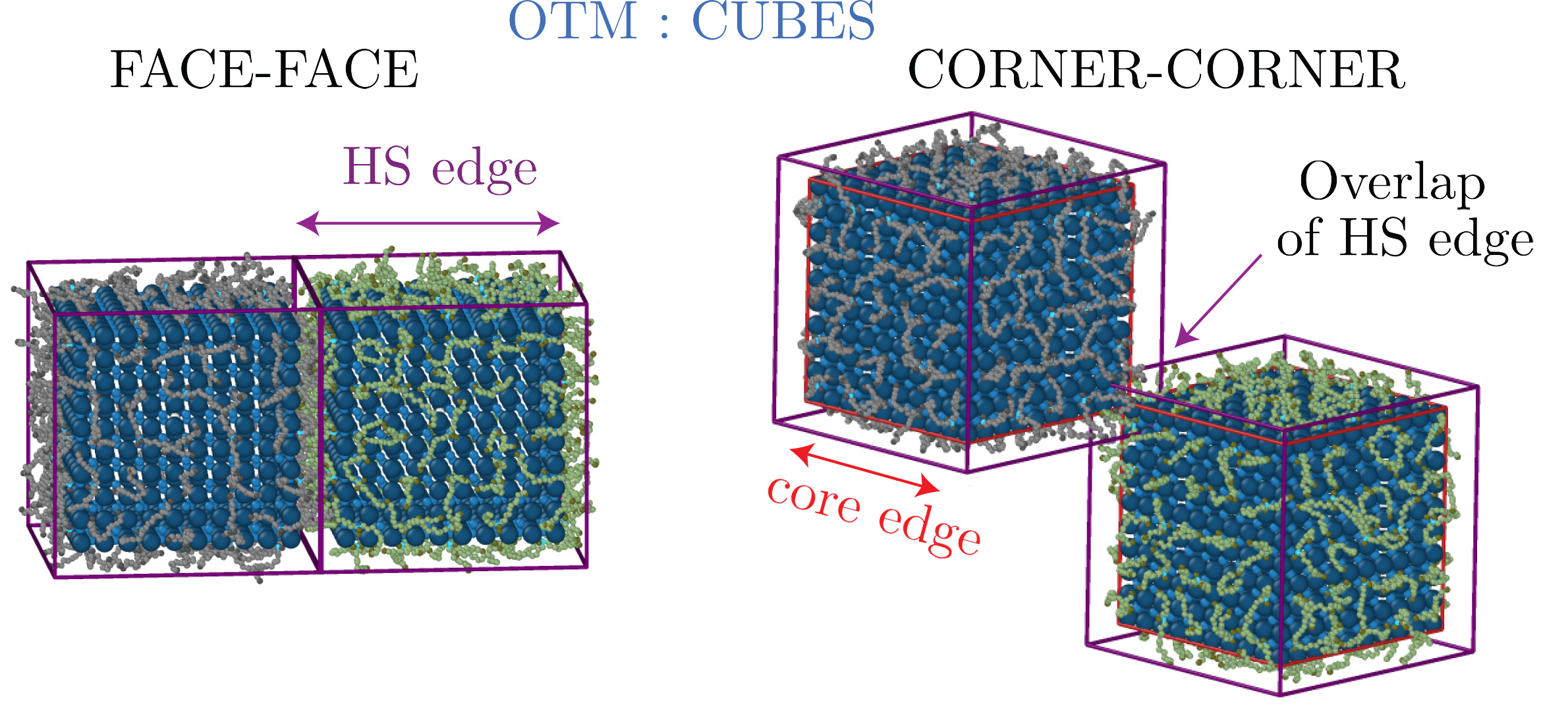}
    \caption{Example of OTM for cubes for the case of face-face, where cubes are described as HS and corner-corner, were the cube vertices of the core are in contact and the HS description breaks down. The ligands are drawn at a reduced scale for visualization. Other cases are discussed in Ref.~\cite{HallstromMe2023}.}\label{fig:otm_cubes}
\end{figure}

The new variable that appears when considering shapes other than spheres is the relative orientation among NCs. Detailed
formulas for other shapes beyond spheres or cubes are not yet available, but the cases of cubes in the relevant limit $S\sim 1$, which I discuss following Ref.~\cite{HallstromMe2023} is general and includes all the elements necessary to generalize to any other shape. If the cubes interact through faces, then vortices are not possible, since the cones, see Fig.~\ref{fig:OTM_example}, become cylinders as there is no curvature, so there is no free volume for neighboring chains to occupy and the NCs are effective \resp{hard} cubes, see Fig.~\ref{fig:otm_cubes}. This is the reason why the \resp{hard shape} prediction works so accurately for single NC cubes assembled into a simple cubic lattice; see Table~\ref{tab:lat_constant}. Therefore, nontrivial OTM solutions will occur when NCs interact through vertices and/or edges. The general cases, including cubes with sphere interactions are described in Ref.~\cite{HallstromMe2023}. I discuss here only the vertex-vertex case, in which the characteristic edge length $\bar{l}$ and center-center separation $\bar{d}$ are given by
\begin{equation}
    \bar{l}=l_{core} \rightarrow \bar{d}=\sqrt{3} \bar{l} = \sqrt{3} l_{core} < \sqrt{3} l_{HS} \ ,
\end{equation}
where it has been assumed that the NC core vertices are in contact. This is illustrated by Fig.~\ref{fig:otm_cubes} from the simulations in Ref.~\cite{HallstromMe2023}. This example serves to illustrate how the HS breaks down for shapes other than spherical, and I refer to~\cite{HallstromMe2023} for other cube orientations and also for systems of spheres and cubes.

\begin{figure}[hp]
    \centering
    \includegraphics[width=1\textwidth]{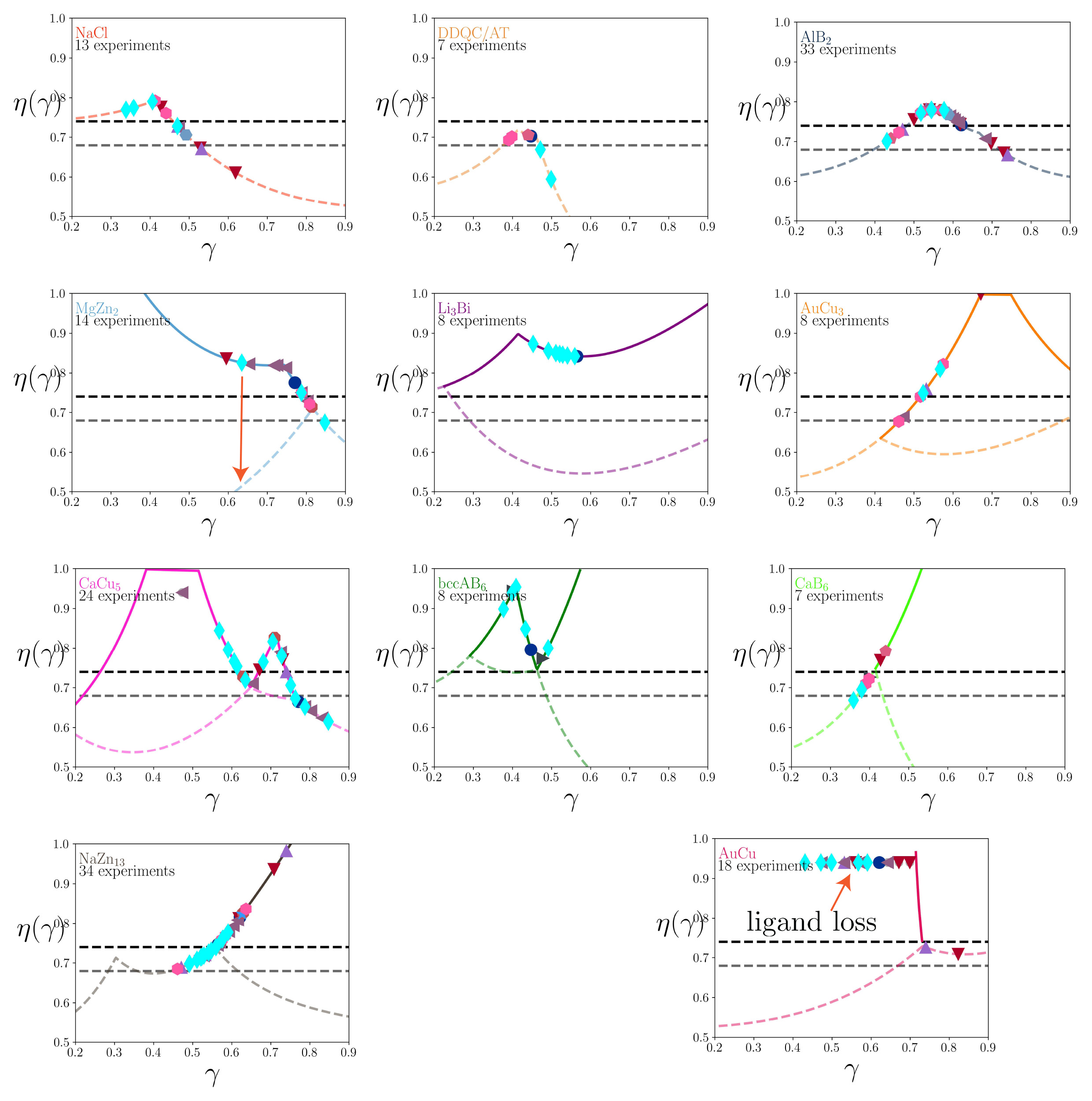}
    \caption{Compilation of almost 200 experiments reporting BNSLs. The data is from the same references as in Fig.~\ref{fig:pf_bnsl_examples} plus Ref.~\cite{Coropceanu2019}, which is drawn with diamond cyan symbols. The AuCu BNSL is special as it corresponds to a situation with ligand loss, discussed in the text.}\label{fig:otm_pf_bnsl}
\end{figure}

As a final note, it would be interesting from a fundamental aspect at least, to functionalize spherical NCs, see Fig.~\ref{fig:spherical} and provide a detailed comparison of assembled structures with shapes such as truncated octahedra, which are well described within the spherical approximation, and in this way establish if there are subtle effects driven from NC shape. 

\subsection{Experimental validation of OTM}

At this point in time, there is no experiment or simulation that contradicts a single OTM prediction. In this subsection, I will emphasize those experiments that enable quantitative verification of OTM and how they fix the HS failures described in SubSect.~\ref{SubSect:HS:Experiments}. Validation of OTM has also been accomplished through simulations, to be discussed in great detail in Sect.~\ref{Sect:Simuls} and also in Sect.~\ref{Sect:Dynamics}.

As a first example, I discuss the results presented in Fig.~\ref{fig:exp_theory_comp}. In all cases except for AuCu, the OTM predictions are in excellent agreement with the experimental results by Boles and Talapin~\cite{Boles2015}(see SubSect.~\ref{SubSect:HS:describe:exp} for a discussion on the high level of accuracy available from experiments). Particularly notable is the agreement for Li$_3$Bi, as the HS prediction is off by almost 100\%.

First, I discuss the MgZn$_2$ BNSL as it has been extensively used in previous discussions, see Eq.~\ref{Eq:otm:mgzn:Eq1}-\ref{Eq:otm:mgzn:Eq4} and Fig.~\ref{fig:mgzn2}. For $\gamma\le \gamma_c$ there is an OTM branch calculated in Eq.~\ref{Eq:otm:mgzn:Eq1} and therein, which is shown in Fig.~\ref{fig:otm_pf_bnsl}. The putative low-PF experiments that were of concern following the discussion in Fig.~\ref{fig:pf_bnsl_examples} no longer exist,  see the orange arrow in Fig.~\ref{fig:otm_pf_bnsl}, and have now become PF larger than the one in fcc. More importantly, the predicted lattice constants (and PF) agree with those from the experiment, as shown in Fig.~\ref{fig:exp_theory_comp}. The observation that all experiments in Fig.~\ref{fig:exp_theory_comp} sit on the left of the maximum PF, also consistent with the newer data from Ref.~\cite{Coropceanu2019}, clearly demonstrates that the larger A NCs do form vortices and that A-NCs within the BNSL are well represented by the tetrahedral eigenshape described in Fig.~\ref{fig:eigen}.

All BNSLs shown in Fig.~\ref{fig:otm_pf_bnsl} include  almost 200 experiments, where I emphasize Ref.~\cite{Coropceanu2019} (in cyan diamonds), as this is the most systematic and complete data available to this date. In some cases: NaCl, AlB$_2$ there are no OTM branches and the NCs are described as HS. The few experiments corresponding to low PFs correspond to old data where the grafting density was not reported, so this is the most likely explanation. In all those other BNSLs, the OTM branches fix the problem of low PFs
and OTM correctly predicts the lattice constants (or PF) whenever they are available, as shown in Fig.~\ref{fig:exp_theory_comp}. The Li$_3$Bi BNSL is interesting as it corresponds to a situation where the breakdown of the HS is maximal. The latest data from Ref.~\cite{Coropceanu2019} included in Fig.~\ref{fig:otm_pf_bnsl}, which also provides detailed measurements of lattice constants, present a very strong validation for OTM.
Applications to \resp{Dodecagonal Quasicrystal/Archimedean Tiling phases} are in qualitative agreement but have not been elaborated.

The failure of OTM to predict AuCu in Fig.~\ref{fig:exp_theory_comp} was discussed in the original references\cite{Travesset2017a,Travesset2017b}, where it became apparent that almost all the values of $\gamma$ for which they were reported correspond to a low HS PF for which there is no OTM branch. All AuCu BNSLs were recognized to have PbS (or PbSe) cores, whose ligands detach due to weak binding\cite{Boles2019}. The reason why OTM does not work is that this BNSL corresponds to an NC in which some facets have lost all their ligands and the actual cores are in physical contact. The assembly of partially grafted NCs has very significant interest\cite{Whitham2016,Lokteva2019,Kavrik2022} as direct contact among the cores allows charge transport throughout the material, so extensions of OTM to these important cases will need to be developed. In either case, it remains an outstanding success that OTM could identify AuCu as inconsistent with a BNSL where NCs are fully grafted, as shown in Fig.~\ref{fig:otm_pf_bnsl}, before the systematic experiments of Ref.~\cite{Boles2019} had been reported.

\begin{figure}
    \centering
    \includegraphics[width=1\textwidth]{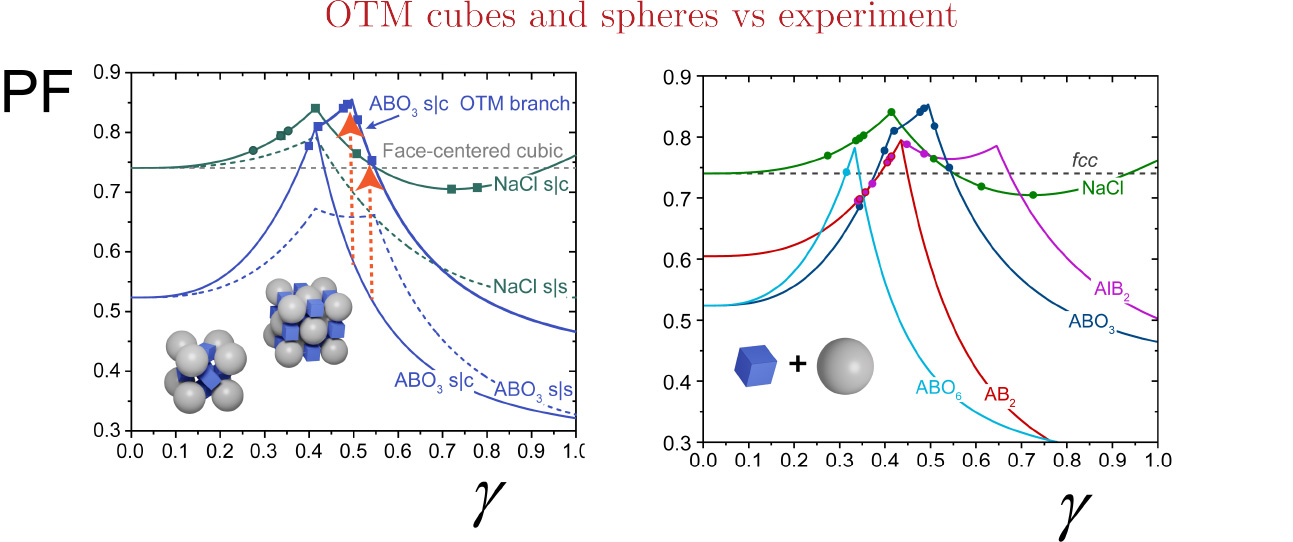}
    \caption{Experiments of 8.6 and 5.3 nm NC cubes functionalized with DDAB (Didodecyldimethylammonium bromodide) with spherical NCs functionalized with oleylamine lead to five BNSL\cite{CherniukhMe2022}. The cases of ABO$_3$ and NaCl are shown in the left, together with the curves describing HS packing for spheres and cubes (s$|$c) and binary spheres(s$|$s). The orange arrows highlight two ABO$_3$ experiments whose HS PF is below 50$\%$. OTM fixes it and predicts the correct lattice constant\cite{Cherniukh2021}. The figure on the right shows the OTM PF for all five BNSL reported.}\label{fig:otm_cubes_experiment}
\end{figure}

Perovskite NCs\cite{Schmidt2014, Protesescu2015, Imran2018} take the form of a cube, which for single component assemble into simple cubic superlattices\cite{Raino2018} (except for small cubes\cite{Boehme2023}) with lattice constants and PF well described by HS as demonstrated in Table~\ref{tab:lat_constant}. Binary systems of cubes and spherical NCs form a very rich phase diagram\cite{Cherniukh2021,Cherniukh2021a,CherniukhMe2022}. Each NC in NaCl has coordination six and NC cubes interact through their faces, therefore, there are no OTM branches and this superlattice is entirely described by \resp{a hard shape model}. In ABO$_3$, however, the NC cubes interact through their vertices, and, as it follows from the discussion in Fig.~\ref{fig:otm_cubes} this leads to the breakdown of \resp{hard cube} description and the emergence of OTM branches. Fig.~\ref{fig:otm_cubes_experiment} highlights two experiments   
(red dashed arrows) whose PF within HS would be less than 50$\%$ and the inclusion of OTM increases the PF above the fcc and also brings the lattice constants in agreement with the experiment. This example illustrates how important the relative NC orientation is in determining whether \resp{the hard shape} description breaks down. Fig.~\ref{fig:otm_cubes_experiment} shows the OTM PF for all five phases encountered (including only 2 NC sizes) and how all have large PF ($> 0.7$). Refer to~\cite{CherniukhMe2022, HallstromMe2023} for more detailed quantitative analysis.

\begin{figure}[htb]
    \centering
    \includegraphics[width=1\textwidth]{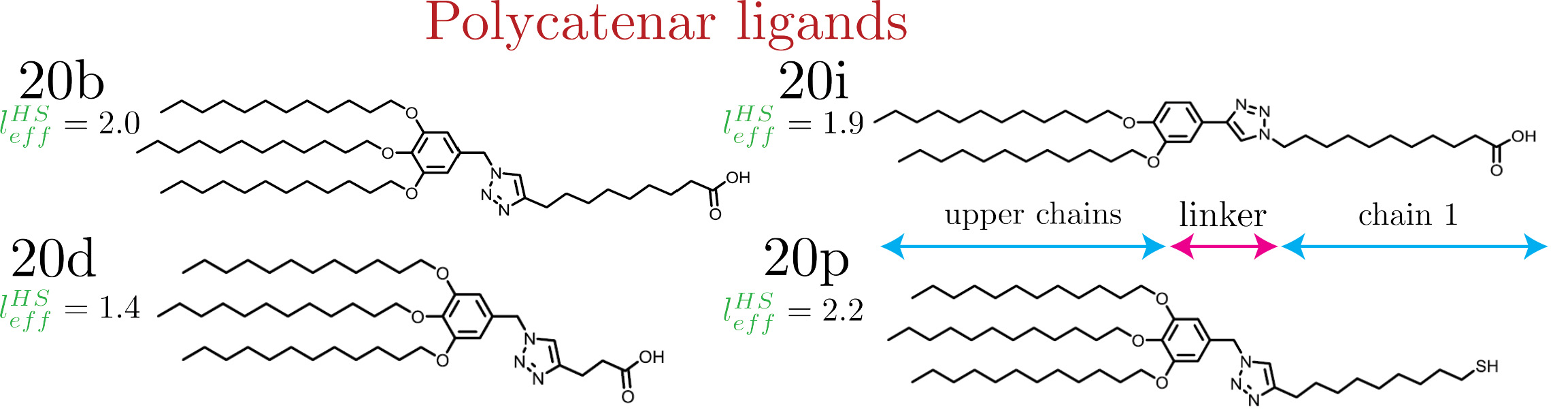}
    \caption{Four representative polycatenar ligands used to assemble BNSLs. $l_{eff}^{DH}$ is the effective size (in nm) of the ligand shell used in the paper Ref.~\cite{Diroll2016}. Notations follow the same convention as the original paper.}\label{fig:polycatenar}
\end{figure}

An interesting more complex example is provided by polycatenar ligands\cite{Diroll2016}, shown in Fig.~\ref{fig:polycatenar}. They consist of a thiol or carboxylic group hydrocarbon chain linked to the NC core, with an intermediate linker with two or three hydrocarbon chains linked through an ether group. The HS diameter of a NC in the original reference was obtained from
\begin{equation}\label{Eq:polycatenar}
    d_{HS}=d_{core}+l_{eff}^{HS} \ ,
\end{equation}
where $l_{eff}^{HS}$ is given in Fig.~\ref{fig:polycatenar} for the four more relevant ligands. Single component systems assemble into bcc and fcc or hcp consistently with previous discussions. 

BNSL phases are also reported. The most notable aspects relevant for this review are the presence of AuCu$_3$ (see discussion after Table~\ref{tab:aucu3}), which is obtained for three different combinations: 6.5nm 20p-Au+2.4nm 20d-CdSe ($\gamma=0.48$), 6.5nm 20p-Au + 5.5nm Oleate-CdSe ($\gamma=0.44$), 5.5nm 20b-PbSe + 2.4nm 20d-CdSe($\gamma=0.55$). There is also an ubiquitous presence of AuCu for many values of $\gamma$ where there is no high PF. In light of our previous discussion, the appearance of this phase is surprising, so newer experiments will be necessary to clarify this case.

For MgZn$_2$, BNSLs were reported for 5.3nm Oleate-CdSe + 2.4 20d-CdSe($\gamma=0.71$), 3.5nm 20i-CdSe + 2.4 20b-CdSe ($\gamma=0.88$), 3.5nm 20b-PbSe+2.7nm 20b-CdSe($\gamma=0.89$). These last two values of $\gamma$ would imply vortices for small B-NCs, which disagree with OTM. I believe this can be explained because $\gamma$ is overestimated by using a fixed ligand shell $l_{eff}^{HS}$ in Eq.~\ref{Eq:polycatenar}, neglecting the curvature effect in the OPM formula, which for small NCs is quite important. A rigorous calculation is beyond the scope here, but a rough estimate can be obtained if we just count the number of actual chemical bonds if the upper ligands were just a single chain and assume an effective maximum grafting density $\xi=1$. This gives $n=29, 31$ for 20b,20i ligands, then using the OPM formula Eq.~\ref{Eq:hs:OPM} for the equivalent hydrocarbon chain, one obtains $d^{HS}=6.8$ nm (3.5nm 20b-PbSe) and $d^{HS}=5.6$ nm
(2.7nm 20b-CdSe), for a $\gamma=0.82$, which is in the expected range according to previous discussion.

In summary, an extensive comparison with single- and binary-NC systems, including spherical and cubic shapes, provides overwhelming experimental verification for OTM. Additional stringent validation can be provided from simulations, as I discuss in Sect.~\ref{Sect:Simuls}.

\begin{figure}
    \centering
    \includegraphics[width=1\textwidth]{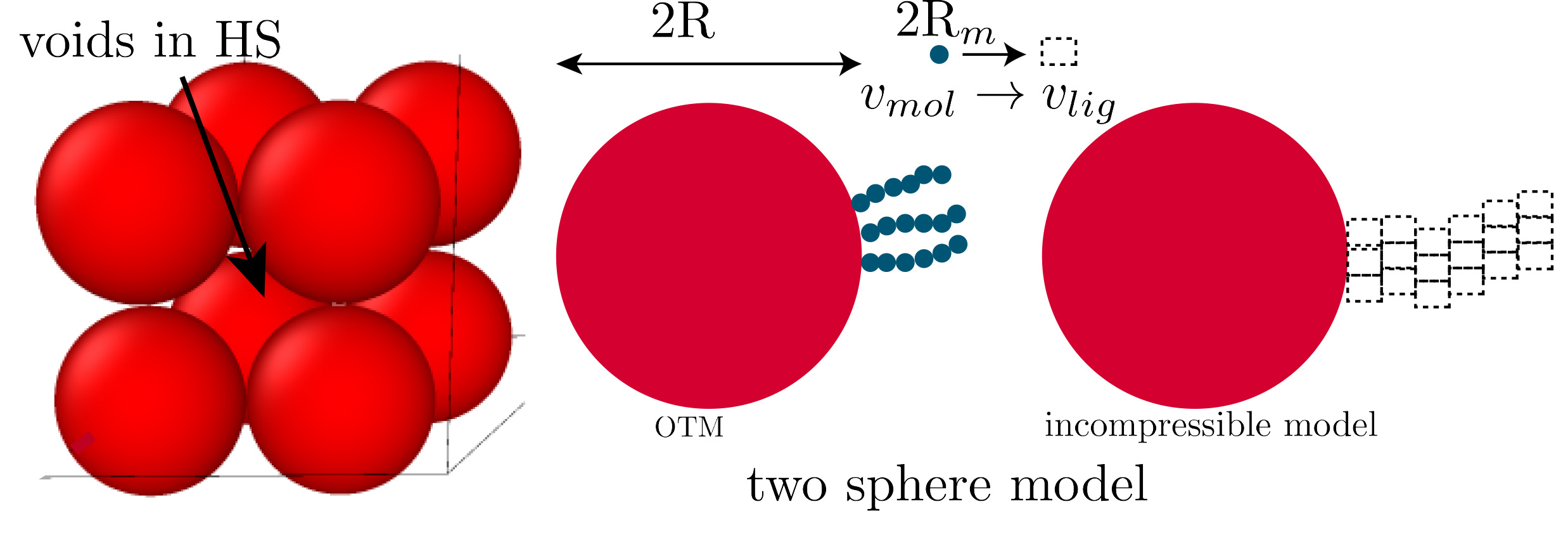}
    \caption{\resp{Example of a HS model illustrating the voids. On the right a two sphere model, where the core is represented by a large sphere of radius $R$ and the ligands consisting of small monomers of radius $R_m$. Also illustrated is $v_{mol}$, the actual molecular volume, and $v_{lig}$, the volume that is used in the incompressibility condition Eq.~\ref{Eq:incompress:sc}, which is roughly the inverse of the number density of the monomer liquid.} } \label{fig:incompres}
\end{figure}

\resp{
\subsection{The incompressibility condition}\label{Subsect:incompressibility}}

\resp{The excellent agreement between the PF from experiments and the OTM, see for example Fig.~\ref{fig:exp_theory_comp}, does show that there is plenty (a fraction $1-\eta$ of the total volume) of space that is not occupied by matter. This may suggest the existence of ``voids'', similar to those present in the HS models; see Fig.~\ref{fig:incompres}. Related systems, such as polymers, surfactants or lipids, for example\cite{Israelachvili2000}, show no evidence of such voids in polymer melts, micelles or bilayers. Instead, what is found is a situation in which the density is constant throughout the phase. Furthermore, our own simulations see Fig.~\ref{fig:simul_sc} or Fig.~\ref{fig:bnsl_sims}, do not show any voids and are consistent with a constant ligand density. The physical expression of a system where the density remains constant is through an incompressibility condition. For a single component NC system within a two sphere model, see Fig.~\ref{fig:incompres}, this condition is
\begin{equation}\label{Eq:incompress:sc}
    V_{WC} = \frac{\pi}{6}(2R)^3+4\pi R^2 \sigma n v_{lig} \ ,
\end{equation}
where $V_{WC}$ is the Wigner-Seitz cell, assuming a Bravais lattice for simplicity, $n$ is the number of monomers within a ligand and $v_{lig}$ is the volume occupied by a single monomer, as illustrated in Fig.~\ref{fig:incompres}. Note that this equation immediately implies that
\begin{equation}\label{Eq:incompress:bcc_to_fcc}
    V_{bcc}=V_{fcc}=V_{WC} \rightarrow  d_{HS} = \left(\frac{3\sqrt{3}}{4\sqrt{2}}\right)^{1/3} d_{HS} = 0.972081 
    \cdot d_{HS} \ ,
\end{equation}
which is Eq.~\ref{Eq:exp_comp:bcc}. This illustrates that the conceptualization of NC assembly in terms of the OTM does not imply the presence of voids, to the contrary, is entirely consistent with the existence of an incompressibility condition imposing a constant monomer density throughout.}

\resp{I note that the incompressibility condition provides a way to generalize the OTM formulas to the swollen case according to
\begin{equation}\label{Eq:incompress:otm}
    V_{WC} = \frac{\pi}{6}(2R)^3+4\pi R^2 \sigma n v_{lig} + N_s v_s \ ,
\end{equation}
where $N_s$ is the number of solvent molecules and $v_s$ is the corresponding solvent molecular volume. If the solvent consists of chains of $n_c$ monomer ligands, for example dodecanethiol as ligand and n-hexane as solvent, then $v_{s}\approx n_{c} v_{lig}$. The consequences of Eq.~\ref{Eq:incompress:otm} are beyond the scope of this study and will be developed elsewhere.
}

\resp{The incompressibility condition Eq.~\ref{Eq:incompress:sc} maybe re-expressed as
\begin{equation}
    V_{WC}=\frac{4\pi}{3} R^3\left(1+3\xi\lambda \frac{n v_{lig}}{A_0 L}\right) \equiv \frac{4\pi}{3} R^3\left(1+3\xi\lambda \frac{v_{lig}}{v_{mol}}\right) ,
\end{equation}
with $v_{mol}$ the molecular volume occupied by a single ligand monomer. Since the volume occupied by matter, Eq.~\ref{Eq:hs:vol_matter} is $V_{matter}=\frac{4\pi}{3} R^3(1+3\xi \lambda)$ the PF becomes
\begin{equation}\label{Eq:incompress:pf}
    \eta = \frac{V_{matter}}{V_{WC}}=\frac{1+3\xi \lambda}{1+3\xi\lambda \frac{v_{lig}}{v_{mol}}} \ ,
\end{equation}
which illustrates that the free space resulting from $\eta<1$ is generally not a large big void, as in HS models, but rather, is distributed among the ligands to provide free volume and enable vibrations and other ligand conformational changes.  Just for proper context, it is of interest to have some estimates. If the ligands are simple hydrocarbon chains, it should be expected that $v_{lig} \approx 30$ \AA$^3$, consistent with a typical density of a hydrocarbon monomer in the liquid state. Then $v_{mol} \approx A_0 \cdot b = 17\cdot 1.2= 20$ \AA$^3$, where $b$ is estimated from Eq.~\ref{Eq:hs:L:estimate} so that
\begin{equation}
    \frac{v_{lig}}{v_{mol}}\approx 1.5 \ .
\end{equation}
}
\resp{Eq.~\ref{Eq:incompress:pf} shows PF close to 1 require $\xi \lambda < 1$, that is, situations where the packing is dominated by the NC core, although if $\lambda$ is too small, ligands cannot reach to fill the far away points of the Wigner-Seitz cell, and voids will occur. In the limit dominated by the ligands 
$\xi \lambda >> 1$, it is $\eta \approx \frac{v_{mol}}{v_{lig}}$, that is, the PF is limited by the assumed spherical monomers, see Fig.~\ref{fig:incompres}. }.

\begin{figure}
    \centering
    \includegraphics[width=1\textwidth]{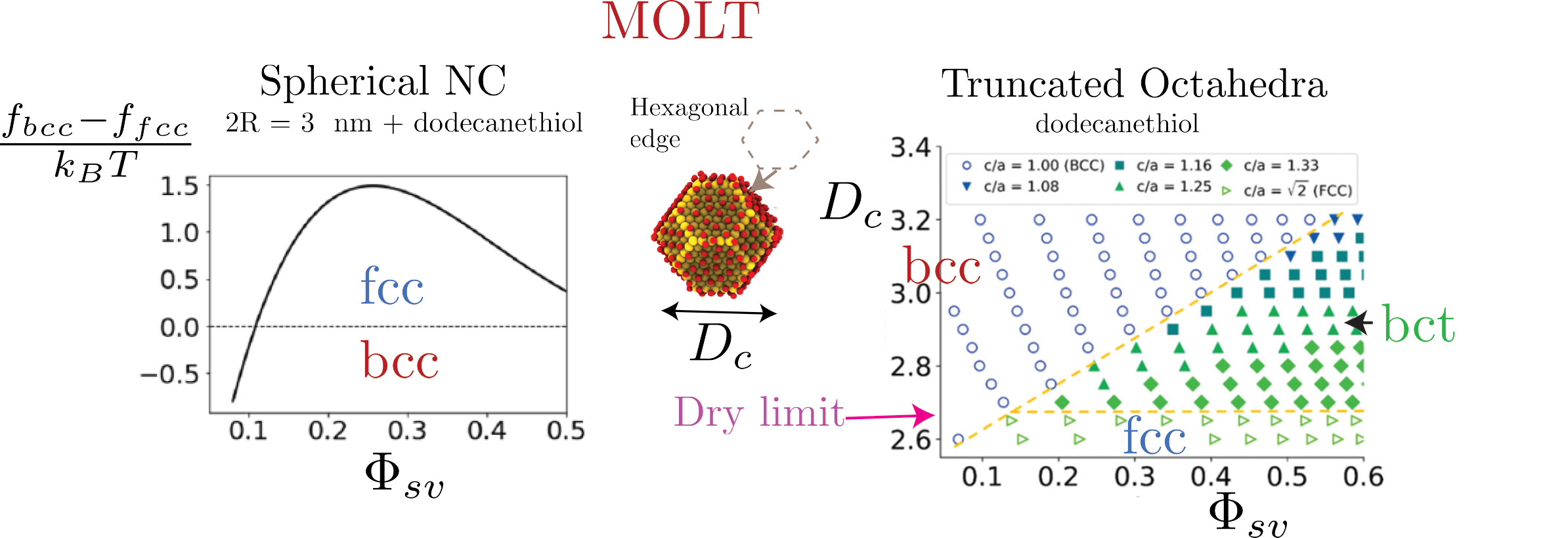}
    \caption{MOLT results from the papers by Missoni and Tagliazucchi\cite{Missoni2020, Missoni2021}. On the left, the case of a spherical NC were only bcc and fcc exist as equilibrium phases. On the right, with truncated octahedra NCs bct becomes stable at higher solvent content. The quantity $\Phi_{sv}$ is the solvent volume fraction defined in Eq.~\ref{Eq:MOLT:phi_s} and has a somewhat different meaning as in the rest of the paper, see discussion in the text. The grafting density is $\sigma= 4$ ligands per nm$^2$. $D_c$ is the actual diameter of the truncated octahedra, where different sizes vary the hexagonal edge, see the original references for additional details.} \label{fig:molt}
\end{figure}

\subsection{Towards an OTM free energy: mean field models}\label{Subsect:MOLT}

OTM provides a detailed prediction of a given BNSL structure but cannot establish which lattice defines the equilibrium state because it lacks an explicit free energy. Such a putative free energy must explicitly include ligand conformations. The Molecular Theory (MOLT) developed in Ref.~\cite{Missoni2020, Missoni2021}, inspired from previous models in lipid bilayers and other soft systems\cite{BenShaul1985, Szleifer1985, Carignano1994, Tagliazucchi2010} is able to incorporate ligand conformations in a minimal and computationally affordable way.

The advantage of MOLT compared with simulations is fast turnaround and no statistical fluctuations. This also enables calculations for processes consisting of a succession of equilibrium states, i.e. quasistatic. A limitation for MOLT is that it requires NC positions and orientations, so going beyond single-component systems requires specifying many degrees of freedom and actual calculations require some prescription for the NC positions and orientation. OTM does provide positions and orientations so MOLT and OTM together enable, in principle, a first-principle determination for any NC configuration.

The original MOLT papers\cite{Missoni2020, Missoni2021} assumed that the \resp{incompressibility condition Eq.~\ref{Eq:incompress:sc}}. Therefore, the volume fraction \resp{occupied by} solvent is
\begin{equation}\label{Eq:MOLT:phi_s}
    \Phi_{sv}=1-\frac{V_{WC}}{V} \ ,
\end{equation}
\resp{see Eq.~\ref{Eq:incompress:sc}}. Applied to single-component systems, MOLT predicts that the equilibrium structure for a NC with a spherical core is either bcc or fcc only, see Fig.~\ref{fig:molt}. However, the difference in free energy 
\begin{equation}\label{Eq:MOLT:df}
    f_{bcc}-f_{fcc} \approx -0.5 k_B T \ .
\end{equation}
at the lowest solvent volume fraction $\Phi_{sv}=0.1$ studied. Fig.~\ref{fig:bnsl_sims} shows that the actual values of the free energy per NC in the dry limit for both $f_{bcc},f_{fcc}$ are of the order of 500$k_BT$. Therefore, calculating Eq.~\ref{Eq:MOLT:df} from numerical simulations requires free energies obtained with a precision of less than 0.1$\%$, which is extremely challenging and more than justifies the efforts to develop methods complementary and alternative to simulations.

When the core is modeled with a more realistic shape, a truncated octahedra, fcc and bcc are still the only equilibrium phases at low solvent fraction $\Phi_{sv} \lesssim 0.1$, but the bct phase becomes stable for larger solvent volume fractions, as illustrated in Fig.~\ref{fig:molt}. These calculations establish that bct is an intermediate during solvent evaporation and not an equilibrium phase in the dry limit, as discussed in SubSect.~\ref{SubSect:HS:Experiments}. Ref.~\cite{Weidman2016} has experimentally shown the presence of bct intermediates. These examples illustrate the power of MOLT, not just to describe equilibrium, but to characterize intermediates during solvent evaporation, a topic that I elaborate on further in Sect.~\ref{Sect:Dynamics}.

An exciting new development is the generalization of MOLT in the grand canonical ensemble, enabling the solvent vapor pressure to be monitored as an external parameter, much like an experiment. \resp{The new theory, dubbed MOLT-CF, allows to verify the incompressibility condition Eq.~\ref{Eq:incompress:sc}, compute free energies in the dry limit and monitor intermediates during solvent evaporation, very much as in a real experiment. It does provide therefore a completely general model to establish what superlattices are equilibrium states.}

\section{Numerical Simulations}\label{Sect:Simuls}

\subsection{HOODLT and technical simulation details}

HOODLT is a software designed to perform calculations with NCs. It follows a design using the best software practices available \cite{Wilson2014,Hunt1999,Wilson2017}, allows full reproducibility of results, simplifies enormously running new simulations and drastically streamlines the complexities of NC calculations. Free energies are calculated either approximately by Dynamic Lattice Theory\cite{Born1954,Hoover1970,Hoover1971,Hoover1972}, or exactly using different algorithms\cite{Calero2016,Zha2018,Waltmann2018a}. HOODLT includes more than 100 NC cores, as described in Refs.~\cite{Lai2019, Lai2019a}), ligands (hydrocarbons, polystyrene, polyethylene glycol, etc.), and solvent (water, toluene, hydrazine, etc.). There are methods to build initial configurations, run simulations through HOOMD-Blue\cite{AndersonMe2008a}, algorithms (identify vortices, motifs, etc.) or data analysis, for example, through WHAM\cite{Kumar1992a} and Bridge Sampling\cite{Shirts2008}. The crystal structure database implements all analytical OTM formulas\cite{Travesset2017} such as Eq.~\ref{Eq:otm:mgzn:Eq1}-Eq.~\ref{Eq:otm:mgzn:Eq4}.
HOODLT is freely available in bitbucket

https://bitbucket.org/trvsst/hoodlt/src/master

see Ref.\cite{TravessetHOODLT}. The force field more commonly used in our simulations is described in great detail in our previous papers~\cite{Waltmann2017, Zha2018, Waltmann2018a, Waltmannn2019, Macias2020, Patra2020}. Atoms within the NC are modeled with rigid body dynamics\cite{Nguyen2011,Glaser2020}. HOODLT\cite{Travesset2014} has been used to implement the force field, prepare simulations, create initial configurations, and analyze the data. Almost all simulations have been run at a nominal temperature of $T_m=387$ K. This temperature is significantly higher than the one most commonly used in experiments, but the main goal of the simulations is to verify the OTM and higher temperatures enable better thermalization. In regards of the parameter $\xi$, see Eq.~\ref{Eq:hs:OPM}, it depends on the maximum grafting density $\sigma_{Max}=\frac{1}{A_0}$ (or the ligand footprint). There are methods to calculate this quantity from simulations\cite{Jimenez2010,Silbaugh2017} and for very small NCs $<2$ nm where the core curvature is large, $\sigma_{Max}$ may depend on NC size\cite{Zha2020}. Herein I concentrate on the simulation results. The simulation technical details are extensively discussed in the original papers.

\begin{figure}
    \centering
    \includegraphics[width=1\textwidth]{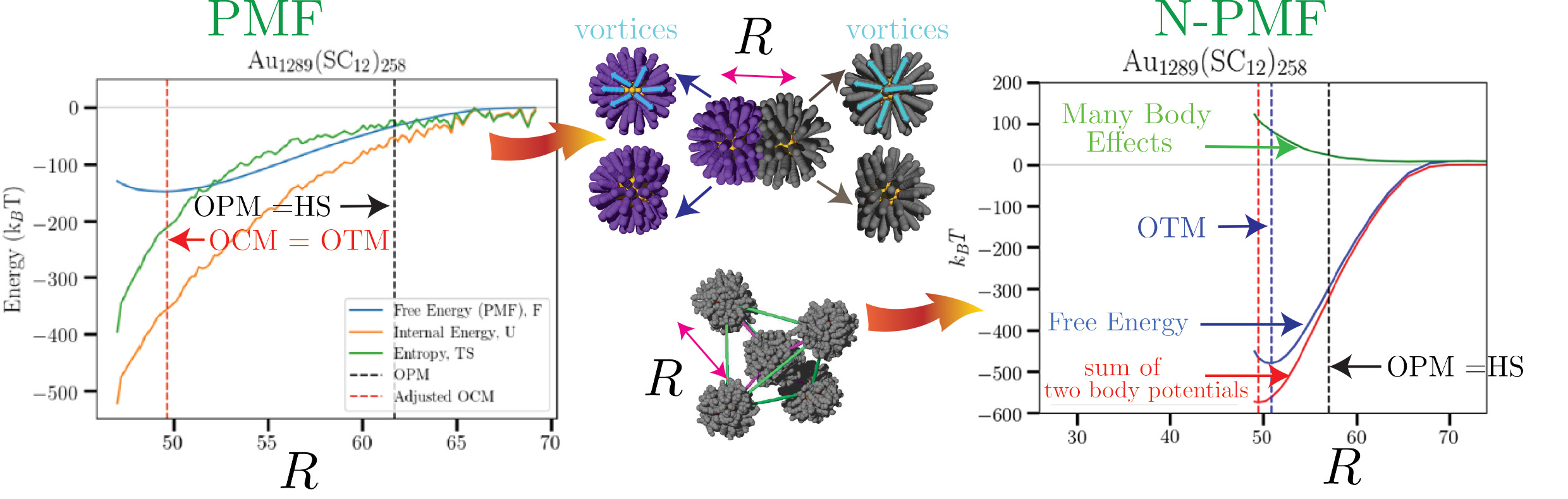}
    \caption{Calculation of the PMF and N-PMF for a tetrahedra with Au$_{1289}$(SC$_{12}$)$_{258}$, which are NC consisting of 1289 Au atoms for a 4.0 nm diameter, with 258 dodecanethiol ligands. The N-PMF is a tetrahedra with a NC at the center, where the edge coordinate and the NC-NC distance are quasistatically decreased. The many body effects $\Delta^{MB}$ defined in Eq.~\ref{Eq:simu:many_body_param}, are obtained by subtracting from the actual free energy the two body free energy: the sum of pairs of potentials of mean force.}\label{fig:simul_clusters}
\end{figure}

The calculations are done by subjecting the NCs with harmonic springs and obtaining different observables as a function of NC separation. There are different ways, as discussed in the original references\cite{Zha2018, Waltmann2018a} to eliminate the possible bias introduced by the springs. The free energy of a superlattice is obtained by integrating the pressure, where the equilibrium lattice constant is obtained from basic thermodynamic identities
\begin{equation}\label{Eq:simul:pressure}
    P=-\frac{\partial{F}}{\partial V} \implies \left(\frac{\partial{F}}{\partial V}\right)_{V=N_{WC}V_{WC}(a)} = 0 \ ,
\end{equation}
so the lattice constant is defined at zero pressure. Here $N_{WC}, V_{WC}(a)$ is the number and volume of Wigner-Seitz cells, where the latter quantity explicitly depends on the actual value of the lattice constant $a$.

\subsection{PMF for pairs and N-PMF for clusters}\label{SubSect:Simuls:Clusters}

The PMF, which here means the free energy of two NCs, is obtained by two methods: 1) {\em Force Integration} of the total forces in the NCs and 2) {\em Weighted Histogram Analysis Method} (WHAM)\cite{Kumar1992a}, with both methods providing equivalent results, as discussed in great detail in Ref.~\cite{UpahTravesset2023}. Fig.~\ref{fig:simul_clusters} shows a concrete example of a 4 nm core functionalized with dodecanethiol at close to maximal grafting density, from which it is possible to establish the following points:

\begin{itemize}
\item At equilibrium (minimum of the PMF) HS breaks down:
NC separation is approximately 10 \AA \ closer than predicted by the HS models, which is defined by OPM  Eq.~\ref{Eq:hs:OPM}.
\item The equilibrium separation is defined by OCM, with some nuances, as described in Ref.~\cite{Waltmann2017}.
\item The PMF minimum is very deep and on the order of -100$k_BT$.
\item The internal energy is 3-4 times higher (in magnitude) than the free energy.
\end{itemize}

Analysis of many other cases establishes that these statements are general. The last statement shows that the free energy is the result of a significant entropy-energy cancellation\cite{Elimelech2020}. The PMF minimum ($F_{min}$) defines the cohesive bonding energy between two NCs. Certainly, $100 k_BTs$ is a large value, which often ignites many passionate discussions about whether these systems are trapped in metastable states and can actually reach thermal equilibrium. This point will be discussed further below in Sect~\ref{Sect:Dynamics}. Here I just emphasize that this large cohesive energy has been consistently obtained in all PMF calculations using different force fields and NC shapes and ligands. Also, cohesive energies grow with NC core size and the length of the ligand. The Vlugt group, for example, summarized their PMF results by the following formula~\cite{Schapotschnikow2009} 
\begin{equation}\label{Eq:simul:f_min_vlugt}
    F_{min} \approx  -1.15 (n+1)^2 k_B T \ ,
\end{equation}
where $n$ is the number of carbon atoms within the alkyl group. For dodecanethiol (n=12) ligands it is $F_{min} \approx -194 k_B T$. Other estimates are entirely consistent with these calculations and can be read in Ref.~\cite{Jabes2014, Yadav2016, Milowska2016, Lange2015, Waltmann2017, Waltmann2018, Bauer2017, Kister2018, LiuLu2019}.

The $N-$ body PMF (N-PMF) has been obtained for many different configurations, such as $P_n$ (NCs in a regular polygon with $n$ vertices surrounding a NC at its center), tetrahedra  shown in Fig.~\ref{fig:simul_clusters} and many others, see Ref.~\cite{Waltmann2018a}. The N-PMF is obtained by holding the NCs with springs and computing the reversible work as the NCs are brought together, as described in Fig.~\ref{fig:simul_clusters}. Many body effects quantify the difference between the actual free energy and the one obtained from assuming two-body interactions only
\begin{equation}\label{Eq:simu:many_body_param}
\Delta^{MB} (T,R) = F_N(T,R) - \sum_{i=0}^N \sum_{i>j} ^N F_2(T, R_{ij}) \ ,
\end{equation}
where $F_N(T,R)$ is the free energy (N-PMF) of the system and $F_2(T, R_{ij})$ is the free energy (PMF) for two NCs separated by distance $R_{ij}$, 

Results for N-PMF are summarized in Fig.~\ref{fig:bnsl_summary} and follow the general conclusions
\begin{itemize}
    \item For coordination less than six, the HS description breaks down, consistent with the OTM assumptions.
    \item Eigenshapes are developed for equilibrium configurations, such as shown in Fig.~\ref{fig:eigen}.
    \item Equilibrium distances increase with coordination: as an example, the N-PMF for a tetrahedra in Fig.~\ref{fig:simul_clusters} clearly shows that the equilibrium, pointed as OTM(dashed blue line) is further away from the minimum of the PMF(dashed red line) thus indicating that with coordination 4 NC are separated further than coordination 2.
    \item There are significant many body effects (marked in green in Fig.~\ref{fig:simul_clusters}), defined by Eq.~\ref{Eq:simu:many_body_param}.
    \item The cohesive free energy is of the order of hundreds of $k_B T$s.
\end{itemize}

These PMFs and N-PMFs provide a rigorous validation of the  assumptions that lead to OTM, as discussed in SubSect.~\ref{Subsect:otm:formalism}. Fig.~\ref{fig:bnsl_summary} summarizes the results of the simulations, clearly showing how the nearest-neighbor distance increases with coordination and approximately saturates for 6, where the results are basically consistent with the HS description, defined by the OPM, see Eq.~\ref{Eq:hs:OPM}. Fig.~\ref{fig:eigen} shows snapshots of representative simulations at the minimum of the N-PMF, with the vortices explicitly drawn for coordination 2 and 4, thus illustrating how NCs spontaneously generate a valence and an effective eigenshape, as discussed in more detail above. At coordination 12 the NCs become isotropic with the corresponding eigenshape of a sphere, in agreement with OTM.

\begin{figure}
    \centering
    \includegraphics[width=1\textwidth]{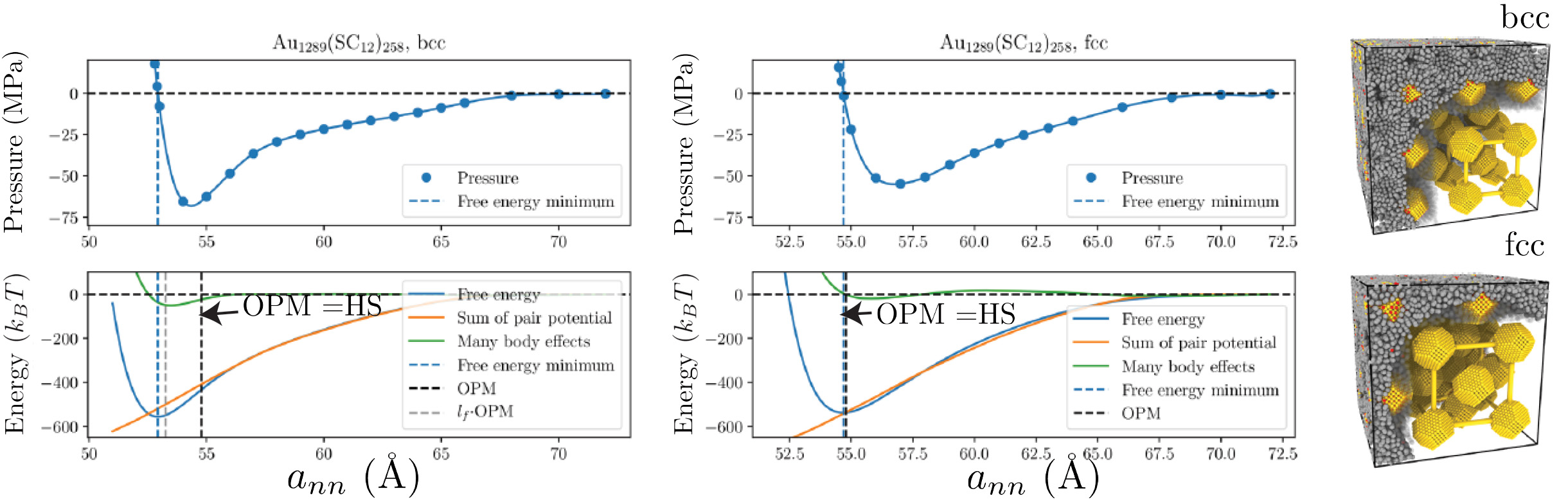}
    \caption{Free energy calculation of a bcc and fcc superlattice with Au$_{1289}$(SC$_{12}$)$_{258}$. The many body effects are obtained by subtracting from the actual free energy the two body free energy: the sum of pairs of potentials of mean force, calculated in Fig.~\ref{fig:simul_clusters}. The lattice constant shown with blue dashed line is the minimum of the free energy, see Eq.~\ref{Eq:simul:pressure}. OPM and $l_f$-OPM are shown in black and grey dashed lines.}\label{fig:simul_sc}
\end{figure}

\subsection{Superlattices}\label{SubSect:Simuls:superlattices}

\begin{figure}[htb]
    \centering
    \includegraphics[width=1\textwidth]{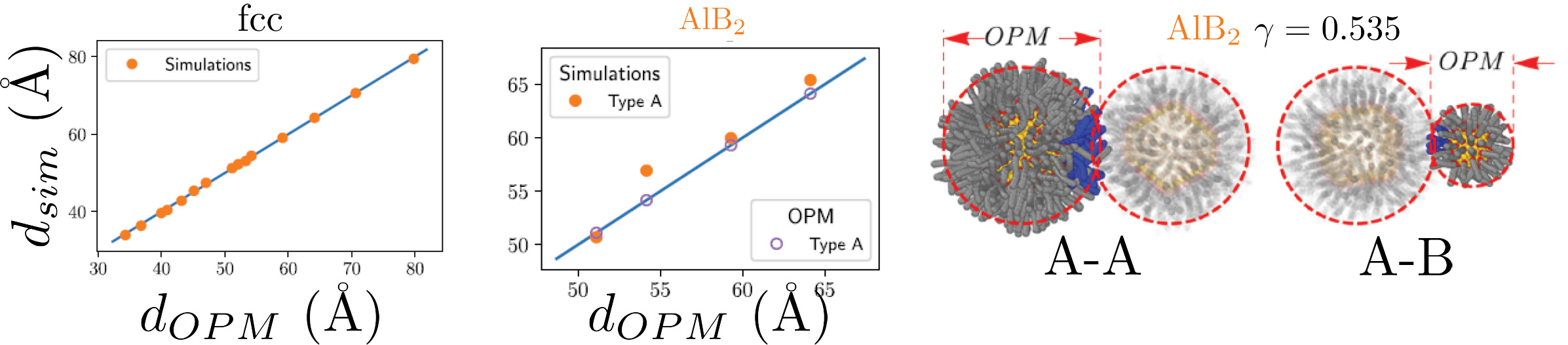}
    \includegraphics[width=1\textwidth]{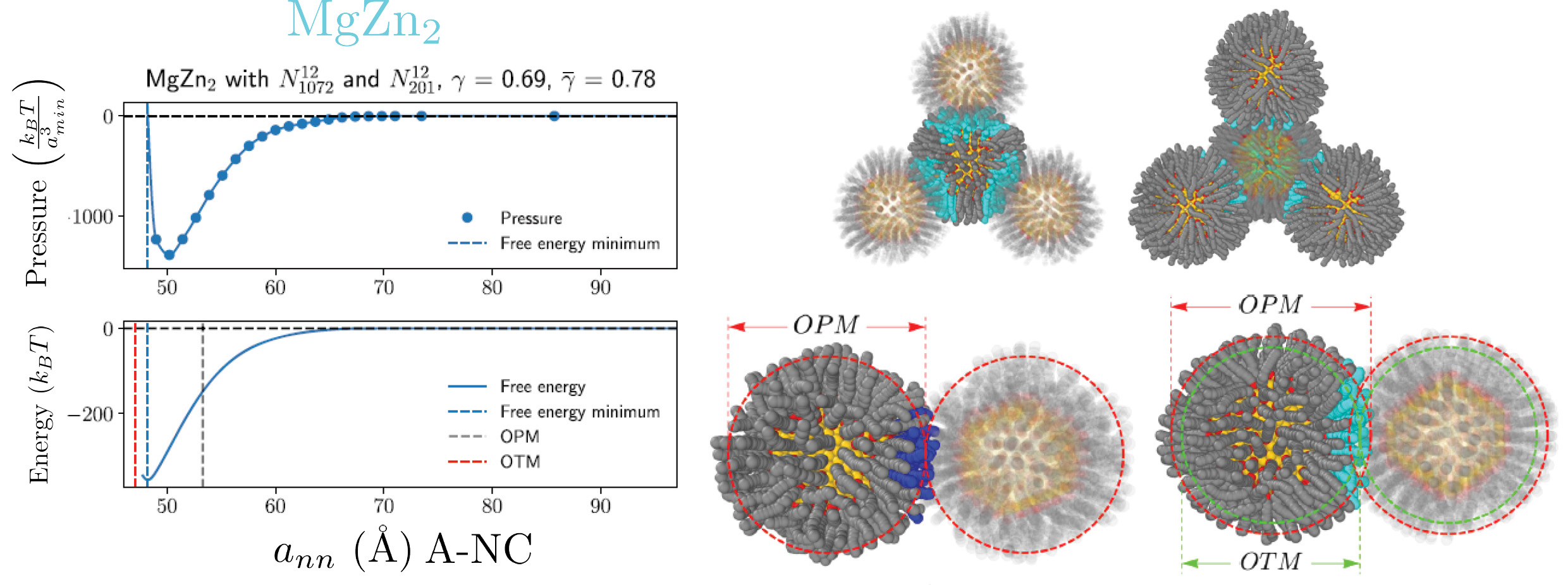}
    \caption{Top right: Comparison of predicted lattice constant for fcc and AlB$_2$ for many different superlattices, showing excellent agreement with HS. Bottom left shows the same calculation as in Fig.~\ref{fig:simul_sc} for MgZn$_2$ with the minimum correctly predicted from OTM and in disagreement with HS. The right side shows a detailed analysis of simulation snapshots at the free energy minimum, with actual nearest neighbor separations predicted from OTM and eigenshapes in agreement with Fig.~\ref{fig:eigen}.}\label{fig:bnsl_sims}
\end{figure}

Simulations of superlattices with ligands containing between 5 and 20 carbons have been reported in Ref.~\cite{Zha2018, ZhaTravesset2021} within all-atom and united models. Typical single-component and binary systems are shown in Fig.~\ref{fig:simul_sc} and Fig.~\ref{fig:bnsl_sims}, where the free energy (in blue, bottom) is obtained by integrating the pressure (in blue, top) as a function of the nearest neighbor distance $a_{nn}$. The equilibrium lattice constant is obtained at zero pressure, i.e. the minimum of the free energy, as described in Eq.~\ref{Eq:simul:pressure}.

OTM predicts that a fcc lattice consists of packing HS, and indeed, Fig.~\ref{fig:simul_sc} shows that the lattice constant is accurately predicted by OPM. Fig.~\ref{fig:bnsl_sims} summarizes the results of many fcc simulations, and all of them align very accurately with the OPM result. For bcc, the OPM is not quite accurate, but this is expected, as the lattice constant or nearest-neighbor distance is given by Eq.~\ref{Eq:exp_comp:bcc}, marked as $l_f$-OPM in Fig.~\ref{fig:simul_sc}. 

In a fcc, the free energy as a function of lattice constant is very well represented by a sum of two body potentials all the way to the equilibrium result. That is, many body effects are very small. For the bcc, there are significant many body effects, see Eq.~\ref{Eq:simu:many_body_param}, near the free energy minimum. The difference in free energy between fcc and bcc is too small, and simulations were not able to establish which lattice describes equilibrium and confirm the derivation (and experimental results) in Eq.~\ref{Eq:exp_comp:bcc_fcc}, see the discussion in Ref.~\cite{Zha2018}. The MOLT\cite{Missoni2020, Missoni2021} mean field model briefly discussed in SubSect.~\ref{Subsect:MOLT} is able to discriminate between bcc/fcc with excellent quantitative agreement with Eq.~\ref{Eq:exp_comp:bcc_fcc}, thus providing an important validation of its predictive power. 

Coarse-grained models have been developed for these problems\cite{FanGrunwald2019}. Particularly relevant are potentials derived from machine learning\cite{CamposVillalobo2022, Giunta2023}. This is an important area for further development, as all atom or united models have obvious limitations that hamper reaching large spatial or temporal scales. An interesting result from Ref.~\cite{Giunta2023} is the negligible effect of many body effects for fcc, in agreement with the results discussed above, see Fig.~\ref{fig:simul_sc}.

Simulations for the MgZn$_2$ BNSL at $\gamma=0.69$ are shown in Fig.~\ref{fig:bnsl_sims}. The minimum (blue dashed line) is clearly off from the OPM result (grey dashed line), thus illustrating the breakdown of the HS description. The OTM result, see Eq.~\ref{Eq:otm:mgzn:Eq3}, predicts the free energy minimum, thus providing yet another cross-check on its validity. Snapshots of the NC configurations at the minimum demonstrate that the large A-NCs develop a four-valence with tetrahedral eigenshapes,  identical to the one obtained in Fig.~\ref{fig:eigen} by direct simulations with tetrahedral clusters, made possible by the presence of vortices. Explicit comparisons show that the OPM sizes overlap and that OTM accurately defines NC contact (or kissing). Fig.~\ref{fig:bnsl_sims} includes a control simulation of a single component with the same A-NCs in fcc where the sizes are described by OPM, nicely demonstrating the OTM prediction where the same NC develops a different valence and eigenshape depending on the environment (the coordination).

Simulations for the AlB$_2$ BNSL are also shown in Fig.~\ref{fig:bnsl_sims}.  All of them are in excellent agreement with OPM, showing that this BNSL consists of packing HS. The equilibrium snapshots for both A-A and A-B contacts are in excellent agreement with the OPM. Thus, the simulations confirm that AlB$_2$ is accurately described by HS. This analysis has been carried out for all other BNSLs reported in the literature and refer to Ref.~\cite{ZhaTravesset2021} for a detailed discussion. The most important point is that they fully validate OTM predictions.

\begin{figure}
    \centering
    \includegraphics[width=1\textwidth]{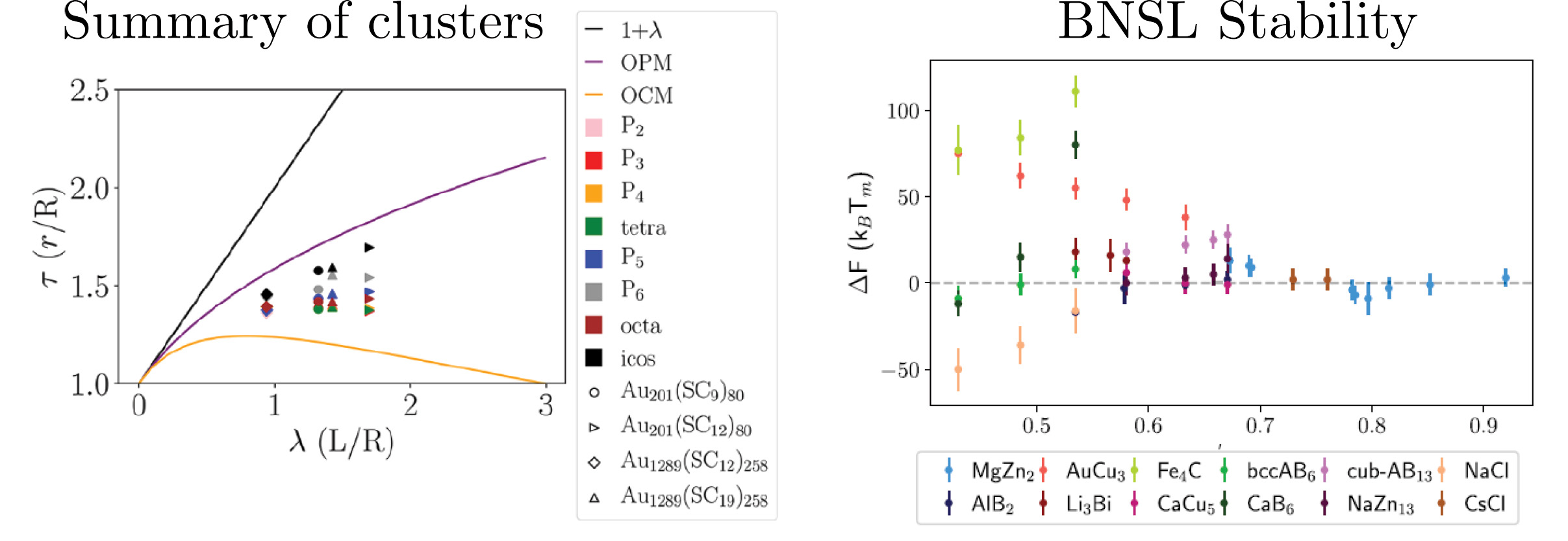}
    \caption{\resp{Left}: Summary of all simulations with clusters. Here $P_n, n=2-6$ are planar clusters with the shape of regular polygons with $n$ vertices. Three polyhedra (tetra=tretrahedra, octa=octahedra and ico=icoshaedra) are also considered. The lattice constant is in between OPM (HS) and OCM, in agreement with OTM and experimental results. \resp{Right}: summary of BNSL stability, Eq.~\ref{Eq:simul:Excess} for all BNSL simulated in Ref.~\cite{ZhaTravesset2021} as a 
 function of $\gamma$, see definition in Eq.~\ref{Eq:hs:def_gamma}.}\label{fig:bnsl_summary}
\end{figure}

A general $A_pB_q$ BNSL, with $p$ A-NCs and $q$ B-NCs within the Wigner-Seitz cell has a free energy $F(A_p B_q)$ normalized by the total number, both $A, B$, of NCs. The quantity $\Delta F$
\begin{equation}\label{Eq:simul:Excess}
	\Delta F = F(A_p B_q)-\frac{1}{p+q}\left(p F(A)+q F(B)\right) \ ,
\end{equation}
where $F(A/B)$ is the free energy of the corresponding single component system, determines whether the BNSL is stable against phase separation (at zero pressure) into two single component superlattices. 

Fig.~\ref{fig:bnsl_summary} shows $\Delta F$ as a function of $\gamma$, see Eq.~\ref{Eq:hs:def_gamma}, for all BNSLs studied in Ref.~\cite{ZhaTravesset2021}. The main conclusions are summarized as:
\begin{itemize}
    \item There are three BNSL: Fe$_4$C, cub-AB$_{13}$ and AuCu$_3$ that are not stable against phase separation for any of the $\gamma$ values for which they were simulated.
    \item NaCl is the only BNSL that is clearly stable, at least for $\gamma< 0.5$.
    \item Stability for all other BNSLs could not be established within the accuracy of the simulation.
\end{itemize}

Regarding the three BNSL that should separate into a single component, Fe$_4$C and cub-AB$_{13}$ were identified in the early days in Ref.~\cite{Shevchenko2006a} and Ref.~\cite{Shevchenko2005} but have not been reported, at least to my knowledge, in more recent experiments. It might be necessary to reanalyze the experimental data to clarify these BNSLs. AuCu$_3$, on the other hand, see Fig.~\ref{fig:otm_pf_bnsl}, has been reported for many different cores and ligands, as shown in the five examples in Table~\ref{tab:aucu3}: PbSe, PbS, Ag and Fe$_3$O$_4$ and they are in the correct range of $\gamma$.  The situation of ligand loss, as is the case for AuCu, does not seem plausible. Although the calculations in Fig.~\ref{fig:bnsl_summary} do not apply, the AuCu$_3$ is also present with polycatenar ligands, as described above. Further work will be required to clarify why free energy calculations predict that AuCu$_3$ is unstable against phase separation.  

\begin{table}[htb]
\centering
\begin{tabular}{c c c c c}
\hline
Reference & \multicolumn{4}{c}{NCs (AuCu$_3$ BNSL)} \\
& core & ligand & $d_{HS}$ &$\gamma$\\ \hline
\cite{BodnarchukKovalenkoHeissTalapin2010} & PbSe(7.7), Pd(3.4) & Oleic, C$_{12}$ & 10.8, 5.3  & 0.49\\
\cite{Wei2015} & Ag(9.6),Ag(4.0) & Oleylamine, C$_{12}$ & 12.9, 6.0 & 0.46\\
\cite{Ye2015} & Fe$_3$O$_4$(13.4), Au(3.8)  &  PS (5.3, 3.0)kDa                   & 20.0, 9.8$^\ast$ & 0.49\\
\cite{WangMin2018} & Fe$_3$O$_4$(10.0), Au(5.0) & Oleic, C$_{12}$ & 15.9, 8.6  & 0.54\\
\cite{Coropceanu2019} & PbS(9.3),Au(5.1) & C$_{18}$,C$_{12}$ & 12.6, 7.2 & 0.57 \\ \hline
\end{tabular}
\caption{Five experimental examples reporting the AuCu$_3$ BNSL. In parentheses is the size of the NC core in nm. PS: Polystyrene. The $d_{HS}$ value $\gamma$ is calculated from OPM in Eq.~\ref{Eq:hs:def_gamma}, except for the one marked with $*$, which is quoted from the reference.}\label{tab:aucu3}
\end{table}

The difficulty in confirming whether BNSL define thermodynamic equilibrium shows that many BNSL are marginally stable at best, raising the possibility that subleading effects, like NC-core interactions, dipoles, and other effects, as discussed in Ref.~\cite{ZhaTravesset2021} may tilt the balance one way or another. It emphasizes why it is so relevant to develop alternative calculations, either with mean-field models like MOLT or from coarse-grained simulations, specially with machine learning potentials with diverse environments included in the training set.

\begin{figure}
\centering
\includegraphics[width=1\textwidth]{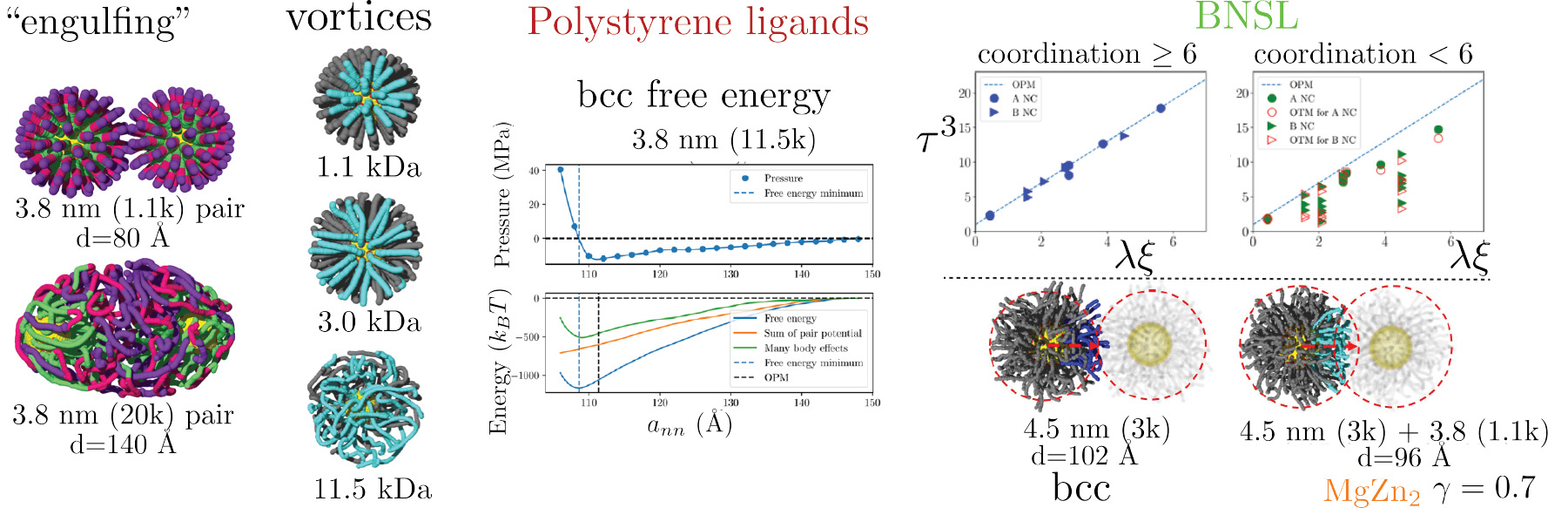}
\caption{Single component and BNSL results from Ref.~\cite{Xia2019a} and Ref.~\cite{Xia2020b}. If polymer ligands are sufficiently large they engulf each other cores, as shown in the figure. Examples of vortices for 
a 3.8 nm core: for short polystyrene ligands the results show nice vortices completely equivalent to the ones with hydrocarbons. \resp{For longer ligands, the vortices are not as well defined but prediction of lattice constants is in agreement with OTM. The top right figure shows that for all BNSLs considered, NCs with coordination $\ge 6$ are well described as HS, while for coordination  $<6$ significant deviations follow, as predicted by OTM, with  similar accuracy as for short ligands Fig.~\ref{fig:simul_sc} and Fig.~\ref{fig:bnsl_sims}}.}\label{fig:bnsl_polymer}
\end{figure}

\subsection{The polymer limit}\label{SubSect:Simuls:polymer}

BNSLs prior to 2015 were obtained with ligands that are relatively short, short enough not to be considered polymers. This situation changed with the experiments in Ref.~\cite{Ye2015}, where Au NCs functionalized with polystyrene with molecular weights in the range of 1.1-20 kDa were reported to assemble into the same BNSLs already obtained with short ligands. The relevance of this development is that it expands the range of NC assembly while connecting it with polymer theory and the abundant literature on polymer brushes\cite{RubinsteinBook2003} in all solvent conditions. Since the focus of this section is mostly on the final assembled structures, i.e dry systems, only the case of poor solvent, mostly non-solvent, is relevant for this discussion. The excellent review in Ref.~\cite{BinderMilchev2012} provides general results for all solvent conditions.

Simulations of NCs with long ligands were carried out in Ref.~\cite{Xia2019a} and Ref.~\cite{Xia2020b} using the coarse-grained model for polystyrene developed in Ref.~\cite{Xia2018}. There are, indeed, new effects when considering ligands that are polymers. An example is given in Fig.~\ref{fig:bnsl_polymer}, where for two NCs, long enough ligands engulf each other NC cores: The ligands from the back of the NC extend and surround the core of the other NC. As the NCs approach and become closer, the minimum of the PMF occurs at a separation significantly lower to that predicted by OPM, so the HS description breaks down. It is possible to define vortices and analyze how they evolve as the polymer size increases.  Fig.~\ref{fig:bnsl_polymer} shows 3.8 nm cores with increasing polystyrene monomer length at the minimum of the PMF: short polystyrene display the same vortices as found with short hydrocarbon chains, thus demonstrating the universality of OTM predictions. For longer polymers, the vortices do not have any obvious geometric structure but, as is elaborated further below, they remain conceptually useful \resp{as, regardless of what their apperance is, OTM formulas still apply.}

Superlattice free energies and lattice constants are determined by the same methods discussed in SubSec~\ref{SubSect:Simuls:superlattices}. Many body effects are quantified by Eq.~\ref{Eq:simu:many_body_param}. However, for long ligands, because of the engulfing effect already mentioned, the PF includes ligand conformations that are not possible with short ligands, so many body effects appear to be very large for both fcc and bcc in Fig.~\ref{fig:bnsl_polymer}. Still, the minimum for fcc is well described by OPM even for a 11.5 kDa polystyrene ligand and longer, see original reference~\cite{Xia2019a}. The bcc minimum is in agreement with Eq.~\ref{Eq:exp_comp:bcc}, see Fig.~\ref{fig:bnsl_polymer}, \resp{which reflects the uniform density of ligands, following the incompressibility condition Eq.~\ref{Eq:incompress:sc}}. Therefore, the evidence from simulations is that the OTM still applies for moderately long ligands.

Extensive simulations of BNSL functionalized with polystyrene have been reported in Ref.~\cite{Xia2019a} for all BNSL experimentally described in Ref.~\cite{Ye2015} (simulated with the same core and ligand)and include explicit calculations of lattice constants and free energies. Fig.~\ref{fig:bnsl_polymer}(right) \resp{summarizes the results for all BNSLs studied with two plots}: The left shows nearest-neighbor separations for all NCs within BNSL with coordination six or larger, while the right shows those with coordination less than six. The left plot follows the OPM result very accurately while in the right the NCs are closer than OPM  and in good agreement with the OTM predictions, which are drawn with empty symbols. Snapshots at equilibrium are shown for the MgZn$_2$ BNSL and demonstrate the expected overlap in the OPM size, with vortices explicitly drawn, illustrating the universality of OTM predictions.

The question on whether OTM applies to moderately long ligands is answered positively by simulations. \resp{As discussed in SubSect~\ref{Subsect:incompressibility}, packing considerations are consistent with ligands described as a melt with uniform density, i.e. an underlying incompressibility condition, which provides the link to polymer theory.}

\begin{figure}
    \centering
    \includegraphics[width=1\textwidth]{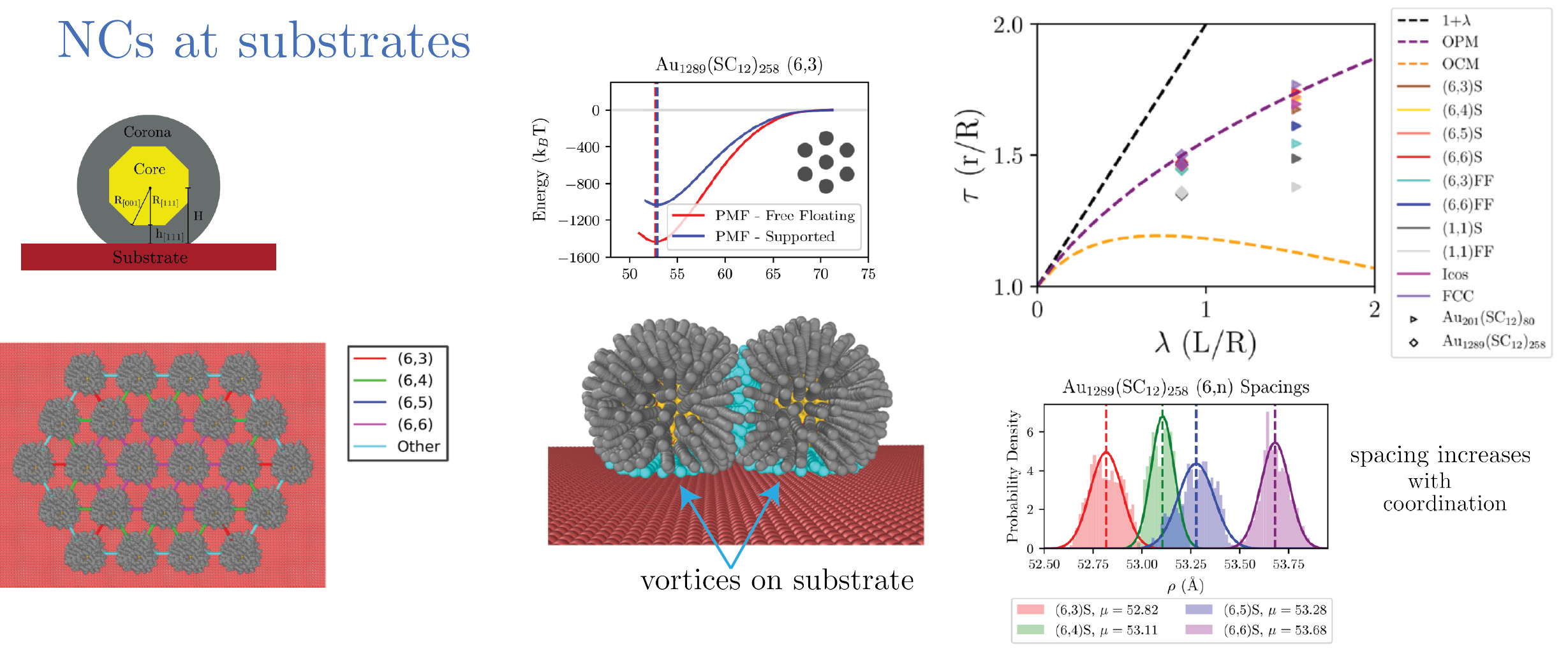}
    \caption{(Left:) Example of a NC on a substrate, explictily showing the role of the relative orientation. Example of a calculation with different bonds, labeled as $(m,n)$ where m,n is the coordination of each NC. (Middle) N-PMF calculated on a supported substrate (S) and Free Floating (FF). Illustration of vortices as induced by the substrate. Right: Summary of all results and example of how nearest neighbor depends on concentration. }\label{fig:substrate}
\end{figure}

\subsection{NCs on a solid support}

NC experiments on substrates, such as those in Fig.~\ref{fig:exp_2d_3d}, enable the test of important properties and effects and have played a very important role in the development of OTM. The presence of a solid support on ligand conformations and eigenshapes needs to be quantified, which provides the motivation for the simulations carried out in Ref.~\cite{Pham2020}. Fig.~\ref{fig:substrate}(Left) shows the setup, where NCs are investigated in different arrangements that include core orientations. Of particular interest are nearest-neighbor distances of the form $(m,n)$ where $m\ge n$ are the coordination of the two NCs involved. A configuration with $(6,n)$ $n=3,4,5,6$ is shown in the figure; see also Fig.~\ref{fig:exp_2d_3d}.

The main physical intuition is that the substrate is akin to a NC with a very large diameter and should induce a vortex on the NC ligands next to it. This intuition is correct, as it is explicitly shown in Fig.~\ref{fig:substrate}(center). Therefore, the effect of the substrate is to add one unit to the coordination or kissing number. The N-PMF shown in Fig.~\ref{fig:substrate}(center) compares a $(6,3)$ configuration without a substrate, i.e. ``free floating'' (FF) and with (S). The free energy is reduced on the latter and there is a small but measurable shift in the minimum, where NCs are further apart because of the increased coordination.

Fig.~\ref{fig:substrate}(Right) at the bottom shows a histogram of $(6,n)$ bonds within the simulation for n = 3,4,5,6 illustrating that the NCs are further away as the coordination increases, in agreement with OTM and the experimental results in Fig.~\ref{fig:exp_2d_3d}. 
Fig.~\ref{fig:substrate}(Right) summarizes all the results where as the coordination is increased, the OPM result is attained. 

\begin{figure}
    \centering
    \includegraphics[width=1\textwidth]{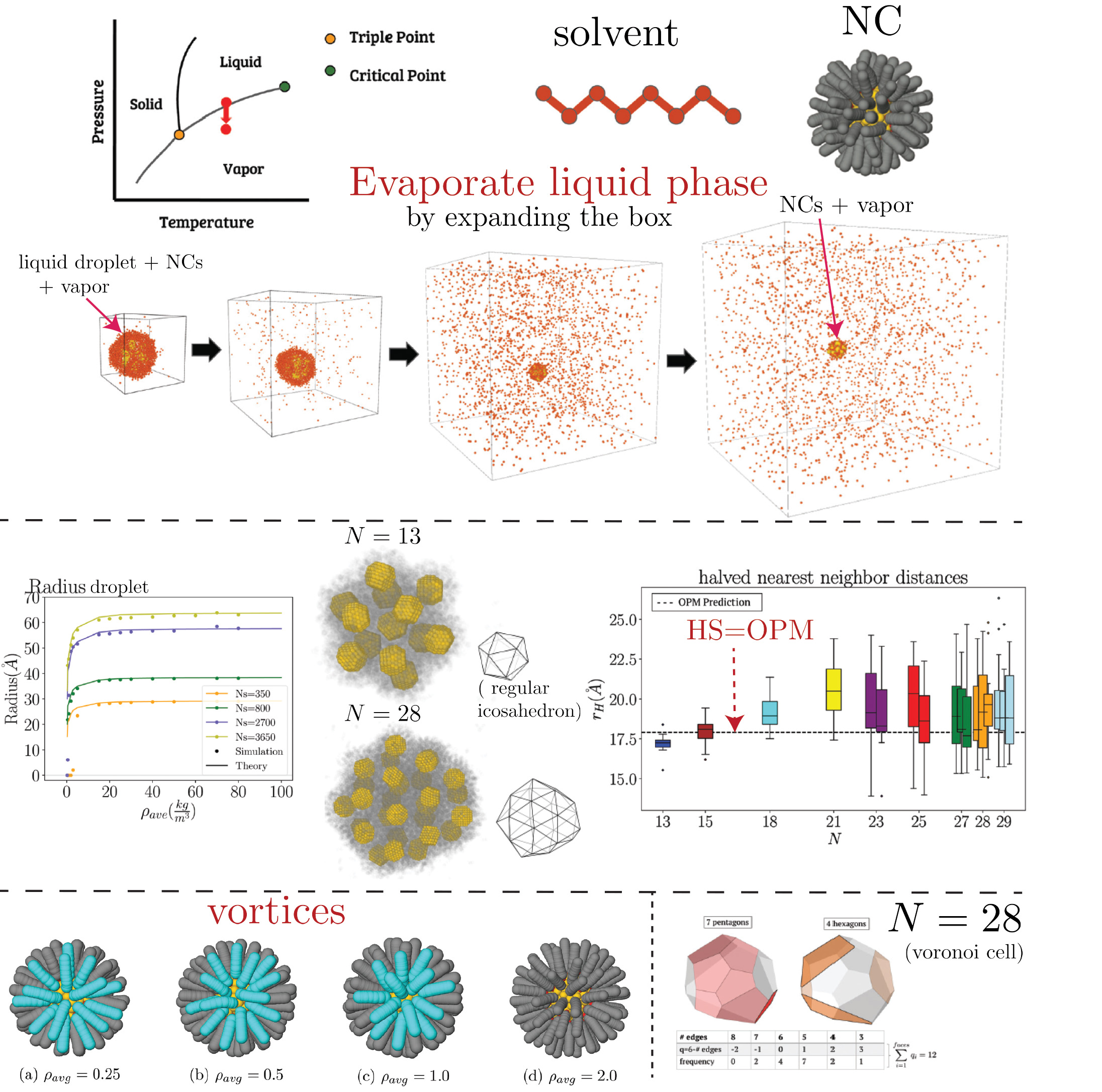}
    \caption{Top: Illustration of the evaporation process: a droplet containing $N$ NCs is quasistatically evaporated until the liquid phase completely disappears, leaving behind an equilibrium NC cluster. Middle: Plot of NC radius as a function of $\rho_{avg}$, defined in Eq.~\ref{Eq:evap:rho_ave}. Example of equilibrium clusters for $N=13, 28$ and summary of nearest-neighbor distance for NCs within the interior of a cluster. Bottom: Vortices as a function of solvent concentration and a Voronoi cell for one of the interior NC at $N=28$.}\label{fig:evaporate}
\end{figure}

\section{Dynamics of NC Assembly}\label{Sect:Dynamics}

\subsection{Solvent evaporation models}

Solvent evaporation for point(structureless) particles has been investigated using implicit \cite{WangBrady2017,HowardNiko2018} and explicit\cite{ChengGrest2012,ChengGrest2013,Howard2018,LiuMidya2019} models. These results are of direct relevance to actual solvent evaporation experiments as there are many aspects that are independent of the microscopic details of the particles. However, there are important effects related to the ligands, which require more fine-grained descriptions.

In order to reveal specific effects caused by the ligands, we have carried out simulations at the united atom level with both  toluene\cite{Waltmannn2019} and octane\cite{Macias2020} as solvent. Due to space and time limitations, these simulations included only a relatively small number of NCs $N<30$. The average density is defined as
\begin{equation}\label{Eq:evap:rho_ave}
    \rho_{avg}=M_{sol} \frac{N_{sol}}{V} \ ,
\end{equation}
where $M_{sol}$ is the molecular weight of the solvent and $N_{sol}$ the number of solvent molecules within a simulation box of volume $V$.

The main process is described in Fig.~\ref{fig:evaporate}. Initially, the system consists of a solvent droplet with $N$ NCs inside, coexisting with the vapor. Because this is a good solvent for the ligands, the solution is stable and NCs repel each other within the solvent. The simulation box is quasi-statically expanded until the liquid droplet is fully evaporated and an equilibrium with $N$ NC cluster is left. These clusters undergo a crystallization process reminiscent of the entropic crystallization in HS as the volume fraction (or PF) increases. Analytical formulas describing the droplet radius as a function of $\rho_{avg}$, defined in Eq.~\ref{Eq:evap:rho_ave}, have been derived\cite{Waltmannn2019} and are in excellent agreement with the simulation results, as shown in Fig.~\ref{fig:evaporate}.

\begin{figure}
    \centering
    \includegraphics[width=1\textwidth]{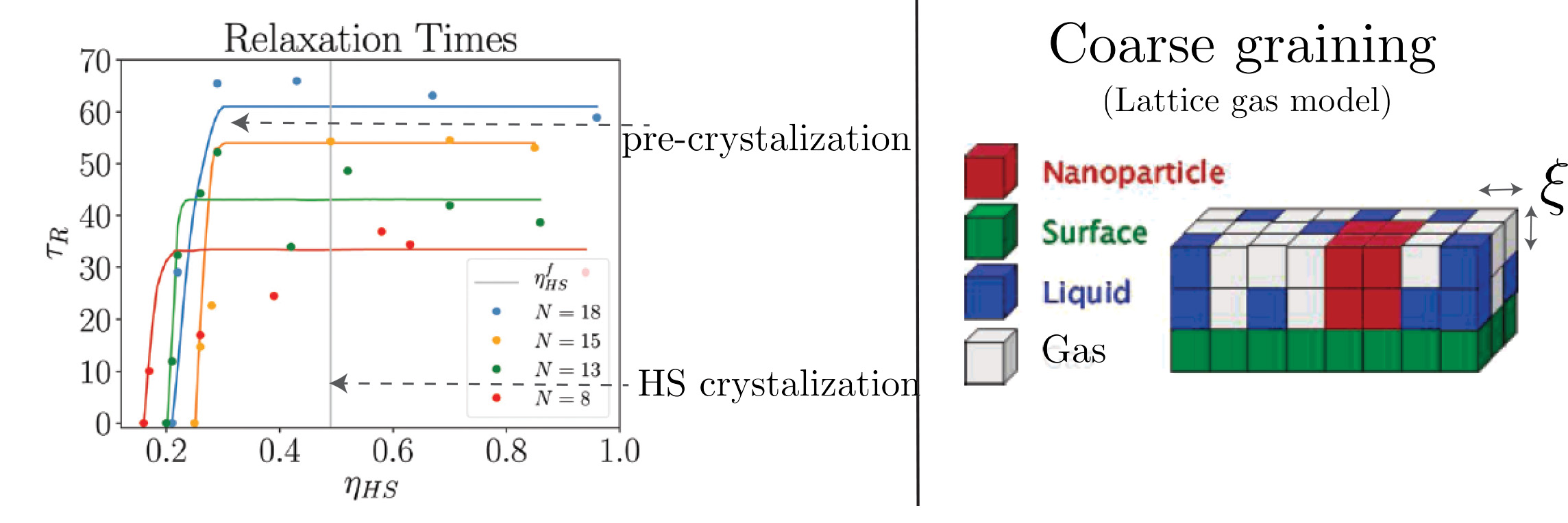}
    \caption{Relaxation times $\tau_R$ for different $N$ clusters as a funcion of PF. A clear transition is apparent at about 0.5 PF. The lattice-gas evaporation model defined in Ref.~\cite{Rabani2003, Sztrum2005} provides a description of the long time scales.}\label{fig:relaxation}
\end{figure}

These simulations allow for a rigorous determination of the relaxation times $\tau_R$ involved in the evaporation process. For a sufficient amount of solvent (more than 50\% volume fraction, similar to HS), the relaxation times are consistent with the Maxwell evaporation model\cite{Fuchs1959}, which assumes that evaporation is a diffusion-limited process of the surrounding vapor phase. However, Maxwell theory breaks down at lower solvent content, where relaxation times increase in a way that is dependent on the total number of NCs within the droplet, as shown in Fig.~\ref{fig:relaxation}.

\subsection{Equilibrium structures from solvent evaporation}

The process described in Fig.~\ref{fig:evaporate} has been carried out for droplets containing up to $N=30$ NCs. Examples of the final, fully evaporated structures are shown for two cases, $N=13, 28$ with the other cases $N<30$ available in the original reference\cite{Macias2020}. For $N=13$ the structure consists of a central NC surrounded by 12 other NCs sitting at the vertices of the icosahedron. For $N=28$, the structure cannot be identified as corresponding to any known polyhedra. It contains 3 NCs inside the cluster and 25 outside. The Voronoi cell for each of these three interior NCs consists of polyhedra that approximate an icosahedron\cite{Macias2020}, as shown in Fig.~\ref{fig:evaporate} where the Voronoi cell contains 7 pentagons of the 12 that form an icosahedral cell. 

The nearest-neighbor distance of interior NCs within the $N$-clusters $N>12$ are shown in Fig.~\ref{fig:evaporate}. Within error bars, their radius is described by HS, marked with a dashed line, as expected for NCs with coordination larger than six. As pointed out in the original reference\cite{Macias2020}, the nearest-neighbor NC distances of the clusters obtained after solvent evaporation are in perfect agreement with the minimum of the N-PMF in completely dry systems, thus proving that at least for small NC systems, solvent evaporation reproducibly leads to the minimum of the free energy and therefore equilibrium, i.e. there is no evidence for metastability.

The evaporation simulations enable to visualize how vortices are formed. This is illustrated at the bottom of Fig.~\ref{fig:evaporate} as the solvent content is reduced, quantified by $\rho_{avg}$ defined in Eq.~\ref{Eq:evap:rho_ave}. The radius of the droplet in Fig.~\ref{fig:evaporate} as a function of $\rho_{avg}$ shows that the appearance of vortices is sudden and occurs in the last stages of the drying process, where the solvent content is minimal, around $\rho_{avg} \sim 1$. Work is in progress to describe similar processes in superlattices and to characterize the intermediates to the final equilibrium structure.

\subsection{Are superlattices in thermal equilibrium?}

The large interaction free energy, of the order of (at least) $100 k_B T$ per NC has prompted many spirited debates on whether dry superlattices are true equilibrium or just metastable. A common argument is that {\em if we had a system of particles whose interaction potential is attractive and of the order of $100 k_B T$ or more, then such a system would agglomerate or quench and never reach equilibrium}. This argument, if true, would apply to the evaporation simulations in Fig.~\ref{fig:evaporate}, but as I have shown, N-clusters are reproducible and minimize the free energy. Basically, the transition from repulsive NCs when solvent is abundant to 100$k_BT$ attraction when solvent is absent occurs slowly enough so that attraction is gradual \resp{and NCs have plenty of time to reorganize and explore all available microstates}. This is implied from Fig.~\ref{fig:relaxation}, where at PF approximating the HS crystallization of 50$\%$, relaxation times increase and there is a weak NC attraction, very far from the 100 $k_BT$ of the dry final state.

The follow-up argument is then that {\em if a metastable structure is reached at some point for sufficiently high PF then the large cohesive energies will prevent the system from reaching the actual equilibrium state}. In other words, the experimentally observed superlattices reflect a metastable state with a low activation energy rather than the true equilibrium phase. The example of bct as an intermediate has been discussed in detail in SubSect.~\ref{Subsect:MOLT}. Experimentally, intermediate states have been characterized\cite{Bian2011,Choi2012,Weidman2016,Smilgies2017,Lokteva2019, WangChen2021}, so this is a real possibility. However, current evidence does not favor metastability, at least for most of the cases that have been investigated in detail; In addition to evaporation on a substrate, microfluidics\cite{Bodnarchuk2011} or emulsification\cite{WangMin2018,YangWang2018,XueYan2019} for example, all lead to the same reproducible structures, despite (or because) the intermediates; see Ref.~\cite{WangChen2021}. I will add that this argument is not specific to NC assembly: it can also be applied to basically any crystallization process: simple carbon, for example, whose cohesive free energy (in diamond or graphene) is 100 times or even larger than $k_BT$ at room temperature.

It is certainly the case that at some point in the evaporation process the cohesive energy becomes too large for the system to experience a phase transition to another crystal structure. However, the main point is that this statement being true does not imply that the system will be in a metastable state. At that solvent content where the cohesive free energy exceeds many $k_BT$ s, it may already have reached a partially swollen equilibrium structure, isostructural to the final dry equilibrium, or the intermediate may have an intrinsic instability at a certain solvent content and undergo a solid-solid transition. This is a difficult problem and will require further studies, such as simulations and mean-field calculations with MOLT\cite{Missoni2020, Missoni2021} shown in in Fig.~\ref{fig:molt} for example.  Still, I insist that this is a general problem intrinsic to many crystallization processes and it is not specific to NC assembly. 

I now make an argument showing that the free energy cohesive energies do not imply irreversible or quenched long time dynamics. The lattice gas model developed in Ref.~\cite{Rabani2003,Sztrum2005} has been shown to accurately model long time and spatial scales, i.e. coarsening, in solvent evaporation processes. It is schematically shown in Fig.~\ref{fig:evaporate}: it approximates the space as a lattice occupied by liquid or gas solvent (blue or gray), nanoparticle (red), and it is possible to include an interface (green) although this is not necessary for the argument here. The Hamiltonian has the form
\begin{equation}\label{Eq:evap:lattice_gas}
    H=-\varepsilon_{l} \sum_{\langle i,j \rangle} l_i l_j -\varepsilon_{n} \sum_{\langle i,j \rangle} n_i n_j -\varepsilon_{n,l} \sum_{\langle i,j \rangle} n_i l_j-\mu \sum_{i} l_i
\end{equation}
where $l_i=0$ states that there is gas and $l_i=1$ liquid at site $i$. The sites occupied by NCs are labeled $n_i=1$ or $n_i=0$ otherwise. The quantity $\varepsilon_{l}$ represents the free energy of attraction between the liquid solvent and $\mu$ is the liquid chemical potential. The coexistence gas-liquid is defined by the condition that the magnetic field term in the Ising model equivalent to Eq.~\ref{Eq:evap:lattice_gas} is zero, which leads to  $\mu=-\frac{z}{2} \varepsilon_l$ (with z=4 for 2d square, z=6 for 3d cubic, etc..). Therefore, simulations are carried in the gas phase, that is $\mu < -\frac{z}{2}$ so that the solvent does evaporate. The parameter $\varepsilon_{np}$ defines the NC-liquid interaction.

The parameter $\varepsilon_{nn}$ defines the NC-NC interaction in the dry state and can therefore be related to the bonding free energy. Typical simulations use $\varepsilon_l = 4 k_BT$ and $\varepsilon_n = 2 \varepsilon_l = 8 k_B T$\cite{Rabani2003} or $\varepsilon_n= 4 k_BT$ \cite{Sztrum2005} . Let us compute the putative free energy of a fcc superlattice. If I assume that $\xi = 1$ nm, see Fig.~\ref{fig:relaxation}, then a 5nm NCs will include 25 of these cubes per face (as opposed to the $2\times2$ in Fig.~\ref{fig:relaxation}). Therefore the cohesive energy will be
\begin{equation}\label{Eq:evap:cohesive}
    F_{cohesive} \approx \frac{12}{2} \times 25 \times \varepsilon_n = 6 \times 25 \times (4-8) k_B T = 600-1200 k_B T \ ,
\end{equation}
where the factor of 12 is the coordination for an fcc lattice.
This result is entirely consistent with the fcc cohesive free energies in Fig.~\ref{fig:simul_sc} although the numbers are somewhat dependent on the actual value of the coarse grained linear length $\xi$ in nm.

The relevance of this discussion and the lattice-gas model in Eq.~\ref{Eq:evap:lattice_gas} is that a large cohesive energy such as in Eq.~\ref{Eq:evap:cohesive} does not imply that the system leads to irreversible dynamics, as the lattice-gas model does not lead to quenched dynamics and random agglomeration. 

\begin{figure}
    \centering
    \includegraphics[width=1\textwidth]{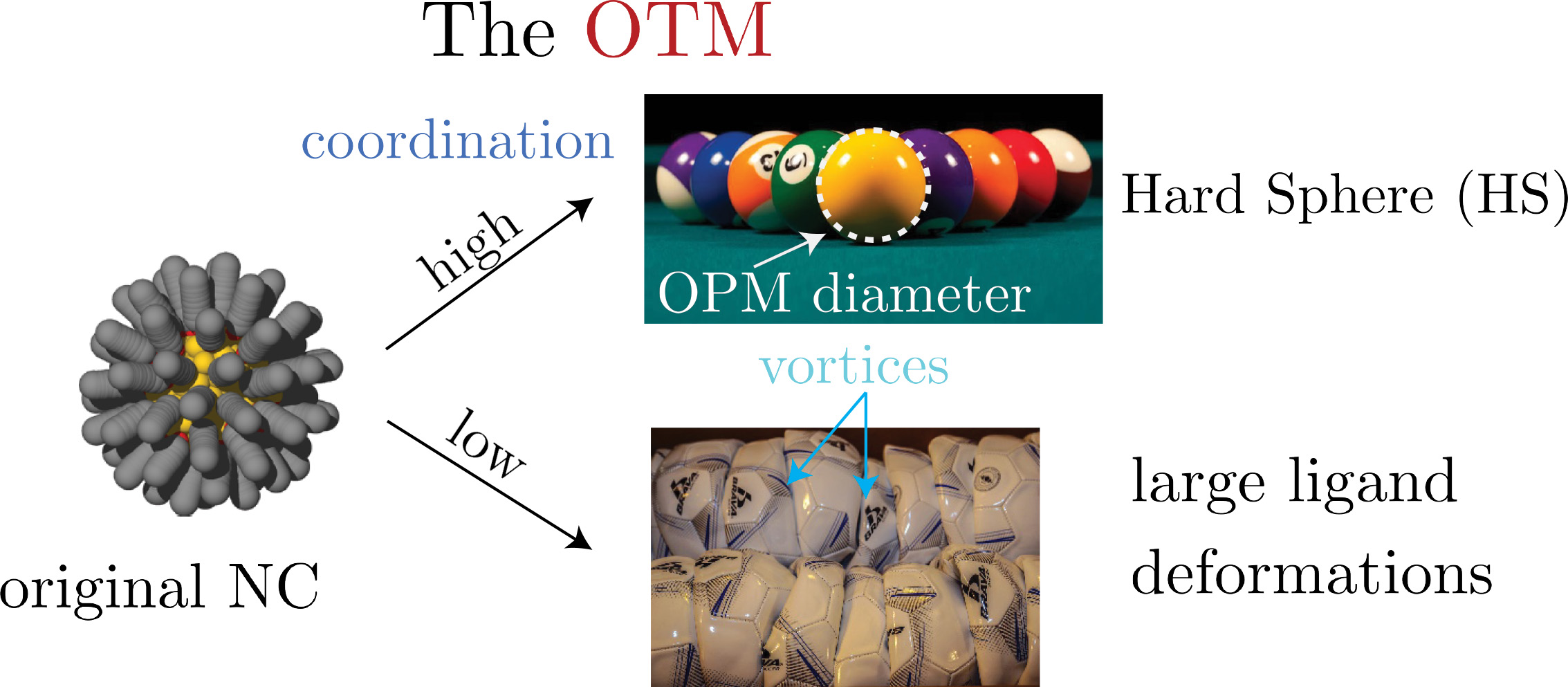}
    \caption{Following the analogy in the introduction with billiard balls, OTM establishes that at low coordination (basically less than six) NCs are more like deflated soccer balls, where vortices naturally define valence and eigenshapes, see Fig.~\ref{fig:eigen}. For large coordination, NCs are HS with diameter defined by the OPM formula Eq.~\ref{Eq:hs:OPM}. }\label{fig:otm_summary}
\end{figure}

\section{Conclusions}\label{Sect:Conclusions}

This article shows that the Orbifold Topological Model (OTM)\cite{Travesset2017a, Travesset2017b} provides a unification of \resp{hard shape} and OXM (X=P,C) models as well as numerical simulations, into a general framework, summarized in Fig.~\ref{fig:otm_summary}.  OTM predictions are independent of the details of the NC interaction (or the force field in numerical simulations) and have been extensively and thoroughly verified throughout the paper; see, for example, Sect.~\ref{Sect:OTM} and Sect.~\ref{Sect:Simuls}, without a single experiment or simulation in disagreement. The most significant limitation is the lack of an explicit free energy, which is obviously needed to predict equilibrium structures. Recent developments with MOLT\cite{Missoni2020, Missoni2021}, see SubSect.~\ref{Subsect:MOLT}, provide a promising direction toward fixing this deficiency.

The paper shows that the many successes of HS models reflect the underlying universality in packing problems, with its precise connection through the OPM model\cite{Landman2004} as discussed in Sect.~\ref{Sect:HS}, see also Ref.~\cite{Zha2020}. Yet, the failures of \resp{hard shape models highlight} that NC systems are far more complex. For example, an fcc phase in HS is possible because it maximizes the entropy, yet in real NC systems, stabilization of any superlattice, fcc or otherwise, involves a precise balance between enthalpy and entropy, see the discussion in Eq.~\ref{Eq:hs_thermo}, Eq.~\ref{Eq:real_thermo} and Sect.~\ref{Sect:Simuls}, with the former almost always larger in magnitude. Generally, NC assembly is not entropically driven (and the largest contribution to the entropy comes from ligand conformations). This point is one of many warning examples against overusing \resp{hard shape} results (and language) in interpreting experimental results.

The OTM is a completely general model that applies to any (dry) NC system with arbitrary ligands. The examples discussed in this paper have extensively used single-component and binary superlattices in two- and three-dimensional environments, mostly with spherical and cubic NCs. There are other cases where more detailed comparisons with experiments will soon become available, such as in quasicrystals\cite{Talapin2010}, three-component superlattices\cite{Dong2011,Boneschanscher2014,Cherniukh2021}, partially grafted NCs\cite{Whitham2016, Boles2019, Kavrik2022}, unbound ligands\cite{Winslow2020}, the beautiful Moire patterns recently reported\cite{Moretti2023} or NCs driven by different stimuli\cite{Grzelczak2019}, among many other examples. Furthermore, systematic studies on the role of NC shape\cite{Weiner2015}, see Fig.~\ref{fig:brush}, should help clarify why the approximation of spherical NCs works so well, even for big, i.e of the order of $8-10$ nm or larger, NCs\cite{Waltmann2018}.

\resp{Simulations of solvent evaporation with all atom and united models, summarized in Sect.~\ref{Sect:Dynamics}, see also Fig.~\ref{fig:evaporate}, are completely reproducible and the configurations obtained in the dry limit, i.e. after all solvent has evaporated, are structurally identical to those obtained after equilibrating NC in vacuum, thus providing clear evidence that equilibrium is reached. Further arguments discussed here make a strong case that solvent evaporation leads to systems in thermodynamic equilibrium. In Ref.~\cite{UpahTravesset2023} it has been shown that despite the large magnitude of the free energies in the dry limit, of the order of several hundred of $k_BT$s see for example Fig.~\ref{fig:simul_clusters}, Fig.~\ref{fig:simul_sc} or Fig.~\ref{fig:bnsl_sims} equilibration is made possible because monomer-monomer interactions, i.e CH$_2$, CH$_3$, S, etc.. are of the order of a fraction of $k_BT$ and therefore, the ligands (and solvent) bind and unbind sufficiently often within the time of a measurement to explore the relevant microstates. Still, it is conceivable that some NC superlattices might be locked in metastable states, although current evidence does not seem to support it. Future research will be needed to establish in which cases metastability may occur in solvent evaporation.}

The role of electrostatic interactions is one of the main topics left out. There are many examples of solvent evaporation with water that contains different salts as a solvent. For example, tetrahedra assembly\cite{ZhouTravesset2022, WangChen2022}
The final structure after solvent evaporation is free of water, but salt remains, which will have a measurable effect on the lattice constant and will modify the OPM formula Eq.~\ref{Eq:hs:OPM}. Additionally, the long-range effect of electrostatic interactions may lead to additional nuances, as I \cite{KimNayak2023, Macias2023} and many others, see for example, Ref.~\cite{kung2010a, Kewalramani2016, Cats2021} have discussed before. Electrostatic driven assembly strategies are a very important topic\cite{Bian2021, KimNayak2023}, as at this point in time it is the \resp{most promising strategy} that leads to the assembly of NC with electrically conducting ligands\cite{Coropceanu2022}.

Another aspect mentioned but not elaborated on is the role of icosahedral motifs, closely related to Frank-Kasper phases\cite{Frank1958,Frank1959}. The simulations presented in Sect.~\ref{Sect:Dynamics} reveal a strong bias of small NC clusters towards icosahedral order. Systematic investigations with much larger clusters including thousands of NCs display icosahedral-order motifs before attaining their bulk, presumably equilibrium, superlattice\cite{Nijs2015,WangDasgupta2020,Mbah2023}. There is also clear evidence that in binary systems icosahedral order clusters have lower activation barriers during superlattice nucleation\cite{Coropceanu2019}, which has significant implications for glass states\cite{Tarjus2005, Turci2017}. Ref.~\cite{Travesset2017} showed that all BNSL experimentally reported so far can be regarded as generalizations of Frank-Kasper phases, i.e quasi Frank-Kasper phases. Ref.~\cite{Serafin2021} has used icosahedral order to characterize the pitch in chiral NC structures. Methods have been developed \cite{Lindquist2018b} to inverse design Frank-Kasper phases. In single-component systems the main obstacle, however, is that HS Frank-Kasper phases have a PF that is too low to become generally stable. This discord between icosahedral order and packing is fixed in binary systems particularly around $\gamma \approx 0.8$\cite{Travesset2017}: all Laves phases, for example, are Frank-Kasper phases. The ubiquitous presence of icosahedral order  and Frank-Kasper phases in NC assembly is an important topic that will definitely require a detailed review on its own.

The discussion in this paper has been restricted to equilibrium and dynamics in dry systems, but, as illustrated in Fig.~\ref{fig:intro}, there are many BNSL phases: AlB$_2$, bccAB$_6$ (also labeled Cs$_6$C$_{60}$) that appear in both dry and DNA assembly\cite{Macfarlane2011}. When ligands are flexible, such as single-stranded DNA\cite{Vo2015} or nanocomposite tectons\cite{ZhangSantos2016} there are fewer phases shared with dry systems. \resp{Currently, OTM has been developed for the dry systems, and a generalization will be required to describe swollen systems, as the ligand conformations are different}, but Fig.~\ref{fig:intro} illustrates that there are important relations and perhaps a more universal description of the equilibrium superlattices.

There is a rich and beautiful literature on polymer melts and brushes\cite{RubinsteinBook2003}.  SubSect.~\ref{SubSect:Simuls:polymer} has presented results showing that OTM also holds for longer ligands that reach the polymer limit. However, in my opinion, there is still a significant disconnect between the methods and language in NC assembly and the many beautiful developments available from polymer physics. \resp{Mean-field results, such as block copolymer melts in the strong segregation limit\cite{Semenov1985, Milner1988, Milner1988a, Olmsted1998, Matsen2005, Dimitriyev2021}, or other models like Ref.~\cite{Ginzburg2017, Midya2021} should describe NC assembly for very long ligands, such as discussed in Ref.~\cite{Midya2021}; It seems reasonable to expect that a NC functionalized with a long polymer ligand is equivalent to a diblock, where one of the components is much shorter than the other, in the strong segregation limit\cite{Semenov1985, Milner1988a} so that the minority component forms a spherical micelle. In that case, the short concave polymer component plays the role of the NC and the longer convex polymer plays the role of the ligand. Also, } NC assembly provides a strategy for homogeneous NC dispersions, a problem of great interest within the polymer community\cite{Kumar2013,
KumarGanesan2017}. \resp{The key aspect for this connection is the incompressibility condition discussed in SubSect.~\ref{Subsect:MOLT}}. Finally, many aspects of Frank-Kasper phases in polymer
melts\cite{Grason2003,Grason2005a,Lee2014,XieLi2014,Reddy2018} expose general geometric principles common to all soft-matter systems that should be relevant for NC assembly where icosahedral order is very present.

NC assembly has experienced extraordinary progress over the past two decades and the future looks even brighter. These are exciting times for both theory and experiment. The combined sophistication of theory and the ever-expanding power of numerical simulations anticipate a future where new NC materials will be led by computation. I hope that one of the contributions of this paper is to help advance towards this very exciting future.

\section{Acknowledgements}

First and foremost, I would like to thank all students in my group who have contributed to this work. This paper would not be possible without them. I also would like to thank many researchers for educating me in this fascinating field, especially: M. Boles, C. Calero, Q. Chen, I. Cherniukh, M. Engel, O. Gang, H. Guo, M. Kovalenko, R. Macfarlane, S. Mallapragada, M. Olvera de la Cruz, E. Rabani, M. Tagliazucchi, D. Talapin, D. Vaknin, W. Wang, X. Ye, S. Zhou and Gerard Wong for inviting me to prepare this paper and to Shan Zhou and Dmitri Talapin for permission to use their experimental figures. \resp{I would like to thank I Cherniukh for a critical reading of the manuscript and for providing many corrections.} Finally, a big thanks goes to all the participants in the KITP Santa Barbara workshop and conference for the amazing environment that motivated this paper. The main funding for this research is from the U.S. Department of Energy (U.S. DOE), Office of Basic Energy Sciences, Division of Materials Sciences and Engineering. Iowa State University operates Ames National Laboratory for the U.S. DOE under Contract DE-AC02-07CH11358. Research is also supported in part by the National Science Foundation under Grant No. NSF PHY-1748958 to the Kavli Institute for Theoretical Physics. This work used EXPANSE, BRIDGES2 and DELTA through allocation MCB140071 from the Advanced Cyberinfrastructure Coordination Ecosystem: Services  Support (ACCESS) program\cite{Boerner2023}, which is supported by National Science Foundation grants 2138259, 2138286, 2138307, 2137603, and 2138296.

\end{document}